\documentstyle[12pt,aaspp4]{article}

\begin{document}

\title{ASCA Observations of Seyfert 1 galaxies: \\ III. 
The Evidence for Absorption \& Emission due to Photoionized Gas}

\author {I.M. George \altaffilmark{1,2}, 
T.J. Turner \altaffilmark{1,2}, 
Hagai Netzer \altaffilmark{3},
K. Nandra\altaffilmark{1,4}, 
R.F. Mushotzky\altaffilmark{1},
T. Yaqoob\altaffilmark{1,2}}

\altaffiltext{1}{Laboratory for High Energy Astrophysics, Code 660,
	NASA/Goddard Space Flight Center,
  	Greenbelt, MD 20771}
\altaffiltext{2}{Universities Space Research Association}
\altaffiltext{3}{School of Physics and Astronomy and the Wise Observatory,
        The Beverly and Ramond Sackler Faculty of Exact Sciences,
        Tel Aviv University, Tel Aviv 69978, Israel.}
\altaffiltext{4}{NAS/NRC Research Associate}

\slugcomment{Resubmitted to {\em The Astrophysical Journal Suppl}
(manuscript 36439)}

\begin{abstract}

We present the results from a detailed analysis of the 0.6 - 10 keV spectra 
of 23 {\it ASCA} observations of 18 objects.
We find that in most cases the underlying continuum 
can be well-represented by a powerlaw with a photon index $\Gamma \sim 2$.
However we find strong evidence for photoionized gas in the line-of-sight
to 13/18 objects. 
We present detailed modelling of this gas 
based upon the {\tt ION} photoionization code. Other 
studies have been made of the 'warm absorber' phenomenon 
but this paper contains the first 
consideration of the importance of the covering-fraction of the 
ionized gas and a direct 
comparison between models of attenuation by ionized versus 
neutral material. 

We find the X-ray ionization parameter for the ionized material is 
strongly peaked at $U_X \sim 0.1$. 
The column densities of ionized material
are typically in the range $N_{H,z} \sim 10^{21}$--$10^{23} {\rm cm}^{-2}$, 
although highly ionized (and hence psuedo-transparent) column densities 
up to $10^{24} {\rm cm}^{-2}$ cannot be excluded in some cases.
We also investigate the importance of 
the emission-spectrum from the ionized gas, finding that 
it significantly improves the fit to many sources with an intensity
consistent with material subtending a large solid angle at the 
central source.
Allowing a fraction of the continuum to be observed without
attenuation also improves the fit to
many sources, and is definitely required in the case of NGC~4151.
A deficit of counts is observed at $\sim 1$~keV in the sources 
exhibiting the strongest absorption features. We suggest this 
is likely to be the signature of a second zone
of (more highly) ionized gas, which might have been seen previously
in the deep Fe $K$-shell edges observed in some {\it Ginga} observations.
We find evidence that the ionized material in NGC~3227 and MCG-6-30-15 
contains embedded dust, whilst there is no such evidence in the other sources

We discuss these results in the context of previous studies and briefly 
explore the implications in other wavebands.

\end{abstract}

\keywords{galaxies:active -- galaxies:nuclei -- galaxies:Seyfert --
X-rays:galaxies}

\clearpage
\section{INTRODUCTION}
\label{sec:intro}

Prior to the launch of {\it ASCA}, the paradigm for Seyfert 1 galaxies,
was that the 2--10~keV regime can be well represented by a 
powerlaw continuum of (photon) index $\Gamma \sim 1.9$ 
(e.g. Nandra \& Pounds 1994). 
Spectra obtained using the Large Area Counter (LAC) onboard 
{\it Ginga} also revealed emission line features 
superimposed on this underlying 'primary' continuum
attributable to Fe $K$-shell fluorescence in the 6.4--7.1~keV 
band. A flattening of the spectra was observed above $\sim$ 10 keV 
thought to be the 'Compton-reflection hump'. 
Evidence for a $K$-shell absorption edge in the 7.1--8.9~keV 
band due to highly-ionized Fe was also reported in a number of cases 
(Nandra \& Pounds 1994). 
The combination of these features offers an explanation as to why 
simple powerlaw models to the data obtained by earlier instruments 
typically revealed flatter ($\Gamma \sim 1.7$) spectra
(e.g. Turner \& Pounds 1989).
There are a few exceptions to this rule, most notably 
NGC~4151 which exhibits a continuum noticeably flatter than the 
average ($\Gamma \sim 1.5$) and absorption within a substantial 
column of gas ($\sim 10^{22}\ {\rm cm^{-2}}$) along the 
line of sight covering most (but possibly not all) of the central continuum
(e.g. Yaqoob, Warwick, Pounds 1993 and references therein).

This picture has been generally supported by {\it ASCA} 
observations, with the addition that the higher spectral resolution afforded by 
the onboard instruments have shown the Fe $K\alpha$ line to be 
significantly broadened (Mushotzky et al. 1995; Tanaka et al 1995).
However, the situation in the soft ($< 1$~keV) X-ray regime is somewhat 
less clear.
Historically, a confusing variety of spectral and temporal properties have 
been seen and/or implied in Seyfert 1 galaxies in the soft X-ray regime.
A number of sources seemed to show a steep and rapidly variable 
'soft excess' which could be characterised by either a low-temperature
thermal component (e.g. Arnaud et al. 1985) 
or a steep second powerlaw 
(e.g. Turner \& Pounds 1989; Turner, George \& Mushotzky 1993). 
Data obtained using the solid-state spectrometer (SSS) onboard the 
{\it Einstein Observatory} showed evidence for a complex spectral form 
in the soft X-ray regime (Turner et al 1991), in some cases this was 
thought to be
due to leakage of the primary continuum through
a patchy absorber (e.g.  Reichert et al. 1985). 

Evidence for absorption by ionized material along the line-of-sight to 
active galactic nuclei (AGN) 
was first obtained in observations of the QSO MR2251-178.
Using data from the Monitor Proportional Counter (MPC) onboard 
the {\it Einstein Observatory},
Halpern (1984) reported the column density in this source to have 
increased by a factor $\gtrsim4$ between two observations separated by 
$\sim1$ yr. The excess flux observed in a subsequent observation 
using the High Resolution Imager (HRI) on the {\it Einstein Observatory}
led Halpern to suggest an explanation in terms 
of material with a column density $\sim 10^{22}\ {\rm cm^{-2}}$, 
but photoionized such as to be transparent below the $K$-edge of 
O{\sc vii} at 740~eV.
Such an interpretation was strongly supported by {\it EXOSAT}
observations of MR2251-178.
Pan, Stewart \& Pounds (1990) found an inverse correlation between 
the inferred column density and the source flux, and demonstrated that this 
was consistent with the behaviour expected from photoionized material
along the line-of-sight.
Similar behaviour was observed in NGC~4151 (Fiore, Perola \& Ramano 1990;
Yaqoob \& Warwick 1991) and MCG-6-30-15 (Nandra, Pounds \& Stewart 1990), 
establishing the effect for low and high 
luminosity sources.
Supporting evidence for such ``warm absorbers'' was 
provided by {\it Ginga} observations of emission-line AGN which revealed
iron $K$-shell features in some sources, suggestive of an origin in
ionized material with columns densities $\gtrsim 10^{23}\ {\rm cm^{-2}}$. 
Nandra \& Pounds (1994) 
find this to be a common occurrence, with 12 out of 27 sources in their
sample of Seyfert 1 galaxies and Narrow Emission Line Galaxies (NELGs) 
showing such a component.
However, the low spectral resolution of {\it Ginga} meant that the results
were likely to be sensitive to the assumptions concerning 
the adjacent Fe $K$-shell emission line.

Confirmation of the existence of warm absorbers
came from {\it ROSAT} Position Sensitive Proportional Counter (PSPC)
observations, which revealed absorption edges attributable to 
ionized oxygen in numerous sources (e.g. MCG-6-30-15, Nandra \& Pounds 1992; 
3C351, Fiore et al. 1993; NGC~3783, Turner et al. 1993a).
Early {\it ASCA}
observations (e.g MCG-6-30-15, Fabian et al. 1994, Reynolds et al. 1995; 
NGC~3783, George, Turner \& Netzer 1995) immediately confirmed the PSPC 
results and showed evidence for warm absorbers in many 
other sources (e.g. NGC~3227, Ptak et al 1994; NGC~4051, 
Mihara et al 1994; Guainazzi et al 1996; EXO~055620-3820.2, 
Turner, Netzer \& George 1996; Mrk~290, Turner et al 1996; 
PG1114+445, George et al. 1997a).
Furthermore Reynolds (1997) finds evidence for such a component in at 
least half of a sample of {\it ASCA} observations of 24 type-1
AGN.
The presence of highly-ionized, absorbing gas instrinsic to Seyfert 1 nuclei 
is further confirmed by the presence of 'associated absorbers' 
seen in the UV in many sources.
Discussion of these observations and 
the attempts to relate the UV and X-ray absorbers 
made to-date is postponed to \S\ref{Sec:XUVabso-disc}.

Comparison of {\it ASCA} data with the predictions of 
detailed photoionization calculations have revealed 
spectroscopic evidence for a second ionized absorber in 
NGC~3516 (Kriss et al. 1996a).
In addition the absorption edges due to ionized oxygen appear to vary in a 
way which is inconsistent with the behaviour of a single absorber
in at least some sources.
In MCG-6-30-15, the depth of the O{\sc viii} edge changes on a timescale 
$\sim 10^4$~s with no corresponding 
change in the O{\sc vii} edge (Reynolds et al 1995; Otani et al. 1996),
whilst the converse is true in NGC~4051 (Guainazzi et al. 1996).

The presence of emission lines at energies $\lesssim 1$~keV has been 
suggested in a number of previous observations (e.g. Turner et al 1991).
A blend of emission lines due to O{\sc vii} (0.568--0.574~keV) 
has been claimed in the {\it ASCA} data from NGC~3783 
(George, Turner \& Netzer 1995) and potentially seen in other sources 
(e.g. IC~4329A, Cappi et al. 1996).
However the reality of such features has since been disputed, 
because of the realization of specific inadequacies in the 
low energy calibration of the {\it ASCA} instrumentation 
(see \S\ref{Sec:calib_uncert}).

Here we present the third in a series of papers describing the X-ray properties
of a sample of 
18 Seyfert 1 galaxies, using data obtained by {\it ASCA}.
The sample is comprised of 23 {\it ASCA} observations of
Seyfert class 1.0--1.5 galaxies (as defined by Whittle 1992) 
performed prior to 1994 May 01. 
In Nandra  et al. (1997a, hereafter Paper~I), we
presented imaging and timing data for this sample. 
In that paper we demonstrate that all of the sources show evidence for 
variability on timescales of minutes-hours, with an 
amplitude anti-correlated with the source luminosity, confirming
previous results. We also showed that for a number of sources the 
variability amplitude is greater in soft X-rays ($\lesssim 2$~keV) than in
hard X-rays (for 10 of the 18 sources on timescales of $\sim1$~hr).
In Nandra  et al. (1997b, hereafter Paper~II), we presented the 3--10 keV 
spectra of the sample sources, and found 14 out of the 18 sources 
to contain an iron ${\rm K}\alpha$ line which is resolved in the 
{\it ASCA} SIS, with mean width $\sigma \simeq 0.4\pm 0.1$~keV for 
a Gaussian profile.
However, we found many of the line profiles to be asymmetric, 
suggesting an origin from the innermost regions of the 
putative accretion disk.

In this paper, 
we present the 0.6-10.0 keV spectra of the sample sources.
The extension of the spectral analysis down to 0.6 keV requires
consideration of the properties of the circumnuclear absorbing material 
as well as any steepening of the spectrum at low energies. We
present detailed modelling of absorption by photoionized 
material and discuss likely emission contributions as well as
combinations of several emission and absorption components.
We present the range of properties applicable to our sample, 
and discuss the characteristics and statistics of the ionized 
absorbers. We also consider the the importance of
allowing a fraction of the underlying continuum to be unattenuated
by ionized gas, and the inclusion of the 
emission features expected from the ionized material.
As will become apparent below, 
there is an obvious variation in signal-to-noise ratio between 
different datasets and hence there 
are variations in detectability of the ionized gas.
As in the two previous papers our intention is to study
the mean properties of the sample and
we only include detailed notes on 
individual sources as an Appendix.

We describe the observations and data reduction in \S\ref{sec:anal}, 
and the photoionization models
in \S\ref{Sec:ion_model}. 
Our spectral fitting procedure is introduced in \S\ref{Sec:procedure},
and the results presented in \S\ref{Sec:basic_models} and 
\S\ref{Sec:additional_models}.
In \S\ref{Sec:disc-multi} we 
discuss the objects for which there 
in more than one observation within our sample. 
Our results are discussed within the general context of the physical 
properties AGN in \S\ref{Sec:Discussion}.
We sumarize our overall findings in \S\ref{Sec:open_issues} along with
outlining a number of important issues that remain to be resolved.

\section{OBSERVATIONS AND DATA REDUCTION}
\label{sec:anal}

The {\it ASCA} satellite consists of four identical, co-aligned X-ray
telescopes (XRTs; Serlemitsos et al. 1995). Two solid-state imaging
spectrometers (known as SIS0 and SIS1), each consisting of four CCD chips, sit
at the focus of two of the XRTs, and provide coverage over the
$\sim$0.4--10~keV band (Burke et al. 1994). Two gas imaging spectrometers
(known as GIS2 and GIS3) sit at the focus of the other two XRTs,
and provide coverage over the $\sim$0.8--10~keV band (Ohashi et al. 1996
and references therein).
Further details on the satellite, its instrumentation and performance can be
found in Tanaka, Inoue \& Holt (1994), 
Makishima et al. (1996), and in the 
{\it ASCA Data Reduction Guide}\footnote{formerly 
known as the {\it ABC Guide to ASCA Data Analysis} and 
available via\\ 
\verb+http://heasarc.gsfc.nasa.gov/+}
and references therein.

\subsection{The Sample}
\label{Sec:sample}

Our sample is listed in Table~1 along with 
the redshift ($z$) of the source, 
the axial ratio ($a/b$) of the host galaxy,
the effective Galactic hydrogen column density along that 
line-of-sight ($N_{H,0}^{gal}$) as derived from 21~cm measurements,
a number of parameters indicative of the multiwaveband spectra of 
the objects, and X-ray luminosity in two bands.
The mean flux at 220~nm ($<f_{220}>$) in the rest-frame of the
source,
and the mean ratio of the flux at 125~nm (rest-frame) to $f_{220}$
for observations carried out within 48 hours of each other
from the {\it IUE} observations were calculated from 
Courvoisier \& Paltani (1992).
The errors quoted for $(f_{125}/f_{220})_{obs}$ and $<f_{220}>$ are 
1$\sigma$ or 10\%, whichever is the larger, and 
datasets flagged 'dubious' by Courvoisier \& Paltani were excluded.
The observed flux at 2~keV ($f_{\rm 2keV}$) in the rest-frame of the
source is from the 
best-fitting model presented in \S\ref{Sec:basic_models}
(in all cases the statistical error is larger than the differences 
in $f_{\rm 2keV}$ between different models). In the case of 
objects for which we have more than one dataset, $f_{\rm 2keV}$
is the exposure-weighted mean.
The radio--to--optical {\it energy} spectral index,
$\alpha_{ro}$, is calculated using $<f_{220}>$ and 
the observed flux at 5~GHz from Veron-Cetty \& Veron (1993).
The optical--to--X-ray {\it energy} spectral index,
$\alpha_{ox}$, is calculated using $<f_{220}>$ and $f_{\rm 2keV}$.
It should be stressed that these values of 
$\alpha_{ro}$ and $\alpha_{ox}$ are based non-simultaneous observations
and hence should be regarded with appropriate caution.
The errors associated $\alpha_{ro}$ and $\alpha_{ox}$ are rather difficult to 
determine without estimates of the typical amplitude of variability 
(on timescales of months--years) in each band.
Unfortunately, such amplitudes are not well determined for most of the objects 
considered here. 
However even assuming a factor 5 variability in each of the three bands 
(which is likely to be an overestimate in the radio and optical bands),
for the typical values of $\alpha_{ro} \sim 0.1$ and $\alpha_{ox} \sim 1.1$
derived for these objects, one obtains relatively small errors of 
$\sim \pm 0.1$ and $\sim \pm 0.2$ respectively.
Finally, Table~1 gives the derived 
luminosities\footnote{$H_0 = 50\ {\rm km\ s^{-1}\ Mpc^{-1}}$ and $q_0 = 0.5$ 
are used throughout} of the underlying 
continuum (i.e. correcting for all line--of--sight absorption) in the 
(rest-frame) 0.1--10~keV ($L_X$) and 2--10~keV ($L_{2-10}$) bands.
As in the case of $f_{\rm 2keV}$, these are calculated from the 
best-fitting model presented in \S\ref{Sec:basic_models}, and are 
exposure-weighted means in the case of objects with multiple datasets.
All 
fluxes and luminosities quoted for the sources will be derived using the 
SIS0 detector.

\subsection{Data Selection}
\label{Sec:data-selection}

The selection criteria and data analysis methods are presented in Paper~I. In
summary, we have used data from the US public archive at NASA/GSFC. Standard
criteria were applied to reject poor quality data and source spectra extracted
using the {\tt FTOOLS/XSELECT} package for each instrument. Background spectra
were extracted from source-free regions of the instrument.

In brief, 
we consider only the gain-corrected (Pulse Invariant or PI) channels for each
instrument\footnote{using \verb+sispi+(v1.0) and the gain-history file released 
on 1994 Jul 28 in the case of the SIS data}.
Furthermore, here we consider only the time-averaged spectrum for each 
instrument during 
each observation in order to maximize the signal-to-noise ratio. The raw
spectra were grouped such that each resultant channel had at least 20 counts
per bin, permitting us to use $\chi^2$ minimization in the spectral analysis.
All such fits were undertaken using the {\tt XSPEC} (v9.01) spectral analysis
package (Arnaud 1996). Appropriate data redistribution matrices were used
(those released 1994 Nov 09 for the SIS, and those released on 1995 Mar 06 for
the GIS). The effective area appropriate for each dataset was calculated using
\verb+ascaarf+\footnote{where, in the case of the SIS we used 
the Gaussian 'fudge', but not the filter 'fudge'} (v2.5). 
This does not include a parameterization of the
azimuthal dependence of the point spread function which can cause
inaccuracies in the normalization, particularly in the case where the source
region is not circular. Uncertainties in determining the source centroid can
also result in normalization errors, and discrepancies remain between the
absolute flux-calibration of the four instruments. These factors lead to
cross-calibration uncertainties between the four instruments, but as the
point-spread function of the XRT is not strongly dependent on energy, we 
do not expect them
to introduce distortions into the individual spectra. We therefore chose to
fit the spectra from all four instruments simultaneously, but allowed the
normalization of that model to be free for each detector.
For the datasets considered here, the two SIS detectors typically 
agree in flux to within
$\sim$1--3\%.
The normalization of
        the GIS detectors are currently forced to agree with the
        average of the SIS detectors for the Crab.

The observing log is presented in Table~2, where the UTC
calendar date is for the start of the observation. As in Papers~I \& II, we
refer to each observation using the name of the source and, in cases where
there are multiple observations, the observation number in brackets (using the
same numbering scheme\footnote{note that observations which
failed to meet our selection criteria are still assigned a number, but not
included here  -- see Paper~I for details}
as in Papers~I \& II).
Table~2 contains
the exposure time ($t_{exp}$), the count rate ($Rate$) in the 0.6--10~keV band,
and the total number of photons in the 0.6--5.0~keV and 7.0--10~keV bands, for
SIS0 (only) after all screening criteria have been applied ($N_{phot}$). 
Also listed in
Table~2 are the the total number of spectral bins ($N_{pts}$;
for all instruments) used in the analysis, and the additional number of
good-quality spectral bins in SIS0 and SIS1 below 0.6~keV 
($\Delta N_{0.6}$).
The latter two quantities enable the reader to judge the goodness-of-fit for
the results of the spectral analysis presented below. Finally in
Table~2 we also provide references to work published
previously on each observation.

\subsection{Calibration Uncertainties}
\label{Sec:calib_uncert}

We wish to fit several spectral features observed in the 0.4--10 keV range.
However, there is known to be some uncertainty in the low energy calibration
of {\it ASCA} that must be addressed first. While the origin of this problem
is currently unclear (Gendreau, p.comm), the 
manifestation of the problem is in the lowest channels of the SIS data. 
Specifically, the {\it ASCA} Guest Observer Facility at NASA/GSFC
reports that SIS spectra from a variety of objects of 
different astronomical classes consistently indicate uncertainties 
in the XRT/SIS effective area at energies
$<0.6$ keV, and especially around 0.5 keV.
This has been mainly due to
the inability to interpret soft X-ray features 
in the spectra from 3C~273 (the SIS/GIS
        cross-calibration target). However, recent simultaneous
        observations of 3C~273 with {\it ASCA} and 
{\it BeppoSAX} may improve this
        situation, and already indicate that the systematic error of
        the SIS relative to the low-energy concentrator system (LECS) on
{\it BeppoSAX} may be no more than $\sim 2 \times 10^{20} \rm \ cm^{-2}$
(Parmar et al 1997).
Further details on the low-energy calibration of the SIS are 
given in Dotani et al. (1996) and from the 
{\it ASCA} GOF at NASA/GSFC\footnote{
for up-to-date information see\\
\verb+http://heasarc.gsfc.nasa.gov/docs/asca/ascagof.html+}

Thus, here we have chosen to ignore all data below
0.6~keV during the spectral analysis, although we do compare the 
extrapolation of the various best-fitting models below 0.6~keV 
as a diagnostic of the quality of the fit 
(see \S\ref{Sec:acceptability}).
For the purpose of this paper, we do not 
use the 5.0--7.0~keV band data in the fitting process, 
the consistency of those data with the fits is discussed later. 
This approach avoids any requirement to include a complex parameterization 
of the iron K-shell line profile in our consideration of the fits to the 
broad-band continuum shape. 
Thus the spectral analysis was performed on simultaneous fits to 
the 0.6--5.0, 7.0--10.0~keV data. 
This provides the best method of determining the broad-band
X-ray continua. 
(Note that in all figures presented below which display data/model ratios, 
the corresponding ratios are included for the 
0.4-0.6 keV and 5.0-7.0 keV ranges 
to illustrate where these data lie with respect to the
corresponding model).

\section{THE PHOTOIONIZATION MODELS}
\label{Sec:ion_model}

The ionized absorber/emitter models used in this paper were generated using
the photoionization code {\tt ION} (Netzer 1993, 1996, version {\tt ION95}) 
to calculate the
physical state of a slab of gas when illuminated by an ionizing continuum.
The code includes all important excitation and ionization processes, full
temperature and radiative transfer solutions, and emission, absorption and
reflection by the gas are calculated self-consistently assuming thermal and
ionization equilibrium. 
Since the gas temperature is well below the equilibrium plasma temperature,
only a few transitions affect the cooling, all of which are included.
We assume an ionizing continuum typical of an AGN with a blue bump. 
Specifically we assume the 'weak IR' case of Netzer (1996), 
with a {\it photon} index $\Gamma_{o} = 1.5$ in the optical/UV band 
from 1.6--40.8~eV, and photon indices 
in the range $1.5 \leq \Gamma \leq 2.5$ in the X-ray band
from 0.2--50~keV. The continuum in the XUV band is 
also assumed to be a powerlaw connecting the fluxes at 40.8~eV and 
0.2~keV, such that the ratio of the fluxes at 250~nm and 2~keV
is always $f_{250}/f_{\rm 2keV} = 8.1\times10^{3}$ 
(corresponding to an optical--to--X-ray {\it energy} index 
$\alpha_{ox} = 1.5$ -- but see below).
Following
Netzer (1996), we assume undepleted 'cosmic abundances' (with 
(He, C, N, O, Ne, Mg, Si, S, Fe)/H =
($10^3$, 3.7, 1.1, 8, 1.1, 0.37, 0.35, 0.16, 0.4)/$10^{-4}$)
and a constant density of $n_H = 10^{10}\ {\rm cm^{-3}}$
throughout the slab.
However it should be noted that as discussed in Netzer (1996), 
for the same illuminating continuum, a wide range of densities
($10^{4} \lesssim n_H \lesssim 10^{10}\ {\rm cm^{-3}}$)
give rise to similar equilibrium solutions (and hence 
similarly shaped absorption/emission models). 
Thus under steady-state conditions
our models are insensitive to the precise value of $n_H$ assumed.
The ionization state of the gas 
is parameterized by the ionization parameter, $U_X$, defined
as 
\begin{equation}
\label{eqn:U_X}
U_X = \int^{10\ {\rm keV}}_{0.1\ {\rm keV}} 
\frac{Q(E)}{4 \pi r^2 n_{H} c} dE
\end{equation}
where $Q(E)$ is the rate of photon emission at energy $E$,
$r$ the distance from the source to the illuminated gas. 
In the past, ionization parameters (defined in various ways) involving an 
integration of the entire UV--X-ray continuum above 13.6~eV (with an upper 
limit often set to 13.6~keV) have been commonly used in the literature.
As pointed out in Netzer (1996), the use of $U_X$ is preferable in the study 
of the effects of ionized gas in the X-ray regime as the level of ionization 
of the dominant species is then directly proportional to $U_X$.
Of course the use of $U_X$ has the additional advantage that it is
independent of the form of the continuum assumed in the 
13.6--100~eV band which is unobservable due to absorption in our 
own galaxy for low redshift objects.
For convenience, the relationships
between $U_X$ and such ionization parameters 
are shown for a number of spectral forms in Fig.~1.
It should be stressed whilst the conversions between $U_X$ 
and these more traditional ionization parameters are obviously 
a strong function of the assumed continuum $< 0.2$~keV, the values derived 
for $U_X$ here are almost completely insensitive to 
such assumptions.

A series of such models were run for various $U_X$ and thicknesses
through an illuminated slab, the latter parameterized by the column density
of the gas, $N_{H,z}$. Two grids were produced for a number of
values of $\Gamma$ in the above range, and converted to a form able to
be used within {\tt XSPEC}. The first grid
consisted of only the absorption (with values between zero and unity) of
the incident continuum taking place within the illuminated slab. The
assumed ionizing continuum was multiplied by this grid during the spectral
analysis, and is referred to as the ionized-absorber below. 
The axes of this grid are spaced logarithmically over the ranges
$3\times10^{20} \leq N_{H,z} \leq 6.3\times10^{23}\ {\rm cm^{-2}}$ and 
$-3 \leq \log U_{X} \leq 1.0$. 
Such a model would be
applicable if the gas only covers the small fraction of the solid
angle surrounding the central source along the line-of-sight (the small
clouds case of Netzer 1993).

The second grid of spectra consisted of the continua
Compton-scattered by the slab (but including the effects of absorption
during its passage through the slab), 
along with the emission lines and recombination
continua produced within the gas. 
The normalizations, $A_{sh}$, of the spectra in this grid are
calculated by integrating the total emission emerging from the
shell. Since to first order this
emission is isotropic for the values of $N_{H,z}$ and $U_X$
considered here, $A_{sh}$ can be compared to $A_{pl}$ to give an estimate
of the fraction of the hemisphere required to be 
present in order to produce the observed emission, and hence
the solid angle subtended by the emitting gas at the ionizing source
$\Omega$ (assuming spherical symmetry).
Thus combining these spectra from the 
'ionized-emitter' with the effects of
the ionized-absorber on the underlying continuum enables us to investigate a
range of physically plausible conditions within, and location and geometry of,
the photoionized gas.

\section{MODEL FITTING}
\label{Sec:procedure}
\label{Sec:models}

The models considered in the spectral analysis include an underlying 
continuum which we assume to be a powerlaw
of photon index $\Gamma$ and normalization $A_{pl}$ at 1~keV.
In all cases the spectrum emerging from the source passes through 
a full screen of neutral material at zero redshift. This screen is
parameterized by an effective column density $N_{H,0}$ of hydrogen, which is
constrained to be $\geq N_{H,0}^{gal}$, the Galactic H{\sc i} column density
along that line-of-sight (Table~1).

We have performed the analysis assuming a 
variety of spectral models of varying levels of complexity. 
For the purposes of clarity, we present
the results in the following order:
\newcounter{model}
\begin{list}%
{\Alph{model}}{\usecounter{model}\setlength{\rightmargin}{\leftmargin}}
\item	The power-law continuum is absorbed by an additional screen 
	of neutral material 
	along the line-of-sight at the redshift of the source 
	(parameterized by an effective column density $N_{H,z}$ of hydrogen).
	This model therefore has 3 interesting parameters:
	$\Gamma$, $N_{H,z}$ and $N_{H,0}$.
\item	As for model {\it A}, but where the absorbing material 
	at the redshift of the source is ionized and parameterized 
	by $N_{H,z}$ and ionization parameter $U_X$ 
	(see \S\ref{Sec:ion_model}).
	This model has 4 interesting parameters:
	$\Gamma$, $N_{H,z}$, $U_X$ and $N_{H,0}$.
\item	As for model {\it B}, but the emission features expected from the 
	ionized gas (with $N_{H,z}$ and $U_X$) are included, 
	parameterized by $\Omega$, the solid angle subtended by the 
	emitting gas as seen at the central continuum source (see 
	\S\ref{Sec:ion_model}).
	This model has 5 interesting parameters:
	$\Gamma$, $N_{H,z}$, $U_X$, $\Omega$ and $N_{H,0}$.
\end{list}
For each model, the spectral analysis is performed under 
two assumptions: 
\newcounter{submodel}
\begin{list}%
{(\roman{submodel})}{\usecounter{submodel}\setlength{\rightmargin}{\leftmargin}}
\item	that the absorbing material at the redshift of the source
	completely covers the source, and that no continuum photons 
	are able to reach the observer via any other light paths.
	These models are labelled {\it A(i)}, {\it B(i)} and {\it C(i)}
	throughout.
\item	that a fraction $D_f$ of the power-law continuum is observed 
	without attenuation at the redshift of the source, whilst 
	$(1 - D_f)$ of the continuum is attenuated by material
        along the line-of-sight at the redshift of the source.
	Of course this is 
	mathematically equivalent to either an absorber within
        the 'cylinder-of-sight' which does not completely cover the central 
	source
	and/or to a geometry where the central source is
	completely covered by the absorber but in which some fraction of 
	the radiation emitted by the central source in other 
	directions is scattered back into the line-of-sight without 
	any spectral change.
	These models are labelled {\it A(ii)}, {\it B(ii)} and {\it C(ii)}
	throughout, and have $D_f$ as an additional interesting parameter.
\end{list}
We have applied all the models to all the 
datasets,
and tabulate all the results
even though the additional parameters in the more complex models 
may not be required by some datasets.
This enables us to present the limits on all the parameters 
with a view to unification schemes.
The results assuming models {\it A(i)}--{\it C(ii)} are presented in 
\S\ref{Sec:basic_models}.
A number of additional models are discussed 
in \S\ref{Sec:additional_models}, and notes on individual sources 
are given in the Appendix.
The ultimate goal is to 
classify objects according to the categories defined above and try to 
establish the fraction of spectra
that require the ionized gas component.

\subsection{Model Acceptability}
\label{Sec:acceptability}

We use a number of criteria to establish whether we consider a model to
be consistent with a given dataset.
During the spectral analysis we use the $\chi^2$-statistic to determine the
best-fitting model parameters. Thus this statistic is used in conjunction with
the number of degrees-of-freedom ($dof$) to determine the goodness-of-fit over
the energy range used during the analysis
({\it ie} 0.6--5, 7--10~keV). To this end, we use
the probability, $P(\chi^2 \mid dof)$, that the statistic should be less
than the observed value of $\chi^2$ under the assumption that the model is
indeed the true representation of the data. Thus $P(\chi^2 \mid dof) = 0.5$
corresponds to a reduced--$\chi^2$ value of unity, values of 
$P(\chi^2 \mid dof) \sim 1$ indicate that the data are poorly represented by
the model, and 
values of $P(\chi^2 \mid dof) \sim 0$ indicate that the model is most
likely an over-parameterization of the data.
For the purposes of this paper, we consider the model to be an adequate 
description of the data (in the 0.6--5, 7--10~keV band) if
$P(\chi^2 \mid dof)\leq 0.95$
(but flag occasions when $P(\chi^2 \mid dof) < 0.05$).

In addition, however, we also require that the best-fitting model extrapolates
into the bands excluded from the fit (i.e. $<0.6$~keV and in the 
5--7~keV band).
In the light of the uncertainties in the calibration of the instrument at
energies $<0.6$~keV (\S\ref{Sec:calib_uncert}), we quote
	the ratio of the increase in the $\chi^2$-statistic to the increase in
the number of data points 
	($\frac{\Delta \chi^2_{0.6}}{\Delta N_{0.6}}$),
and 
	the weighted mean of the data/model residuals
($\overline{R_{0.6}}$),
when the best-fitting model is extrapolated below 0.6~keV.
The rationale behind these two parameters is that whilst the calibration 
is suspect at these energies, it is considered unlikely to be in error by
$\gtrsim20$\%. The use of $\frac{\Delta \chi^2_{0.6}}{\Delta N_{0.6}}$ and
$\overline{R_{0.6}}$ therefore allows us to identify models, which are 
deemed acceptable $>0.6$~keV, but in which 
the extrapolation to energies $<0.6$~keV is inconsistent with 
the suspected size of the calibrations uncertainties.
For the purposes of this paper, we consider the model to 
extrapolate to energies $<0.6$~keV in an acceptable manner if
{\it either}
	$\frac{\Delta \chi^2_{0.6}}{\Delta N_{0.6}} \leq 2.0$
{\it or}
	$\overline{R_{0.6}}$ lies in the range 
	$0.8 \leq \overline{R_{0.6}} \leq 1.2$.
A number of the models include line emission, and clearly we 
also require that the model does not predict more Fe $K$-shell
emission than observed in the 5--7~keV band. The few occasions 
where this is the case are flagged in the text and tables.
We use the $F$-statistic to determine whether the inclusion of an 
additional free parameter significantly improves the fit. 
For the number of $dof$ typically in these datasets 
$F \gtrsim 3.8$ and $\gtrsim 6.6$ correspond to improvements at 
$> 95$\% and $>99$\% confidence respectively.

To summarize, the formal criteria for acceptibility used 
throughout are that 
$P(\chi^2 \mid dof)\leq 0.95$ {\it and}
either
	$\frac{\Delta \chi^2_{0.6}}{\Delta N_{0.6}} \leq 2.0$
or
	$\overline{R_{0.6}}$ lies in the range 
	$0.8 \leq \overline{R_{0.6}} \leq 1.2$.
However, with the menagerie of objects under investigation,
these formal criteria for acceptability were found to be potentially 
misleading in a number of cases. In such cases 
somewhat more subjective criteria were used to determine whether 
the model provided an acceptable representation of the data, but all
such cases are noted in the text.
For convenience, a summary detailing which models are deemed 
acceptable for each dataset along with various other notes
is provided 
in Table~12.

\section{BASIC MODELS}
\label{Sec:basic_models}

\subsection{Model {\it A(i)}: Single power law and neutral absorption}
\label{Sec:zwabspo}

The results of the analysis assuming a single underlying powerlaw 
absorbed by a neutral gas at the redshift of the source, and a neutral 
gas at zero redshift 
are listed in Table~3.
As well as the best-fitting parameters, Table~3
also lists the total $\chi^2$-statistic for each fit ($\chi^2_{Ai}$),
and
the parameters described in 
\S\ref{Sec:acceptability}
used to determine the acceptability 
of the fits.
Only 
4 datasets fulfill all the 
formal criteria for acceptability:
3C~120,
NGC~6814(1), 
Mrk~509 and 
NGC~7469.
However
inspection of the data/model ratios
reveals 4 additional datasets 
(Fairall~9, NGC~4593, Mrk~841(2) and MCG-2-58-22)
which appear to be well represented by the simple powerlaw model.
In the case of NGC~4593 and Mrk~841(2), model {\it A(i)} 
provides an adequate description of the data over the 0.6--5, 7--10~keV
band, but the best-fitting solution predicts a slight 
excess of counts ($\overline{R_{0.6}} \sim 1.3$)
when the best-fitting model is extrapolated $<$0.6~keV.
In the case of Fairall~9 and MCG-2-58-22 the data/model residuals 
indicate that the best-fitting models fail our criteria for acceptability 
primarily due to residuals in the 2--4~keV band beyond that 
expected from statistical considerations alone.
This region of the {\it ASCA} bandpass contains a number of instrumental 
features, primarily the Au $M$-edge complex due to the coating on the XRTs, 
suggesting these residuals may arise as a result of slight miscalibration 
of the Au $M$-edges and/or gain of the detectors at time of these observations
(see also Fig.~13).
However, since the residuals appear to have little effect on the best-fitting
parameters, we consider model {\it A(i)} most likely does indeed provide 
an adequate description of these datasets.
A similar problem occurs when a number of the other models are compared 
to these and other datasets, and all such occurrences are also
indicated in Table~12.

It can be seen that the low signal-to-noise ratio of the NGC~6814(1)
dataset results in an extremely low value of $\chi^2_{Ai} = 160$
corresponding to reduced $\chi^2$ value of $\chi^2_{\nu} = 0.78$
for 205 $dof$.
As reflected in the low value of $P_{Ai}=0.01$ listed in 
Table~3, this indicates that 
for the observation of NGC~6814(1) even this simple model 
can be considered an over-parameterization of the available data.
A similar problem occurs when a number of the other models are compared 
to other datasets, and all such occurrences are indicated in 
Table~12.
However, it should be noted that the fact that a 
number of the datasets are 
adequately described by a simple powerlaw model in the 0.6--6.0~keV 
band is not purely an
artifact of a low signal-to-noise ratio of the observations. 
Whilst a low signal-to-noise ratio may provide an explanation in 
the cases of NGC~6814(1) \& Mrk~841(2), it certainly 
does not provide an explanation in the case of the
3C~120 \& Mrk~509 datasets which have the {\it highest} 
signal-to-noise ratios in our sample (as illustrated by these sources having 
the highest values of $N_{phot}$(SIS0) in Table~2).
Thus, it is already apparent that some Seyfert 1 galaxies are indeed adequately 
described by a simple powerlaw, even though the majority are not.
(We note, however, that there is evidence for spectral curvature in 
3C~120 at energies $\gtrsim6$~keV -- see \S\ref{Sec:complex_cont}.)

Histograms showing the distributions of the photon index ($\Gamma$)
and intrinsic column density ($N_{H,z}$) for these datasets are shown in 
Fig~2a and 
Fig~3a respectively. 
For these and all the others presented here, the histograms are generated by
assigning a rectangle of equal area to each dataset ({\it not} each object).
The width of the rectangle is determined from the 68\% confidence 
range for
that parameter as listed in the relevant table (and hence its height derived).
We chose such a method as datasets for which the parameter is ill-determined
then make wide, flat contributions to the total, whilst datasets for which
the parameter is determined very accurately make thin, high contributions (which
can visually dominate the total). We feel the resultant plot is a 
better representation of the results.
In all cases the hatched region shows the histogram for the datasets 
which satisfy 
all the criteria for acceptability (\S\ref{Sec:acceptability}). The 
open histogram shows the 
effect of also including the datasets which just fail to meet 
our formal criteria but for which the model is considered likely to be
applicable
(as described in the text and summarized in Table~12).
As can been seen from Fig~2a, all the datasets for which 
a simple powerlaw provides an adequate fit 
have photon indices in the range
$1.5 \lesssim \Gamma \lesssim 2$.
Furthermore all of these datasets have intrinsic column densities 
$N_{H,z} \lesssim$ few $\times 10^{20}\ {\rm cm^{-2}}$
(Fig~3a) and Galactic column densities
($N_{H,0}$) consistent with those  derived from 21~cm measurements 
(Table~3), although we cannot 
distinguish between $N_{H,z}$ and $N_{H,0}$ in low-redshift, 
low-$N_{H,z}$ sources in this sample.

\subsection{Model {\it A(ii)}:Partial-absorption by neutral material}
\label{Sec:zpcfpo}

The results of the analysis assuming model {\it A(ii)} are listed in
Table~4. This model assumes a fraction $D_f$ of the
observed continuum is a bare powerlaw whilst a fraction $(1 - D_f)$ is a
powerlaw attenuated by neutral gas at the redshift of the source. As
stated in \S\ref{Sec:models}, such a model is applicable for geometries either
where there is an absorber along the line-of-sight which does not completely
cover the central source or where the central source is completely covered by
the absorber but in which some fraction of the radiation emitted by the
central source in other directions is scattered back into the line-of-sight.
As can be seen such a model provides a vast improvement in the fits as
measured by the reduction in the $\chi^2$-statistic for the majority of the
datasets. Indeed, from the $F$-statistic (listed as $F(\frac{Aii}{Ai})$ in
Table~4) it can be seen that the addition of $D_f$ as a
free parameter leads to an improvement over the results obtained assuming
model {\it A(i)} significant at $>$95\% confidence for all but 1 dataset
(NGC~6814(1)), 
and at $>$99\% confidence for all but that dataset and Mrk~841(1,2).

A total of 10 datasets satisfy all our formal criteria for acceptability
assuming model {\it A(ii)}. 
These include the 4 datasets for which our criteria for acceptability were
also satisfied by model {\it A(i)}, 
along with the datasets of 
Mrk~335, NGC~3227, Mrk~766, NGC~4593, Mrk~841(2) and MCG-2-58-22.
The histogram of the best-fitting photon indices from the datasets for which 
model {\it A(ii)} provides an adequate description is shown in
Fig~2b. 
It can be seen that the distribution 
is extended over the range $1.5 \lesssim \Gamma \lesssim 2.5$,
but as for model {\it A(i)} contains clear evidence for a 
peak at $\Gamma \sim 2$.
The best-fitting solutions to all the datasets have Galactic column
densities
($N_{H,0}$) consistent with that derived from 21~cm measurements
(Table~3) with the exception of 
3C~120 ($\Delta N_{H,0} \simeq 6 \times 10^{20}\ {\rm cm^{-2}}$)
and 
NGC~7469(2) ($\Delta N_{H,0} \simeq 3 \times 10^{20}\ {\rm cm^{-2}}$).

The derived model spectra and data/model ratios for the 8 datasets for which 
model {\it A(ii)} 
is acceptable and offers an improvement at $>$99\% confidence over 
model {\it A(i)} are shown in Fig~4.
For 2 of the remaining datasets (Fairall~9 and IC~4329A)
we find model {\it A(ii)} to offer a significant 
improvement over model {\it A(i)}, yet no significant 
further improvement in the goodness--of--fit is obtained assuming
subsequent models presented in \S\ref{Sec:basic_models}.
These two datasets are also shown in Fig~4, 
despite our formal criteria for acceptibility not being satisfied.
As can be seen from Table~4, for many of the datasets 
(Fairall~9, 3C~120, Mrk~841(2), NGC~6814(2), NGC~7469(2) \& MCG-2-58-22) 
the best fitting 
solution occupies the region of high-$N_{H,z}$,high-$D_f$ parameter-space.
This is also clearly apparent from the histograms of $N_{H,z}$ and $D_f$ 
for these datasets are shown in Figs~3b 
and~5a (respectively), and gives rise to a relatively subtle 
upturn in the observed spectrum $\gtrsim 7$~keV in these sources
(Fig~4).

It is important to appreciate that given the {\it ASCA} bandpass (and typical
signal-to-noise ratios) we are only sensitive to and/or can constrain
$N_{H,z}$ and $D_f$ over a limited range of parameter space. Clearly 
the model is 
insensitive to $N_{H,z}$ in all models where $D_f \simeq 1$ (models with
$D_f=1$ being identical to a powerlaw). The lower limit on the SIS
bandpass (along with calibration uncertainties) leads to us being insensitive
to $N_{H,z} \lesssim 2\times 10^{20}\ {\rm cm^{-2}}$, and the upper limit on
the {\it ASCA} bandpass leads to us being insensitive to $N_{H,z} \gtrsim
10^{24}\ {\rm cm^{-2}}$. Two datasets are consistent with $D_f=1$ at 
68\% confidence: Mrk~841(2) \& NGC~6814(1). These sources are consistent
with a single powerlaw, and thus $N_{H,z}$ is effectively undefined (as 
illustrated by the errors 'pegging' at the limits 
in Table~4). In addition it should be noted that Fairall~9
is consistent with $D_f=1$ at 90\% confidence and statistically acceptable
solutions to the Mrk~509 dataset obtained at 90\% confidence at both high and
low values of $N_{H,z}$.
Thus 
model {\it A(ii)} does not provide a unique explanation to the datasets 
which exhibit a subtle upturn at the highest energies
(i.e. those with a high-$N_{H,z}$,high-$D_f$ solution).
Indeed, this might be an indication of a real curvature in the underlying
continuum (see \S\ref{Sec:complex_cont}), although it might also 
arise as a result of
spectral variability during the observation or (less likely) the result of
incomplete background-subtraction.
However,
it is worth noting that the best-fitting models for 4 datasets
(Mrk~335, NGC~3227, Mrk~766 \& NGC~4593) are in regions of parameter-space
where the {\it ASCA} data are sensitive to the model.
These sources have 
$10^{21} \lesssim N_{H,z} \lesssim$few$\times 10^{22}\ {\rm cm^{-2}}$ 
and 
$0.3 \lesssim D_f \lesssim 0.7$.
Nevertheless yet superior fits are obtained with 
subsequent models in all 4 cases
(Mrk~335 -- \S\ref{Sec:2plaw}; 
NGC~3227, Mrk~766 \& NGC~4593 -- \S\ref{Sec:ion}).

\subsection{Model {\it B(i)}: Complete covering by ionized material}
\label{Sec:ion}

The results of the analysis assuming the powerlaw continuum passes through 
ionized material, completely covering the source are listed in
Table~5. It can be seen that for the majority of the datasets
such a model provides a vast improvement over model {\it A(i)} in which the
material responsible for the absorption was assumed to be neutral.
Indeed, from the $F$-statistic ($F(\frac{Bi}{Ai})$ in 
Table~5) it can be seen that the addition of $U_X$ as a 
free parameter leads to an improvement over the results obtained 
assuming 
model {\it A(i)} (\S\ref{Sec:zwabspo}) significant at 
$>$95\% confidence for all but Fairall~9 and NGC~6814(1), and 
at $>$99\% confidence for all but Fairall~9, 3C~120, NGC~6814(1) and Mrk~841(2).
A direct comparison of $\chi^2_{Bi}$ in Table~5 
with
$\chi^2_{Aii}$ in Table~4 (both models having the 
same number of $dof$) indicates model {\it B(i)}
also offers a superior description of the data than model {\it A(ii)} 
for the majority of the datasets.
The notable exceptions to this are the observations of 
3C~120, IC~4329A and (especially) all 3 observations of NGC~4151.
It should also be noted that 
	$\frac{\Delta \chi^2_{0.6}}{\Delta N_{0.6}}$
and 	$\overline{R_{0.6}}$ 
(Table~5)
are both greatly improved 
compared to those derived assuming model {\it A(ii)}
(Table~4) for a number of datasets, 
indicating model {\it B(i)}
provides a far superior description of the data $<$0.6~keV.
This is true even for a number of datasets that do not satisfy our criteria for 
acceptability, and 
in most cases the result of a 'recovery' of the absorbed spectrum
at low energies (below the oxygen $K$-shell edge)
due to carbon and nitrogen being fully ionized and hence 
the gas pseudo-transparent at these energies.

A total of 14 datasets satisfy all our criteria for acceptability assuming 
model {\it B(i)}. These include the 10 datasets for which 
our criteria were also satisfied by model {\it A(ii)}, 
along with NGC~3783(1,2), NGC~5548 and Mrk~841(1).
The derived model spectra and data/model ratios for the 8 datasets for which 
model {\it B(i)} 
is acceptable and offers an improvement at $>$99\% confidence over 
both models {\it A(i)} and {\it A(ii)}
are shown in Fig~6. 
We note that $P_{Bi} = 0.03$ in the case of 
Mrk~766, indicating this model (and all subsequent models) could be considered
an over-parameterization of the data.
For 2 of the remaining datasets (NGC~3516 and MCG-6-30-15(1))
we find model {\it B(i)} to offer a significant 
improvement over both models {\it A(i)} and {\it A(ii)}, yet no significant 
further improvement in the goodness--of--fit obtained assuming
subsequent models presented in \S\ref{Sec:basic_models}.
These two datasets are also shown in Fig~6, 
despite our formal criteria for acceptibility not being satisfied.

The superior description of the data at low energies afforded by 
model {\it B(i)} compared to that of model {\it A(ii)} 
is illustrated by comparison of panels for 
NGC~3227, Mrk~766, NGC~4593 and Mrk~509 in 
Figs~4 and~6,
and attributable to the reduction in opacity of the ionized gas at low energies.
Inspection of the data/model residuals of the remaining datasets 
reveal that both observations of MCG-6-30-15 fail to satisfy our 
criteria for acceptability primarily due to a deficit of counts 
around $\sim$1~keV. 
As noted in Table~12, such a deficit is also evident 
in the data/model residuals for NCG~3516 (along with an apparent excess 
scatter across the entire energy range), and in NGC~3783(1,2).
The feature may be the result of 'excess' absorption by O{\sc viii}, 
and is further discussed in \S\ref{Sec:2ion}.

The distribution of the photon index from these fits, shown in
Fig~2c, covers a similar range and remains peaked 
at $\Gamma \sim 2$ as for models {\it A(i)} and {\it A(ii)}.
The distribution of the column density of the ionized material shows 
no clear preference, with a broad variety of values over the 
range $10^{21} \lesssim N_{H,z} \lesssim 10^{23}\ {\rm cm^{-2}}$.
The distribution for the ionization parameter ($U_X$), plotted in
Fig~7a, is extended over the range 
$10^{-2} \lesssim U_X \lesssim 10$, but exhibits a clear peak at 
$U_X \sim 0.1$.
It is worth noting that the datasets with ionization parameters in the
range $1 \lesssim U_X \lesssim 10$ lie in regions of 
$N_{H,z}$--$U_X$ parameter-space where the ionized gas is largely 
transparent in the {\it ASCA} band. Thus the best-fitting spectra 
for these datasets (e.g. Fairall~9, 3C~120, NGC~7469(2), MCG-2-58-22)
are essentially powerlaws, consistent with the findings above.
In Fig~8a we plot the derived values for 
$N_{H,z}$ and $U_X$ for the 11 datasets for which 
model {\it B(i)} offers an adequate description of the data {\it and}
leads to an improvement at $>99$\% over model {\it A(i)}.
The corresponding values for NGC~3516 and both 
datasets of MCG-6-30-15 are also shown in Fig~8a
for comparison.

Finally we note that 
all the datasets for which model {\it B(i)} provides an acceptable 
solution have best-fitting $N_{H,0}$ consistent with that derived 
from 21~cm measurements (Table~3) with the exception of 
3783(1) ($\Delta N_{H,0} \simeq 2 \times 10^{20}\ {\rm cm^{-2}}$)
along with 
3C~120 
and 
NGC~7469(2), both of which have excess column densities similar to those 
found in \S\ref{Sec:zpcfpo}.

\subsection{Model {\it B(ii)}: Partial-covering by an ionized absorber}
\label{Sec:ion_pc}

The results of the analysis assuming model {\it B(ii)} 
are listed in Table~6. 
The relationship of this model to model {\it B(i)} is the 
same as the relationship between models {\it A(i)} and {\it A(ii)}:
identical except that now a fraction $D_f$ of the observed continuum
is a bare powerlaw whilst a fraction $(1-D_f)$ 
is attenuated by line-of-sight material at the redshift of the source.
(see \S\ref{Sec:models}).
It can be seen that model {\it B(ii)} provides a further substantial 
reduction in the $\chi^2$-statistic for many of the datasets.
From the $F$-statistic ($F(\frac{Bii}{Aii})$ in Table~6)
we find that the addition of $U_x$ as a free parameter leads to a
improvement over the results obtained assuming model {\it A(ii)}  
for 17 (16) datasets at $>$95\% ($>$99\%) confidence\footnote{See 
the note to Table~6 for an explanation of the apparent 
increase in $\chi^2$ for the observations of 
Fairall 9, 3C~120 \& IC~4392A.}.

A total of 15 datasets satisfy all our criteria for acceptability assuming 
model {\it B(ii)}. These include 13 out of the 14
datasets for which model {\it B(i)} was 
found formally acceptable\footnote{The exception being Mrk~841(1), which 
falls outside our formal limit of $P_{Bii} \leq 0.95$ by just 
$\Delta P_{Bii} = 0.001$}, along with NGC~4151(2,4).
As was the case for model {\it B(i)}, inspection of the data/model residuals 
reveals that a deficit of counts at $\sim$1keV is the primary reason why 
several of the remaining datasets fail to satisfy our criteria for 
acceptability (namely NGC~3516, NGC~4151(5) and MCG-6-30-15(1,2)).

Interestingly, we find
that the inclusion of the additional free parameter $D_f$ compared 
to model {\it B(i)} improves the fit at $>$99\% confidence for 
10 of the 23 datasets (as shown by the values of 
the $F$-statistic $F(\frac{Bii}{Bi})$ listed in Table~6).
Furthermore, 8 of these datasets satisfy our criteria for acceptability,
indicating the spectrum observed from a significant 
fraction of the datasets may indeed contain a component of the underlying 
continuum which does not experience attenuation by the ionized gas.
The best-fitting solutions for all 5 datasets for which 
model {\it B(ii)} is acceptable and offers an improvement at 
$>$99\% confidence over both models {\it A(ii)} and {\it B(i)}
have $D_f>0$, as illustrated by the 
derived model spectra and data/model ratios 
shown in Fig~9.
For one of the remaining datasets (NGC~4151(5))
we find model {\it B(ii)} to offer a significant 
improvement over previous models,
yet no significant 
further improvement in the goodness--of--fit obtained assuming
subsequent models presented in \S\ref{Sec:basic_models}.
This dataset is also shown in Fig~6, 
despite our formal criteria for acceptibility not being satisfied.

In Fig~8b we plot the derived values for 
$N_{H,z}$ and $U_X$ for the 10 datasets for which 
model {\it B(ii)} offers an adequate description of the data {\it and}
leads to an improvement at $>99$\% over model {\it A(ii)}.
The corresponding values for NGC~3516, NGC~4151(5), MCG-6-30-15(1,2) 
and Mrk~841(1) are also shown in Fig~8b
for comparison.

Again the limitations imposed on parameter-space by the {\it ASCA} bandpass 
and the signal-to-noise ratios are reflected in a number of the fits.
Obviously, the comments made in \S\ref{Sec:zpcfpo} regarding solutions 
with $D_f \simeq 1$, and in \S\ref{Sec:ion} regarding the transparency of
the ionized gas in the low-$N_{H,z}$, high-$U_X$ region of parameter space 
are also applicable to model {\it B(ii)}.
These effects prevent useful constraints being placed on $N_{H,z}$
in a number of cases.
Furthermore, the gas for models in the high-$N_{H,z}$,low-$U_X$ 
region of parameter space is close to neutral, giving rise to 
spectra identical to those in model {\it A(ii)}.
As for model {\it A(ii)}, the best-fitting models for a number of datasets 
exhibit a subtle upturn at the highest energies.

As might be expected, the distribution of the photon index 
(Fig~2d) covers a similar (but slightly 
broader) range as that derived from previous models, and is still 
peaked at $\Gamma \sim 2$.
As for model {\it B(i)}, there appears to be no preferred value of 
the intrinsic column density (Fig~3d), with the 
derived values (if errors are also considered) 
covering over three decades in $N_{H,z}$.
The datasets for which $D_f>0$ 
are gathered in the range $0.3 \lesssim D_f \lesssim 1.0$, with the 
exception of the 3 observations of NGC~4151 which have 
$0.03 \lesssim D_f \lesssim 0.1$ (Fig~5b).
The distribution in ionization parameter 
is extended over the range $10^{-2} \lesssim U_X \lesssim 1.0$, but as 
for model {\it B(i)}, there is a clear concentration around
$U_X \sim 0.1$ (Fig~7b).

Finally, we note that 
the best-fitting solutions to 13 of the 15 datasets 
considered acceptable have Galactic column densities 
($N_{H,0}$) consistent with that derived from 21~cm measurements
(Table~3). Again the exceptions are
3C~120 
and 
NGC~7469(2), both of which have excess column densities similar to those 
found in \S\ref{Sec:zpcfpo}.

\subsection{Model {\it C(i)}: Complete covering by an ionized absorber plus 
an ionized emitter}
\label{Sec:ion_emis}

The results of assuming model {\it C(i)} 
are listed in Table~7. 
This model assumes the powerlaw continuum passes through ionized 
material, completely covering the source (i.e. as for model {\it B(i)}),
but in addition contains the emission and scattered spectrum expected 
from the ionized gas assuming it subtends a solid angle $\Omega$
at the central source (see \S\ref{Sec:models}).
We see that the such a model provides further improvement for 
many of the datasets. Using the 
$F$-statistic ($F(\frac{Ci}{Bi})$ in Table~7) 
it can be seen that the addition of $\Omega$ as a free 
parameter leads to an improvement over the results obtained
assuming model {\it B(i)} for 15 (11) datasets at
$>$95\% ($>$99\%) confidence.
A direct comparison of $\chi^2_{Ci}$ in Table~7 
with $\chi^2_{Bii}$ in Table~6 (both models having the 
same number of $dof$) indicates model {\it C(i)} also offers a significant 
improvement over the results obtained assuming model {\it B(ii)} for 
NGC~3783(1,2), NGC~4051 and MCG-6-30-15(2).
We note, however, that model {\it C(i)} provides an inferior fits
compared to model {\it B(ii)} for the observation of 3C~120
and all 3 observations of NGC~4151.
This indicates that consideration of the covering fraction of the ionized
absorber is more important than its emission in these cases.
Model {\it C(i)} offers only a modest improvement or slightly inferior fit
to the remaining datasets.

A total of 15 datasets satisfy all our criteria for acceptability assuming 
model {\it C(i)}. As summarized in 
Table~12, this 
includes all 14 datasets for which our criteria for acceptability were 
also satisfied by model {\it B(i)}, along with 
MCG-6-30-15(2).
As was the case for models {\it B(i)} \& {\it B(ii)}, 
inspection of the data/model residuals 
reveals that a deficit of counts at $\sim$1keV is the primary reason why 
the datasets of NGC~3516, MCG-6-30-15(1)
fail to satisfy our criteria for 
acceptability.
The spectrum emitted by the ionized gas includes Fe lines in the 
6.4--7.1~keV band (rest-frame). Since the bulk of this region is excluded 
from our spectral analysis, it is possible that the best-fitting 
model will be inconsistent with the observations within this band.
However for all 15 datasets which 
satisfying our criteria for acceptability, we find the 
best-fitting solutions lie in a region of $N_{H,z}$--$U_X$--$\Omega$
parameter-space such that the predicted line-strength is less than that
observed.
The derived model spectra and data/model ratios for the 4
datasets for which
model {\it C(i)} is acceptable and offers an improvement at
$>$99\% confidence over both models {\it B(i)} and {\it B(ii)}
are shown in Fig~10.

In Fig~8c we plot the derived values for 
$N_{H,z}$ and $U_X$ for the 13 datasets for which 
model {\it C(i)} offers an adequate description of the data {\it and}
leads to an improvement at $>99$\% over model {\it A(i)}.
The corresponding values for NGC~3516 and MCG-6-30-15(1)
are also shown in Fig~8c
for comparison.

The inclusion of the emission spectrum does not globally
change the distributions of the $\Gamma$, $N_{H,z}$ and $U_X$ for
these datasets as shown Figs~2e, 3e,
and 7c (respectively).
In general the parameters cover similar ranges as for models {\it B(i)} \& 
{\it B(ii)},
with the datasets exhibiting a range of photon indices (but peaked at $\Gamma
\sim 2$), column densities and ionization parameter (again peaking at $U_X
\sim 0.1$).
As can be seen from Fig~11a, the solid angle of the ionized
emitter is poorly constrained in most cases, but there is a clear
peak at $\Omega \sim 4\pi$ (i.e. consistent with emission from a complete
shell).

Again, the best-fitting solutions to all the datasets 
considered acceptable have Galactic column densities 
($N_{H,0}$) consistent with that derived from 21~cm measurements
(Table~3), with the exceptions of
3C~120 
and 
NGC~7469(2), both of which have excess column densities similar to those 
found in \S\ref{Sec:zpcfpo}.

\subsection{Model {\it C(ii)}: 
Partial-covering by an ionized absorber plus an ionized emitter}
\label{Sec:ion_pc_emis}

The results of the analysis assuming model {\it C(ii)} are listed 
in Table~8. As described in 
\S\ref{Sec:models}, this model is the same as model {\it C(i)}
except that only a fraction $(1-D_f)$ of the observed continuum is 
attenuated by the ionized gas. 
(The spectral component
due to the emission and scattered spectrum expected from the ionized 
gas is unaffected and parameterized by $\Omega$, just as in model 
{\it C(i)}.)
It can be seen that model {\it C(ii)} provides a significant 
reduction in the $\chi^2$-statistic for several datasets.
From the $F$-statistic ($F(\frac{Cii}{Bii})$ in Table~6)
we find that the addition of $\Omega$ as a free parameter leads to a
improvement over the results obtained assuming model {\it B(ii)}  
for 14 (8) datasets at $>$95\% ($>$99\%) confidence.
Alternatively, we find that the addition of $D_f$ as a free parameter 
leads to model {\it C(ii)} being preferred over model {\it C(i)} for 9 (8) 
datasets at $>$95\% ($>$99\%) confidence.
However, we note that model {\it C(ii)} offers an improvement at 
$>$99\% confidence over both models {\it B(ii)} {\it and} {\it C(i)} 
for only 4 datasets (NGC~4051, NGC~4151(4), NGC~5548 and Mrk~509).
The derived model spectra and data/model ratios of these are shown
Fig~12.
It should be noted, however, that in the case of NGC~4051, the best-fitting 
solution assuming model {\it C(ii)} fails to meet our formal criteria for 
acceptability. This is mostly likely due to (at least in part)
spectral variability within the observation
giving rise to spectral curvature (see \S\ref{Sec:complex_cont}).
In addition, the best-fitting solution to NGC~5548 over-predicts the 
number of counts $<0.6$~keV ($\overline{R_{0.6}} \sim 1.3$).
Finally it should be noted that the best-fitting solutions for 
NGC~5548 and Mrk~509 lie in the 
high-$N_{H,z}$,high-$U_X$,high-$D_f$ region of parameter 
space. The resultant spectra are therefore essentially powerlaws 
with subtle upturns at both the very lowest and very highest energies
(Fig~12). The meaningfulness of such 
solutions is therefore questionable, as it may simply be indicative 
of spectral curvature (see \S\ref{Sec:complex_cont}).

A total of 16 datasets satisfy all our criteria for acceptability assuming 
model {\it C(ii)}. 
These include 14 out of the 15
datasets for which model {\it C(i)} was
found formally acceptable (the 
exception being NGC~5548, which just fails our criteria for
extrapolation $<$0.6~keV as noted above), along 
with 2 of the 3 observations of NGC~4151.
As was the case for previous models, inspection of the data/model residuals 
reveals that a deficit of counts at $\sim$1keV is the primary reason why 
the datasets from NGC~3516, NGC~4151(5) and MCG-6-30-15(1)
fail to satisfy our criteria for 
acceptability.
We also note that for 5 of the 16 datasets satisfying our criteria for 
acceptability for this model
(Mrk~335, 3C~120, NGC~5548, NGC~7469(2) and MCG-2-58-22), the
best-fitting solutions over-predict the strength of the Fe $K$-shell lines 
in the 6.4--7.1~keV band.
In a number of cases the 90\% confidence region in 
$N_{H,z}$--$U_X$--$\Omega$ parameter-space does include solutions 
consistent with observed line strengths.
However, as will be shown in \S\ref{Sec:complex_cont}, in all cases 
the solution obtained assuming 
model {\it C(ii)} is most likely an artifact of spectral curvature 
in the underlying continuum.

In Fig~8d we plot the derived values for 
$N_{H,z}$ and $U_X$ for the 11 datasets for which 
model {\it C(ii)} offers an adequate description of the data {\it and}
leads to an improvement at $>99$\% over model {\it A(ii)}.
The corresponding values for NGC~3516, NGC~4151(5), MCG-6-30-15(1) 
and NGC~5548 are also shown in Fig~8e
for comparison.

As for model {\it C(i)}, we find 
the inclusion of the emission spectrum does not globally
change the distributions of the $\Gamma$, $N_{H,z}$, $D_f$ and $U_X$ for
these datasets as shown in Figs~2d, 3d,
5c and 7b (respectively).
In general the parameters cover similar ranges as for 
previous models.
In most cases, the emission from the ionized gas 
forms a significant fraction of the observed spectrum 
only at low energies, and is dominated by emission lines due to
$K$-shell processes of C, N \& O.
However, it should be noted that in a number of cases
(e.g. NGC~4151(4) \& Mrk~509) the 
Fe $K$-shell emission line and scattered continuum do contribute to the 
spectrum observed above $\sim$3~keV.

Again we find the best-fitting solutions to all the datasets 
considered acceptable have Galactic column densities 
($N_{H,0}$) consistent with that derived from 21~cm measurements
(Table~3), with the exceptions of
Mrk~509 (with $\Delta N_{H,0} \simeq 5 \times 10^{20}\ {\rm cm^{-2}}$),
and
3C~120,
NGC~3783(1) 
and 
NGC~7469(2), all of which have excess column densities similar to those
found previously.

Finally, in Fig.~13
we show the mean data/model ratios (in the {\it observer's} frame) 
in the SIS and GIS for the 16 datasets satisfying our criteria 
for acceptability assuming model {\it C(ii)}. 
This plot was constructed by taking the error-weighted average of 
the individual data/model ratios for each detector from 
each of the appropriate observations.
These averages were then rebinned for display purposes. 
Caution is urged when interpreting such plots.
They potentially contain artificial features introduced by time-dependent
errors in the calibration of the instruments (such as slight offsets
in the gain of the detectors at the time of each observation), 
along with real features due to errors in the calibration of the 
instrument, as well as real features of astrophysical origin.
Furthermore such plots are dominated by the datasets with the 
highest signal-to-noise ratio.
Nevertheless, we have found Fig.~13 helpful 
in judging the reality of features/noise in individual datasets.

\section{ADDITIONAL MODELS}
\label{Sec:additional_models}

In the previous section we made the premise that the 'primary' X-ray 
continuum was a powerlaw. We then compared the observed spectrum with that 
expected assuming a number of (probably simplistic) assumptions regarding 
the geometry and physical conditions within the circumnuclear material.
We found that one or more of this family of models to be consistent with 17 
of the 23 datasets considered within our formal criteria of acceptability
(\S\ref{Sec:acceptability}).
Of the 6 remaining datasets, the primary reason why 
the criteria of acceptability are not achieved is 
due to excess scatter in the residuals 
within the 2--4~keV band for Fairall~9
(see Fig.~4, the observed spectrum otherwise being 
consistent with model {\it A(ii)}), 
a deficit of counts at $\sim$1~keV in the case of 
NGC~3516, NGC~4151(5) and MCG-6-30-15(1) (see 
Figs~6 \& ~9), 
and a more complex spectral form in the case of 
NGC~4051 and 
IC~4329A
(Figs~12 \& ~4
respectively).
In this section we extend our spectral analysis to address a number of 
issues raised in the previous section.

\subsection{The redshift of the ionized gas}
\label{Sec:redshift}

In \S\ref{Sec:basic_models} it was assumed that the ionized gas 
along the line-of-sight was situated close to (and not in motion 
relative to) the active nucleus (i.e. at the redshift of the host galaxy,
$z_{gal}$).
The sharp features imprinted by the ionized material (primarily O{\sc vii} 
and/or O{\sc viii} absorption edges for most datasets, with rest-frame 
energies of 740 and 871~eV) facilitate testing such an hypothesis.
The spectral resolution of the SIS is a complex function of energy, 
time and various other instrument effects. However 
at the epoch of the observations, 
$\Delta E/E  \simeq 0.07$--$0.09$
in the 0.6--1.0~keV band 
(e.g. Dotani et al 1996).
Useful
constraints on the redshift of the ionized gas $z_{ion}$ are only 
likely to be obtained for those datasets with the 
strongest absorption and/or emission features.
Nevertheless, we have repeated the spectral analysis for 
models {\it B(i)}--{\it C(ii)} 
allowing the redshift of the ionized gas ($z_{ion}$) to be a free parameter.
However, in the vast majority of cases we found the 90\% confidence range
for $z_{ion}$ to encompass both $z_{ion}=0$ and $z_{ion} = z_{gal}$. 
We find  the most stringent constraints on $z_{ion}$ to be provided by 
model {\it C(i)}
applied to 3783(1), where $z_{ion} = z_{gal} - 0.002^{+0.015}_{-0.012}$
(where $z_{gal}=0.009$).
The formal limits exclude $z_{ion}=0$ and/or $z_{ion} = z_{gal}$ for a 
few dataset/model combinations, but further investigation reveals 
that in each case the result is due to fitting 
small statistic fluctuations in the data.
Given the recent detection of ionized gas at the redshift of the 
host galaxy in another AGN (the high-redshift quasar PG~1114+445; 
George et al. 1997a)
and lack of evidence to the contrary here, 
we conclude our assumption to be valid.

\subsection{Evidence for a second, ionized absorber}
\label{Sec:2ion}

During the analysis presented in 
\S\ref{Sec:basic_models}, a number of datasets were observed to 
consistently exhibit a deficit in their data/model ratio at 
$\sim$1~keV. This feature was most evident in
both datasets of NGC~3783 (e.g. Figs~6,
9 \&~10)
and MCG-6-30-15 (Figs~9 \&
~10), and in 
the single dataset from NGC~3516 
(Fig~6).
As noted in the introduction, the $\sim$1~keV deficit 
has also been seen by 
a number of other workers, and since it appears close to the $K$-shell
edges of O and Ne, it has been suggested that it is the result
of absorption by a second cloud/screen of ionized gas along the
line-of-sight
(e.g. Kriss et al. 1996a; Otani et al. 1996).

As will be discussed in \S\ref{Sec:disc-2ion} there are a number of 
explanations for such a feature, and a detailed investigation 
of these are beyond the scope of the current paper.
However, in Table~9
we list the results obtained when an absorption edge of optical depth 
$\tau_{ONe}$ is added to model {\it B(i)}.
The (rest-frame) energy of the edge, $E_{ONe}$,
was allowed to vary over the range 0.568--1.362~keV, thus covering 
the edges due to O{\sc vi}--Ne{\sc x}.
As can be seen from the values of the $F$-statistic $F_{ONe/Bi}$, 
the addition of this component
improved the goodness-of-fit at 
$>99$\% confidence for all 5 datasets.
A comparison with Table~5 shows that the effect of including 
such an additional edge is to reduce 
the best-fitting values of $U_X$ and $N_{H,z}$.
Furthermore, 
in all but one case (NGC~3783(1)) we find the best-fitting energy for 
the edge to be consistent with that due to
O{\sc viii}.
We also tested the addition of such an edge to model {\it C(i)} 
and obtained similar results, except that the edge in NGC~3783(1) 
is now also consistent with that due to O{\sc viii}. 

Finally, it should be stressed that such a deficit is {\it not} seen in several 
of the other datasets presented here, and hence unlikely to be
the result of a calibration error. For example, in the case of the high 
signal-to-noise, powerlaw dataset 3C~120, we find the additional of
an O{\sc viii} 
edge to model {\it B(i)} makes an insignificant improvement in 
the goodness-of-fit ($\Delta \chi^2 < 1$), with 
$\tau_{ONe} \lesssim 0.02$ at 90\% confidence (for one interesting 
parameter).

We consider these results to be indicative of a second absorbing system in
some objects. This possibility, along with the fact that the 
1~keV deficit is present in the datasets with the strongest 
absorption features, is discussed in
\S\ref{Sec:disc-2ion}.

\subsection{Implications for Fe $K$-shell band}

\subsubsection{Fe $K$-shell emission}
\label{Sec:Fe-line}

In 
Paper~II we presented an analysis of the spectra of all 23 datasets
in the 3--10~keV band. The approach used was to approximate the 
underlying continuum with a powerlaw in order to 
perform a detailed investigation the profile of the Fe $K\alpha$ line.
In the present work, we have ignored the 5--7~keV band, which contains
the bulk of the Fe $K\alpha$ line emission. Clearly it is inappropriate 
to repeat here the full analysis presented in Paper~II. 
However there are two consistency checks which should be 
performed.

First, 
in \S\ref{Sec:basic_models} we have shown that the 0.6--10~keV continuum
in these sources can be well represented by a powerlaw imprinted with 
(primarily absorption) features due to ionized gas along the line-of-sight.
Comparison of Tables~3--6
with the corresponding results from Paper~II reveal consistent values of 
$\Gamma$. Thus the approximation of the local continuum 
used in Paper~II is indeed valid, and hence the intensity and profile 
of the line emission presented in Paper~II is also valid.
This is illustrated in Fig~14 in which we 
show the mean profiles obtained for models {\it B(ii)}, 
{\it C(i)} and {\it C(ii)} compared to that presented in Paper~II.

Second, models {\it C(i)} and {\it C(ii)}
contain the emission spectrum for the ionized gas, which 
includes significant Fe $K$-shell features in 
certain regions of $N_{H,z}$--$U_X$ parameter-space.
We have therefore repeated the analysis for these models, but including
the 5--7~keV band and an emission line 
as an additional spectral component.
The profile of the emission line assumed is that expected from the 
innermost regions of an accretion disk under the influence of the strong 
gravitation effects of the central black hole.
For convenience, we have followed Paper~II and 
adopted the parameterization of Fabian et al (1989), 
fixing the 
inner and outer radii of the disk at 
$R_i = 6 R_g$ and $R_o = 10^3 R_g$ (respectively, where $R_g$
is the gravitational radius), the axis of rotation with respect to 
the line-of-sight of $i = 30^{\circ}$, and the (rest-frame) 
energy of the line to be $E_{K\alpha} = 6.4$~keV.
The emissivity of the line is parameterized as a powerlaw as a function 
of radius ($\propto R^{-2.5}$), leaving 
only the equivalent width of the line, $W_{K\alpha}$, as a free 
parameter.
As may be expected from Fig~14, 
for both models {\it C(i)} and {\it C(ii)}, we find results 
consistent with those presented in Paper~II, with 
mean equivalent widths of the diskline
$\overline{W_{K\alpha}} \simeq 290\pm30$~eV,
compared with $\overline{W_{K\alpha}} \simeq 270\pm14$~eV
derived from Table~5 of Paper~II.

\subsubsection{Fe $K$-shell absorption}
\label{Sec:Fe-absorption}

{\it Ginga} observations indicated a large fraction ($\gtrsim 40$\%) of 
type-1 AGN might possess a significant Fe absorption edge in the 
7.1--8.9~keV band (Nandra \& Pounds 1994).
However given the spectral resolution of the LAC and the presence of the 
strong Fe $K$-shell emission in the 6--7~keV band, the parameters of such a 
feature could not be unambiguously determined by {\it Ginga}. 
Nevertheless, Nandra \& Pounds did
find a preferred energy of $\sim 7.9$~keV (equivalent to Fe{\sc xviii})
and effective hydrogen column densities 
$N_{H,z} \sim$few$\times10^{23}\ {\rm cm^{-2}}$.

Unfortunately the end of the effective bandpass of the SIS and GIS
at $\sim 10$~keV and the low signal-to-noise ratio of most datasets
in the 7--10~keV band makes the study of such a feature extremely 
difficult with {\it ASCA}.
We have not attempted such a study here, especially since we
employed an analysis technique whereby we explicitly excluded the 
Fe $K$-shell emission (in the 5--7~keV band) from 
the majority of the fits.
However as can be seen in 
Figs~6,
9, 
10
\& 
12, 
our models for the photoionized gas do {\it not} predict 
significant Fe $K$-shell absorption edges for 
the values of $U_X$ and $N_{H,z}$ typically derived 
in \S\ref{Sec:basic_models} (the {\it ASCA} results being 
primarily based on 
O $K$-shell, Ne $K$-shell and Fe $L$-shell absorption features).
If confirmed, the absorption features seen by {\it Ginga} 
imply an additional, large column density of 
highly-ionized gas along the cylinder-of-sight, possibly related 
to that suggested as responsible for the '$\sim 1$~keV deficit'
described in \S\ref{Sec:2ion}.
However we note that there is no obvious evidence for such features 
in Fig~14.

\subsection{Constraints on a more complex continuum form}
\label{Sec:complex_cont}

Given the diversity in the observed spectra presented here, the 
assumption that the 
continuum is well represented by a single powerlaw over the 
entire 0.4--10~keV band can be questioned.
Indeed, based on both observational and theoretical arguments, in the past 
a number of more complex spectral forms have been suggested for the 
underlying X-ray continuum in Seyfert 1 galaxies.
These can be 
divided into those in which the underlying emission mechanism 
gives rise to a continuum with a more complex form 
(a simple case case a continuum represented by the sum of two powerlaws),
and those in which a 'primary' powerlaw continuum is augmented by other 
sources of emission (both thermal and non-thermal, 
inside and outside the active nucleus).

In \S\ref{Sec:basic_models} we found a number of datasets 
consistently satisfied by best-fitting parameters which could imply a 
gradual flattening of the observed spectrum to higher energies
(e.g.  Mrk~335, NGC~4051, IC~4329A).
The data/model residuals of a number of other datasets also imply 
a 'soft excess' of counts at the lowest energies and/or a 'hard tail' 
at the highest energies
(e.g.  NGC~3516, 3C~120).
In the light of this, we have investigated two forms of the 
continuum commonly assumed for Seyfert 1 galaxies: the addition of a 
'Compton Reflection' component, and an underlying continuum represented by 
the sum of two powerlaws. We have also investigated the inclusion of a 
thermal emission component at the softest energies.

It should be remembered, however, that spectral curvature can of course also
arise as the result of spectral variability within an observation. Throughout
this paper we consider only the mean spectra. However in Paper~I we presented
evidence for variability in many of the datasets. In most datasets any {\it
spectral} variability was of relatively low amplitude, but a number of cases
(primarily the low luminosity sources, and especially NGC~4051), the amplitude
could be at a level to start affecting our analysis of the mean spectra.

\subsubsection{The effects of including 'Compton Reflection'}
\label{Sec:reflection}

'Compton-reflection' of the 'primary' continuum by optically-thick, neutral 
material out of the line-of-sight is often invoked as an explanation of the 
high equivalent width of the Fe $K\alpha$ line emission and 
flattening of the spectrum $\gtrsim 10$~keV 
in many AGN (e.g. Nandra \& Pounds 1994). 
The presence of
an unresolved reflection component can flatten the observed continuum, 
so the underlying continuum could be slightly steeper than 
observed, which could have implications on the best-fitting 
values of the parameters associated with the ionized gas.
We have therefore repeated the analysis for model {\it C(ii)} 
including the effects of the 'Compton-reflection, which 
we parameterize by the addition of the enhancement factor,
${\cal F}$, of the reflected continuum (only) compared to that
expected from a semi-infinite plane illuminated by a point source
viewed face-on.

We find fixing ${\cal F} = 1$ does not improve 
the goodness--of--fit at $\geq 99$\% confidence 
for any of the datasets. As expected, in most cases the derived spectral 
index is slightly steepened by $\Delta \Gamma \lesssim 0.15$,
and in all cases best-fitting values of $N_{H,z}$ and $U_X$ are consistent 
with those quoted in Table~8.

Interestingly, allowing ${\cal F}$ to be a free parameter in the analysis, 
we do find a significant improvement in the goodness--of--fit for 
9 datasets (Mrk~335, NGC~3783(2), NGC~4051, NGC~4593, MCG-6-30-15(1,2), 
IC~4329A, NGC~5548, Mrk~509), with ${\cal F} \sim$3--5.
Such values have been suggested for some of these objects by previous studies 
using different instrumentation (e.g. Nandra \& Pounds 1994; 
Molendi et al, 1997).
However, since the {\it ASCA} bandpass effectively ends at 10~keV, 
such a component is difficult to constrain and weaker Compton-reflection
components would be undetectable.
For ${\cal F} \lesssim 10$, the predicted equivalent width of the 
Fe $K\alpha$ from the reflecting material can be approximated by
$W_{K\alpha} \sim 150 {\cal F} (\frac{A_{\rm Fe}}{3.3\times10^{-5}})$~eV,
where $A_{\rm Fe}$ is the abundance of Fe relative to hydrogen
(e.g. George \& Fabian 1991).
The most recent estimations of the Fe abundance are 
$A_{\rm Fe} \sim 4.7\times10^{-5}$ (Anders \& Grevesse 1989) 
and 
$\sim 3.2\times10^{-5}$ (Feldman 1992), 
and from observations we find 
$\overline{W_{K\alpha}} \simeq 290\pm30$~eV (\S\ref{Sec:Fe-line})
for these 9 datasets.
Thus these best-fitting solutions imply a moderate 
under-abundance of Fe in these Seyferts compared to solar values. 
We do not consider this unreasonable {\it per se}, but note these solutions 
may also be indicative of curvature in the underlying continuum shape, rather 
than the presence of a strong reflection component. 
Indeed alternative explanations are available in the case of
Mrk~335, NGC~4051, IC~4329A (see \S\ref{Sec:2plaw}).
In any event, we find the inclusion of the reflection component typically 
only steepens the underlying spectral index by $\Delta \Gamma \lesssim 0.2$
in these 9 datasets,
and in all cases again find the best-fitting values of 
$N_{H,z}$ and $U_X$ are within the error ranges 
quoted in Table~8.

\subsubsection{Double powerlaw continua}
\label{Sec:2plaw}

The form of the continuum considered in this section 
is represented by the sum of two powerlaws\footnote{i.e. not 
the somewhat artificial 'broken-powerlaw' model (which has also been
applied in the past to AGN spectra by some workers) in which the 
continuum is assumed to be a steep powerlaw below some energy at which it 
sharply breaks to a flatter powerlaw}, 
with photon indices 
$\Gamma_s$ and $\Gamma_h$ (where $\Gamma_s > \Gamma_h$).
The ratio of the normalization of the soft powerlaw
divided by that of the hard powerlaw at 1~keV (in the rest-frame of
the source) is denoted by $R_{s/h}$.
For computational expediency we restrict parameter space 
such that 
$1.5 \geq \Gamma_s \geq 5.0$
and
$0.0 \geq \Gamma_h \geq 2.5$.
Such a model has been applied to all the datasets considered here.
Here we concentrate on the results when 
this additional powerlaw component is added 
to model {\it C(i)} as they are representative of the rest.
Clearly such a model has the same number of interesting parameters 
as the model discussed in \S\ref{Sec:reflection}.

We find this model leads to an improvement over the results
obtained assuming model {\it C(ii)} for 10 of the 23 datasets 
at $>$99\% confidence. 
The best-fitting solutions for 4 of these datasets have $R_{s/h}\sim 1$
with both powerlaws of similar importance over a 
sizable fraction of the {\it ASCA} bandpass.
In the case of the remaining 6 datasets, 
the bulk of the {\it ASCA} bandpass is dominated by 
a single powerlaw, with the second powerlaw component making a significant 
contribution only at either the very lowest energies (the 
'ultra-soft excess' datasets) 
or highest energies (the 'hard tail' datasets).

We consider first the 3 datasets (3C~120, NGC~4593, Mrk~509)
for which this model implies a 'hard tail'. Naturally we find
the powerlaw which dominates the observed spectrum at the highest 
energies to be relatively weak ($R_{s/h}>>1$) and flat (although its 
photon index is poorly constrained within the range 
$0.2 \lesssim \Gamma_h \lesssim 1.8$).
In each case the steeper powerlaw, dominating the observed spectrum 
over most of the {\it ASCA} bandpass, has a photon index steeper by 
$\Delta \Gamma \sim 0.2$ than that found assuming model {\it C(ii)}
(in fact all 3 datasets have $\Gamma_s \simeq 2.2$).
However, it should be noted that the underlying continua derived 
for these 3 datasets assuming this model exhibit only a 
relatively subtle curvature across the {\it ASCA} bandpass.
It is therefore not that surprising that in all cases the best-fitting values
of the parameters associated with the ionized gas ($U_X$ and $N_{H,z}$) are
similar to those found when the underlying continuum is approximated by a
single powerlaw.

Turning now to the 3 datasets for which this model implies an
'ultra-soft excess' (NGC~3516, MCG-6-30-15 (1,2)), in all cases we find 
the powerlaw which dominates the observed spectrum over most of 
the {\it ASCA} bandpass has a photon index flatter by $\Delta \Gamma \sim 0.1$
than that found for model {\it C(ii)}.
The best-fitting solutions imply 
a very steep second powerlaw component (pegging at our upper limit of
$\Gamma_s = 5$ for all 3 datasets) dominating the observed spectrum only 
at the very lowest energies
(with $R_{s/h} \sim0.3$--0.5). 
However, none of the best-fitting models extrapolate $<$0.6~keV in a 
satisfactory manner (even given best-fitting values of 
$N_{H,0}$ significantly in excess of $N_{H,0}^{gal}$), and 
all imply an $\alpha_{ox}$ far steeper than that observed.
Further investigation of such a component is extremely difficult with 
{\it ASCA} data.
However, again,
the best-fitting values
of the parameters associated with the ionized gas ($U_X$ and $N_{H,z}$) are
similar to those found when the underlying continuum is approximated by a
single powerlaw.                   

Finally, we consider the 
remaining 4 datasets for which this analysis indicates both powerlaws 
are of similar importance over a sizeable fraction of the {\it ASCA}
bandpass
(Mrk~335, NGC~4051, IC~4329A, NGC~5548).
The best-fitting parameters of for these datasets are listed in 
Table~11, and the 
derived model spectra and data/model ratios 
shown in Fig~15.
As further discussed in the Appendix, the form of the continuum
implied for Mrk~335 is in good agreement with previous observations.
However any imprints due to ionized gas are weak in Mrk~335, 
allowing no useful constraints to be placed 
on $N_{H,z}$, $U_X$ or $\Omega$.
The best-fitting solution for the other 3 datasets are somewhat 
similar, with a relatively steep ($\Gamma_s \gtrsim 3$) powerlaw 
dominating the spectrum $\lesssim 1$~keV along with evidence for 
absorption and emission from ionized gas.
However, in all 3 cases the reality of such a continuum can be questioned. 
Significant spectral and flux variability occur on timescales much shorter 
than the 
observation period in the case of NGC~4051 (Paper~I). 
Thus we consider short-timescale spectral 
variability may be partly responsible for the apparent curvature of the 
underlying continuum in NGC~4051.
As is apparent from Table~11, 
this model fails to satisfy our criteria for acceptability 
in the case of IC~4329A. As further discussed in the Appendix, this 
source appears to have a particularly complex spectral form, and 
hence it is difficult to judge the robustness of the values derived 
for $N_{H,z}$, $U_X$ or $\Omega$.
In the case of NGC~5548, an acceptable fit is achieved assuming this
model. However it should be noted 
that the steep ($\Gamma_s \sim 3.2$), soft component derived
does appear to be a relatively poor representation of the 
data $< 0.6$~keV, implying a neutral column density 
far in excess of $\geq N_{H,0}^{gal}$
($\Delta N_{H,0} \sim 1.5\times10^{21}\ {\rm cm^{-2}}$)
and giving $\frac{\Delta \chi^2_{0.6}}{\Delta N_{0.6}} \sim 14$ 
(though $\overline{R_{0.6}} \sim 1$).
Finally, 
as shown in \S\ref{Sec:reflection}, 
a single powerlaw with a strong Compton-reflection component can provide 
a comparable fit for all 3 datasets.	

\subsubsection{The effects of including a low-temperature, thermal component}
\label{Sec:thermal}

It has long been suggested that Seyfert galaxies are likely to contain several
sites which give rise to thermal emission at temperatures consistent
with soft X-rays. These sites occur on different size-scales and arise as a
result of processes both directly related to and independent of the AGN
phenomenon. Examples include regions of active star formation, 
the hot-phase of the narrow emission-line region (NELR), 
the halo of the galaxy and the shock- or X-ray heated ambient 
interstellar medium (ISM).
Thermal emission closely associated with the active nucleus is also expected
as a result of the accretion process itself and as a result of the 
absorption (followed by thermalization) of the continuum.
The relative importance of such components probably differs 
from object to object. In the case of NGC~4151, {\it Einstein} HRI and 
{\it ROSAT} HRI observations have revealed spatially resolved soft X-ray 
emission (Elvis, Briel \& Henry 1983; Morse et al 1995).
{\it ASCA} observations of normal galaxies have also revealed soft X-ray 
emission with temperatures in the range $kT \sim 0.3$--0.6~keV, and luminosities 
$L_{kT} \sim 10^{38}$--$10^{40} {\rm erg\ s}^{-1}$ (e.g. Makishima 1994; 
Serlemitsos, Ptak \& Yaqoob 1996). 
Thus one could question whether the existence of such a component might 
effect our conclusions regarding the photoionized gas -- particularly since the 
thermal emission offers an alternative to 
the decreased opacity and emission features at the lowest energies
implied in previous models.

We have investigated this by including a component due to an optically-thin
plasma (with cosmic abundances, parameterized by a temperature $kT$ and 
unabsorbed luminosity $L_{kT}$ in the 0.1--3.0~keV band) to model 
{\it C(ii)} (but located  {\it outside} the photoionized absorber).
We find that the goodness-of-fit is indeed improved ($\Delta \chi^2 \sim 9$)
in the case of all 3 datasets of NGC~4151, with $kT \sim 0.66$~keV and 
$L_{kT} \sim 2^{+3}_{-2} \times 10^{41}\ {\rm erg\ s}^{-1}$, 
consistent with the luminosity of the extended gas measured by the 
{\it ROSAT} HRI observation (see \S\ref{Sec:multi-4151}
and Appendix). 
However, it should be noted that in the {\it ASCA} bandpass, 
the dominant features from such gas are Fe $L$-shell lines in 
the 0.7--1.5~keV regime. Thus whilst $L_{kT}$ is a function of 
$A_{\rm Fe}$, the addition of this optically-thin component does not 
significantly affect the derived values for all other of the parameters
(although the size of the allowed parameter space is obviously increased --
e.g. see Fig~16).

We find no other dataset is significantly improved by the inclusion of 
emission from thermal gas at such a temperature. This is hardly surprising 
since only NGC~4151 occupied a region of the $N_{H,z}$--$U_X$ plane 
in which emission of similar strength would not be swamped by the 
underlying continuum. Thus we consider the the omission of such a component 
is generally not important in our results. 

\subsubsection{Summary}

In summary, we find evidence for significant curvature in the 
{\it observed} continuum in approximately half
of the datasets considered. 
We have shown that for the majority of the datasets, such curvature
can be accounted for by the inclusion of a Compton-Reflection and/or a second 
powerlaw component. 
However, 
in the majority of cases, the ionizing continuum can be well
represented by a single powerlaw (with $\Gamma \sim 2$) close to 
the 
O{\sc vii} and O{\sc viii} absorption edges.
These edges provide the main diagnostics for our photoionization models.
Thus we consider that the best-fitting 
parameters associated with the ionized 
gas derived in models {\it B} and {\it C} are generally
reliable. 
Allowing the presence of a low-temperature, thermal component improved the fit
only in the case of NGC~4151. However, again we found it to have no
significant effect on the best-fitting parameters associated with the ionized
gas.

\section{OBJECTS WITH MULTIPLE OBSERVATIONS}
\label{Sec:disc-multi}

Our sample contains
more than one observation of NGC~3783, NGC~4151, MCG-6-30-15 and 
Mrk~841. We have found evidence for absorption by ionized material in all 
four sources, enabling us to compare and contrast the behaviour of the 
ionized gas.

\subsection{NGC~3783}
\label{Sec:multi-3783}

Two datasets of NGC~3783, separated by 4 days, are contained within our 
sample from observations carried out in 1993 December.
As shown in \S\ref{Sec:ion}--\ref{Sec:ion_pc_emis}, there is 
clear evidence for ionized gas in this source, with
models {\it B(i)}--{\it C(ii)} providing acceptable fits to both datasets.
The observed count rate was $\sim25$\% higher for the second observation
(e.g. Table~2), which is reflected 
in an increase in the derived luminosity of the ionizing 
continuum (for model {\it C(i)}, 
after correcting for absorption)
from $L_{X} \simeq 4.6\times10^{43}\ {\rm erg\ s^{-1}}$
to $\simeq 5.5\times10^{43}\ {\rm erg\ s^{-1}}$.
Both observations are consistent with $\Gamma \sim 1.8$, 
$D_f = 0$,
$U_X \sim 0.1$--0.15 and
$N_{H,z} \sim 2\times10^{22}\ {\rm cm^{-2}}$
(Table~8).
However, as shown in Fig~16, the two 
datasets actually occupy slightly different regions of the $N_{H,z}$--$U_X$ 
plane, consistent with an increase in $U_X$ by $\lesssim 40$\% and/or a 
decrease in $N_{H,z}$ (by $\lesssim 5\times10^{21}\ {\rm cm^{-2}}$)
between the 2 epochs.
The former offers the 
simplest interpretation of these data: 
a screen of ionized gas completely covers the cylinder-of-sight in NGC~3783, 
becoming more highly ionized as the photoionizing flux increases.
Both observations are consistent with the presence of emission from 
the ionized material, with $L_e \sim 2$--$6\times10^{42}\ {\rm erg\ s^{-1}}$
in the 0.1--10~keV band (after correcting for absorption).
Unfortunately however, the signal-to-noise ratio of these datasets is 
too low to 
unambiguously determine whether the intensity of this emission 
responded to the increase in the ionizing continuum.

Interestingly, a '$1$~keV deficit' was also 
evident for both datasets in the data/model residuals 
of all the models considered in \S\ref{Sec:basic_models}.
This can be modeled by an additional absorption edge 
($\tau_{ONe} \sim 0.5$) at an energy close to that of O{\sc viii} 
(\S\ref{Sec:2ion}) and most likely due to an additional screen of 
more-highly ionized gas covering part, or all, the cylinder-of-sight.

\subsection{NGC~4151}
\label{Sec:multi-4151}
\label{Sec:bolX}

Three datasets of NGC~4151 fulfilled the criteria defining this 
'sample' (Paper~I). These were an observation performed 
1993 Nov 05 (NGC~4151(2)), an observation performed
exactly 30 days later (NGC~4151(4)), and an observation performed
2 days thereafter (NGC~4151(5)).
The observed count rate for NGC~4151(4) was only $\sim2$\% higher than 
that for NGC~4151(2), but that for NGC~4151(5) was 
$\sim15$\% higher.
As shown in Table~12, we find a fit satisfying 
our criteria for acceptability is achieved for the first two observations 
(NGC~4151(2,4)) only when the absorbing material is allowed to be
ionized {\it and} a fraction of the underlying continuum is allowed to
escape without suffering any attenuation 
(i.e. models {\it B(ii)} \& {\it C(ii)}).
Such an hypothesis is strongly supported by the third observation 
(NGC~4151(5)), even though none of the models in \S\ref{Sec:basic_models} 
provide a solution which formally satisfies our criteria for acceptability
(see Fig.~9).

As can been seen from 
Tables~6 and~8, 
all 3 datasets are consistent with $D_f \sim 3$--4\% and $\Gamma \sim 1.5$.
However the parameters associated with the ionized gas vary in an 
interesting manner (see also Fig~16).
In the month between the NGC~4151(2) and NGC~4151(4) observations, there is 
an increase in the absorbing column density 
($\Delta N_{H,z} \sim 2\times10^{22}\ {\rm cm^{-2}}$) 
and increase in ionization parameter ($\Delta U_X \sim 30$\%), whilst the 
derived luminosity of the ionizing continuum (in the 0.1--10~keV for model 
{\it C(ii)}, after correcting for absorption) increases by 
$\lesssim 10$\% (from $L_{X} \simeq 1.3$ to 
$1.4\times10^{43}\ {\rm erg\ s^{-1}}$).
Two days later, at the time of the NGC~4151(5) observation, 
$N_{H,z}$ is similar to NGC~4151(4). However the 
ionization parameter has decreased ($\Delta U_X \lesssim 15$\%)
despite an increase in the luminosity of the ionizing continuum 
(to $L_{X} \sim 1.8\times10^{43}\ {\rm erg\ s^{-1}}$).
Such behaviour is difficult to understand in terms of a simple 
model, unless the $N_{H,z}$ and/or the number density of the gas $n_H$,
(but not $D_f$)
are allowed to vary on timescales $\lesssim 2$ days, or unless the 
photoionized gas is out of equilibrium. 
A simple explanation of such behaviour is that the absorbing gas 
completely covers the cylinder--of--sight but is 
inhomogeneous (with differing $N_{H,z}$ and/or $n_H$) on a scale
comparable or greater in size to that of the central source, 
whilst a constant fraction $D_f$ of the underlying continuum
follows an alternate path.
If the transverse velocity of the absorbing material is sufficiently high, 
then the movement of such clumps across the cylinder--of--sight would 
then give rise to variations in $U_X$, $N_{H,z}$ seemingly decoupled from 
variations in the intensity of the underlying source.
Assuming only Keplerian motion, the length-scale $l$ passing through a given 
line--of--sight in a time $t$ is 
$l \simeq 2\times10^{8} (f_{bolX}/f_{Edd})^{0.5} L_{X44}^{0.5}
r_{ld}^{-0.5} t\ {\rm cm}$ 
at a distance $r_{ld}$~light-days from a central object with a bolometric 
luminosity $L_{bol} = f_{bolX} L_{X44}$ emitting at a fraction $f_{Edd}$ of 
its Eddington luminosity, where 
$t$ is in seconds, 
$L_{X44} = 10^{44} L_{X}\ {\rm erg\ s^{-1}}$ 
and $L_{X}$ defined over the 0.1--10~keV band.
Equating $t$ with the timescale $t_{abs}$ on which significant 
variations in $N_{H,z}$ are seen, and putting $l \geq c t_{var}$, 
the size of the central source, 
we obtain 
\begin{equation}
\label{eqn:r_ld}
	r_{ld} \lesssim 4.5\times10^{-5} (f_{bolX}/f_{Edd}) L_{X44} 
(t_{abs}/t_{var})^2\ {\rm light-days}
\end{equation}
The ratio $f_{bolX}/f_{Edd}$ is generally considered 
to be $\sim$few, and 
from our analysis we find $L_{X44} \simeq 0.15$ for NGC~4151. 
Considering the variability between the 4151(2) and 4151(4) datasets 
and thus setting $t_{abs} \sim 30$~days, 
we obtain the limit $r_{ld} \lesssim 45 (f_{bolX}/f_{Edd})$~light-days
assuming $t_{var} \sim 10^{3}$~s. The limit if the 
same explanation is proposed to explain the 
variations between the 4151(4) and 4151(5) datasets ($t_{abs} \sim 2$~days)
is a factor $\sim200$ smaller.
We stress that these limits are based only upon the Keplerian motion, 
which may not be the major component governing the dynamics of the 
photoionized clouds.

In \S\ref{Sec:thermal} we found that there was evidence for an additional
spectral component, which we tentatively identified with the extended soft 
X-ray emission seen in this source and modeled as an optically-thin plasma. 
We found that all 3 datasets are consistent with such gas with 
$kT \sim 0.7$~keV and 
$L_{kT} \sim 2^{+3}_{-2} \times 10^{41} {\rm erg\ s}^{-1}$. 
Such a component alone (but absorbed by $N_{H,0}^{gal}$) would give rise to 
a count rate in the range $10^{-3}$--$0.07\ {\rm count\ s^{-1}}$ for the 
{\it ROSAT} HRI, and hence is consistent with that reported for the extended 
emission in this source ($0.06\ {\rm count\ s^{-1}}$, Morse et al 1995).
However, the inclusion of this component did not significantly affect the 
values derived for the ionized gas
(dashed contours in Fig~16), 
and hence the behavior of the source 
between the 3 observations still suggests clumps of gas of differing 
$N_{H,z}$ traversing the cylinder--of--sight.
These results are further discussed in relation to previous studies in the 
Appendix.

\subsection{MCG-6-30-15}
\label{Sec:multi-mcg6}

Two datasets of MCG-6-30-15 are contained within our sample from observations 
carried out in 1993 July, separated by $\sim3$ weeks, with 
the observed count rate $\sim52$\% higher for MCG-6-30-15(1)
(Table~2).
As shown in 
\S\ref{Sec:ion}--\ref{Sec:ion_pc_emis}, there is clear evidence for ionized 
gas in this source, with models {\it B(i)}--{\it C(ii)} providing acceptable 
fits. In all cases a 
'$1$~keV deficit' 
was also evident, 
which can be modeled by an additional absorption edge 
(\S\ref{Sec:2ion}).
The best-fitting continuum for MCG-6-30-15(1) has $\Gamma \sim 2.1$, whilst 
that for MCG-6-30-15(2) is slightly flatter ($\Delta \Gamma \sim 0.1$).
Both observations are consistent with $D_f = 0$.
As shown in Fig~16, 
there is clear evidence for an increase in $N_{H,z}$
(by $\sim 1$--$6\times10^{21}\ {\rm cm^{-2}}$) 
over the 3 weeks between the observations.
However, the ionization parameter remains constant, or perhaps 
even increases (but with $\Delta U_X \lesssim 50$\%), between 
the 2 observation despite a decrease in the 
luminosity derived for the ionizing continuum 
from $L_{X} \simeq 4.5\times10^{43}\ {\rm erg\ s^{-1}}$
to $\simeq 3.0\times10^{43}\ {\rm erg\ s^{-1}}$ (for model {\it C(ii)}, 
after correcting for absorption).
Such behaviour has been reported previously for these datasets by 
Fabian et al (1994), 
and as in the case NGC~4151, is difficult to reconcile with that expected in 
the simplest picture whereby a uniform shell of gas reacts to 
changes in the intensity of the photoionizing continuum.
Again appealing to only Keplerian motion to 
move inhomogeneities through the cylinder--of--sight
between the 2 observations (i.e. setting $t_{abs} \lesssim 3$~weeks),
and again setting the variability $t_{var} \sim 10^{3}$~s,
from Eqn~\ref{eqn:r_ld}
one obtains the rather weak constraint 
$r_{ld} \lesssim 5\times10^2 (f_{bolX}/f_{Edd})$~light-days. 
Again we note that these limits are based upon only the Keplerian motion, which 
may not be the major component of the dynamics of the ionized gas.

However, there is evidence in a subsequent {\it ASCA} observation that 
whilst the depth of the O{\sc vii} edge 
appears to be constant on long timescales in this source, the depth of the
O{\sc viii} edge varies on a timescale $\sim 10^{4}$~s (Otani et al 1996).
Such behaviour has been recently confirmed in {\it BeppoSAX} LECS 
observations (Orr et al 1997).
Setting $t_{abs} \sim 10^{4}$~s in Eqn~\ref{eqn:r_ld} one obtains 
$r_{ld} \lesssim 4 (f_{bolX}/f_{Edd})$~light-hours.
However, since the photoionization and recombination timescales for these ions
are similar, such behaviour is incompatible with 
the observed O{\sc vii} and O{\sc viii} edges being produced in the 
same region of ionized gas.
Otani et al suggested the most natural solution was a two-zone model
in which the gas giving rise predominately to the O{\sc vii} edge
is stable, whilst that responsible for the O{\sc viii} edge undergoes 
rapid variations. Such an hypothesis is of course supported by our 
observations of a (relatively small)
'$1$~keV deficit' 
in the data/model residual of all the models considered in 
\S\ref{Sec:basic_models}.
In \S\ref{Sec:2ion} we showed that
this can be modeled by an additional absorption edge 
($\tau_{ONe} \sim 0.14$) at an energy close to 
that of O{\sc viii} (see \S\ref{Sec:disc-2ion}).

Both observations of MCG-6-30-15 are consistent with the presence 
of emission from 
the ionized material, with $L_e \lesssim 10^{42}\ {\rm erg\ s^{-1}}$ 
in the 0.1--10~keV band (after correcting for absorption).
However, the signal-to-noise ratio of these datasets is 
too low to determine unambiguously whether the intensity of this emission
responded to the decrease in the ionizing continuum.

\subsection{Mrk~841}
\label{Sec:multi-841}

Two datasets of Mrk~841 are contained within our sample from 
observations carried out in 1993 August and 1994 February.
The observed count rate was ($\sim33$\%) higher for Mrk~841(1)
(e.g. Table~2). 
In \S\ref{Sec:basic_models} we
found evidence for ionized gas in this source. 
For models {\it B(i)}--{\it C(ii)}, the underlying continuum for Mrk~841(1) has 
$\Gamma \sim 1.9$, but the values derived for $U_X$ and $N_{H,z}$
are rather model dependent and often poorly constrained.
In contrast, the data from Mrk~841(2) are consistent with a simple powerlaw, 
preventing any strong constraints being placed on $U_X$ and $N_{H,z}$.
In all cases, the region of the $U_X$--$N_{H,z}$ plane occupied 
by Mrk~841(1) for a given model is consistent with the limits 
from Mrk~841(2) assuming that model.
However the underlying continuum for Mrk~841(2) does 
appear to be significantly flatter ($\Gamma \sim 1.7$).
Dramatic spectral variability has been observed previously in this source
(see Appendix and references therein).
The luminosities derived for the ionizing 
continuum 
are $L_{X} \simeq 1.6\times10^{44}\ {\rm erg\ s^{-1}}$
and $\simeq 1.1\times10^{44}\ {\rm erg\ s^{-1}}$
for Mrk~841(1) and Mrk~841(2) respectively (for model {\it C(ii)}, 
after correcting for absorption).
In view of the poor constraints which can be placed on the ionized gas 
from these data, few further comments can be made.

\section{DISCUSSION \& CONCLUSIONS}
\label{Sec:Discussion}

As discussed in the introduction, it is becoming increasingly 
clear that the spectra of a number of individual AGN contain features 
indicative of absorption by ionized gas within the cylinder-of-sight.
It seems highly likely that this gas is photoionized by the intense radiation 
field produced by the central object. 
The aim of the work presented here was to determine the 
frequency and characteristics of such gas based upon new,
self-consistent modelling using the {\tt ION} photoionization code. 

In the preceeding sections we have presented the results from 
23 {\it ASCA} observations of 18 objects. When considered together
these objects certainly do not constitute a complete sample.
Nevertheless, in at least some aspects, we do believe our results to 
be representative of the variety of properties exhibited by the type-1 
AGN as a whole.

In \S\ref{Sec:basic_models} we considered models assuming the underlying 
continuum in the 0.6--10~keV band of our sources is well-represented by a 
single powerlaw.
A number of models were considered, starting with the case in which 
the spectrum emerging from the source passes through
two screens of neutral material fully covering the source: one at 
the redshift of the source and a second at zero redshift
(with column densities $N_{H,z}$ and $N_{H,0}$ respectively).
Models of increasing complexity were then considered 
by allowing a fraction ($D_f$) of the underlying continuum 
to be observed {\it without} attenuation by $N_{H,z}$, 
allowing the gas represented by $N_{H,z}$ to 
be photoionized (with an ionization parameter $U_X$), and finally 
by also 
including the emission spectrum expected from the ionized gas
(assuming it subtends a solid angle $\Omega$ at the central source).

In \S\ref{Sec:additional_models} the results of further spectral analysis 
were presented, including models with additional spectral components,
with the primary goal of testing the robustness of the derived 
characteristics of the ionized gas.

\subsection{General Considerations}
\label{Sec:disc-general}

All models considered in this work are single-zone models, and given the
assumed gas density $n_H = 10^{10}\ {\rm cm^{-3}}$, the gas reacts 
instantaneously to continuum variations. However, the models are intended to 
represent a wide range of conditions including cases with much lower values 
of $n_H$. In such cases, the level of ionization represents the response of 
the gas to the long-term, average flux of the illuminating source. We have 
not considered all such cases here since we do not consider the data yet 
warrant this level of discussion. 

In \S\ref{Sec:basic_models} we found that one or more of the models considered 
offer an explanation of 17/23 datasets within the formal criteria outlined 
in \S\ref{Sec:acceptability}. The criteria used to determine the acceptability 
of model as a description of the data is of course a somewhat subjective 
decision. The primary criterion used here that 
$P(\chi^2 \mid dof)\leq 0.95$ (i.e. that the model is a true representation of 
the data with a probability of better than only 1 chance in 20) might be 
considered relatively lax by some readers. Furthermore, as can been seen from
Tables~3--8, a 'perfect' 
parameterisation of the data ($P(\chi^2 \mid dof) \simeq 0.5$ corresponding to 
a reduced--$\chi^2$ value of unity) is only obtained in a small number of cases.
Relatively high values of $P(\chi^2 \mid dof)$ imply that there are
residuals in the data not modelled adequately. However, whilst some of these 
residuals might be carrying physical information associated with the source, 
others might be the result of inaccuracies in the calibration 
of the instruments (see also Fig.~13). 
The latter is currently under intense study by the instrument teams 
in anticipation of being able to assign unmodelled residuals to the former 
explanation with a higher degree of confidence.

However our intention in this paper is to explore just how well 
the data from the sample can be described by our relatively simple models 
for the ionized gas.
We believe this approach compliments that taken by some previous workers 
where the absorption features introduced by the ionized material are 
modelled as a series of absorption edges.
Throughout the analysis we have not assumed any systematic 
errors, and 
have implicitly assigned the relatively poor values of 
$P(\chi^2 \mid dof)$ to be the result of problems with the calibration in 
the case of 4 additional datasets in \S\ref{Sec:basic_models} 
(Table~12).
Furthermore, beyond the limited number considered in 
\S\ref{Sec:additional_models}, we have not explored more complex models
in detail.
For instance we have not considered models in which the 
ionized material has a significant ($\gtrsim 10^4\ {\rm km\ s^{-1}}$) velocity
dispersion along the line-of-sight. Given the spectral resolution of the SIS
detectors, dispersions smaller than this value are impossible to constrain with
the signal-to-noise ratio of the data presented here (see also
\S\ref{Sec:redshift}). We have not explored different abundance ratios, or
multi-component models in those cases where we have indications for their
existence (see \S\ref{Sec:disc-2ion}).
Our intention in the present work is 
merely point the reader to the need of such analysis and defer it 
for future discussion.        

\subsection{The form of the underlying continuum}
\label{Sec:disc-2plaw}

It is well established that many Seyfert~1 galaxies exhibit so-called 
'soft-excesses' and 'Compton-reflection' tails (e.g. Turner \& Pounds
1989; Nandra \& Pounds 1994).
After accounting for these effects, and those of the ionized-absorber,
{\it Ginga} observations showed that 
the underlying X-ray continua can be generally well represented by a 
powerlaw of 
photon index $\Gamma \sim 1.9$--2.0 (e.g. Nandra \& Pounds 1994). 
In \S\ref{Sec:basic_models} we obtained acceptable fits for most of the 
datasets by one or more of the models considered, with the distribution 
of the photon index peaked at $\Gamma \sim 2$ (Fig.~2).
In \S\ref{Sec:complex_cont} we found significant improvements in the 
goodness--of--fit were obtained for many of the datasets by the inclusion of 
additional spectral components. 
However, in the majority of cases, we again found the bulk of 0.6--10~keV 
bandpass to be consistent by an underlying powerlaw with $\Gamma \sim 2$.
It should be noted that in a number of cases the best-fitting models 
imply that very little of the underlying continuum is observed directly
(e.g. see NGC~3783(1,2) in Fig~6; 
NGC~4151(2,4,5) in Fig~9).
In passing we note that in such cases, $\Gamma$ can still be well constrained
by the data presented here as the shape of the 'recovery' in the 
absorption features in the 1--5~keV band is well defined in the models and
well constrained by {\it ASCA} data.
For comparison with other work, we have calculated the expectation value
and dispersion of the distribution of spectral indices, employing
the method of Maccacaro et al. (1988). Considering those datasets 
for which we obtained acceptable fits, 
we find $<\Gamma> = 1.94 \pm 0.08$, with a significant, intrinsic dispersion
of $0.18\pm 0.07$. 
These values are in excellent agreement with
those found by {\it Ginga} and in Paper~II, when the effects of Compton
reflection are taken into consideration.

In a number of cases, complex continua do seem to be present in the
{\it ASCA} bandpass.
Mrk~335 does indeed appear to possess real curvature in the underlying 
continuum (Fig~15).
Evidence was also found for curvature in the observed continuum 
in NGC~4051, IC~4392A \& NGC~5548. However in at least 
NGC~5548
we consider the evidence for curvature in the {\it underlying} continuum 
to be questionable and/or 
alternative explanations are available (see \S\ref{Sec:2plaw}).
The observed spectrum of NGC~4151 has long been considered 
somewhat idiosyncratic amongst Seyfert 1 galaxies. However we do find 
acceptable solutions and consider the underlying continuum to indeed 
be a powerlaw, albeit flatter ($\Gamma \sim 1.5$) 
than that observed for the majority of sources
(see \S\ref{Sec:disc-multi} and \S\ref{Sec:disc-Df}).

\subsection{The frequency of neutral absorbers}
\label{Sec:disc-freq-neut}

Unlike Seyfert 2 galaxies, Seyfert 1 galaxies are generally not believed to 
contain significant column densities of neutral gas detectable in
the 0.6--10~keV band (ie. within the range 
$10^{20} \lesssim N_{H,z} \lesssim  10^{25}\ {\rm cm^{-2}}$).
In the case of models {\it A(i)} and {\it A(ii)}, 
the absorption at the redshift of the source 
(with column density $N_{H,z}$) is assumed to be due to 
neutral material.
As described in \S\ref{Sec:zwabspo}, significant absorption is not detected 
in any of the datasets for which model {\it A(i)} offers an acceptable 
solution (with $N_{H,z} \lesssim$ few $\times10^{20}\ {\rm cm^{-2}}$;
see Fig~3a).
Model {\it A(ii)}, in which a fraction $D_f$ of the underlying 
continuum is able to escape without passing through $N_{H,z}$,
does offer an improvement in the goodness--of--fit for many 
datasets, and a range of derived column densities 
for those in which an acceptable solution was obtained
(Fig~3b).
However, as discussed in \S\ref{Sec:zpcfpo}, these solutions 
are most likely due to a residual problem in the instrument 
calibration (e.g. 3C~120), lie in a region of the 
$N_{H,z}$,$D_f$ indicative of (relatively-subtle) spectral curvature 
(e.g. NGC~7469),
and/or superior solutions are obtained assuming subsequent 
models (e.g. Mrk~335). 
We note that 
it has long been suggested that NGC~4151 might contain 
a significant column density attenuating $\sim 95$\% of the continuum.
However as shown in \S\ref{Sec:basic_models}, we find acceptable 
solutions only for models in which the gas is ionized
(see also \S\ref{Sec:disc-multi} and \S\ref{Sec:disc-Df}).

We have not explicitly tested for {\it additional}
intrinsic column densities of neutral material in any of our models which 
contain ionized gas (i.e. in models {\it B(i)}--{\it C(ii)}).
Such a situation might 
arise for example in the case where the primary continuum 
passes through a region of ionized gas, and then a further 
screen of neutral gas during its passage through the host galaxy.
Nevertheless, all our models do contain a neutral 
column at zero redshift, $N_{H,0}$, to account for the 
Galactic absorption (and hence expected to be $\simeq N_{H,0}^{gal}$).
At the redshift of most of the sources in our sample, 
any absorption by neutral material intrinsic to the source would be 
indistinguishable from absorption at zero redshift. Thus any 
such absorption would result in a derived value of 
$N_{H,0} >> N_{H,0}^{gal}$.
Generally we find no such evidence in our sample,
with the only reliable detection in the case of IC~4329A 
(with a neutral column density $\sim 4\times10^{21}\ {\rm cm^{-2}}$).
The host galaxy is this AGN is observed edge-on so this result is not 
surprizing (see Appendix).

\subsection{The frequency of ionized absorbers}
\label{Sec:disc-freq}

In \S\ref{Sec:basic_models} we found 16/23 datasets (11/18 objects) are 
improved at $>99$\% confidence over models {\it A(i)} \& {\it A(ii)}
if the absorbing material is assumed to be photoionized.
Of these, 12/16 datasets (9/11 objects) can be adequately described by
one or more of models including ionized gas presented in 
\S\ref{Sec:basic_models} (i.e. models {\it B(i)}--{\it C(ii)}), under 
the assumption of a single power-law continuum. 
Of the remaining 4/16 datasets,
3 datasets (only one of which being from an additional object: NGC~3516) 
show strong evidence  for ionized gas (with the 
statistical poverty of the fit due to further spectral complexity).
Thus we find clear evidence for absorption by ionized gas in 
15/23 datasets (10/18 objects).
However the presence of ionized gas is also strongly suspected 
in 3 additional datasets (3 additional objects: NGC~4051, 
IC~4329A, NGC~7469) from the analysis presented in 
\S\ref{Sec:additional_models}. 
\begin{itemize}
\item	{\it Thus we conclude there is evidence for ionized gas in 
	18/23 datasets (13/18 objects) in our sample}.
\end{itemize}

Reynolds (1997) has also presented results from a sample of 24 type-1 
AGN using {\it ASCA} data, using the same datasets as presented here 
for 15 sources (Table~2).
Based on a search for O{\sc vii} and O{\sc viii} edges, 
Reynolds found strong evidence for ionized gas in 12/24 objects.
These 12 include 9 sources also in our sample 
(Table~12), along with 
Mrk~290, 3C~382 and MR~2251-178.
Reynolds also fitted these data with a single-zone photoionization
model (based on the photoionization code \verb+CLOUDY+
which is similar in approach and scope to \verb+ION+ used here),
assuming a photoionizing continuum consisting of a single
powerlaw over the entire 0.0136--13.6~keV band. 
Converting their
quoted ionization parameters ($\xi$) to $U_X$ using Fig.1b, 
gives 
$1.5 \lesssim \Gamma \lesssim 2.0$,
$10^{21} \lesssim N_{H,z} \lesssim 2\times10^{23}\ {\rm cm^{-2}}$
and
$0.02 \lesssim U_X \lesssim 0.09$, 
broadly in agreement with the distributions found 
here (Figs~2, 3 and
7 respectively).
A detailed source-by-source comparison between our results 
and those of Reynolds (1997) and other workers is 
provided in the Appendix.

\subsection{Column density and ionization parameter of the ionized gas}
\label{Sec:disc-hagai}

We find no preferred value of $N_{H,z}$ under any of the models tested,
with clear differences from object--to--object (e.g. Fig.~3).
However, in each of the models including photoionized gas, we find the 
distribution in the ionization parameter strongly peaked at $U_X \sim 0.1$ 
(Fig.~7). Considering those datasets for which we obtained 
acceptable fits (and again employing the method of Maccacaro et al. 1988), 
we find $<\log U_X> = -0.92 \pm0.21$. While the values cluster around this 
value, there is a highly significant intrinsic dispersion in 
$\log U_X$ of $0.21^{+0.32}_{-0.09}$.
Nevertheless this relatively narrow range in $U_X$ 
is our most interesting and unexpected result. As such it deserves 
some further discussion as the whether it is purely the result of 
selection effects, or has physical significance.

We considering first values of $U_X >> 0.1$. 
With the signal--to--noise ratio typically afforded by {\it ASCA} data, 
the presence of ionized gas is most readily inferred by deep 
($\tau \gtrsim 0.2$) absorption edges which it introduces into the observed 
X-ray spectrum. For a given value of $N_{H,z}$, there is a maximum value of 
$U_X$ above which the gas is unstable and goes to the high electron 
temperature ($T_e  \sim 10^6$~K) branch of the two-phase curve 
(e.g. see Fig.~2 in Netzer 1996). 
The value of $U_X$ at which this thermal instability occurs 
is a function of the gas density and composition, along with the form of the 
ionizing continuum. Under the assumptions made here, the instability occurs for 
$U_X > 10^{-22} N_{H,z}$. Such gas is completely transparent, and the 
observed spectrum will be identical to the underlying (powerlaw) continuum, 
consistent with the data from a minority of objects in our sample.
Therefore the presence of such material along the cylinder--of--sight is
impossible to prove (or disprove) using the current {\it ASCA} data alone, 
offering a potential explanation of the apparent lack of objects with 
$U_X >> 0.1$ in Fig.~7.

For $U_X << 0.1$, different ions contribute different continuum
opacity, depending on the gas composition and level of ionization. For
$10^{21} < N_{H,z} < 10^{24}\ {\rm cm^{-2}}$ and
$0.03<U_X<0.3$, most of the opacity is due to O{\sc vii} and O{\sc viii}
(Fig~8).
For lower values of $U_X$, 
carbon, nitrogen and lower ionization states of oxygen are more important. 
However, we note that none of the objects presented here lay in the region of 
$U_X$--$N_{H,z}$ parameter-space where $U_X <0.03$, 
$N_{H,z}> 10^{22}\ {\rm cm^{-2}}$. 
There is no reason {\it a priori} why there should not be objects 
with ionized material in this region.
However, there is a potential selection-effect here whereby if such 
material lays along the cylinder--of--sight to the broad emission line region 
(BELR) and contains embedded dust, the broad emission lines will be 
absorbed by this material. This may lead to such objects being classified 
as something other than Seyfert 1 galaxies, and hence being excluded 
from the sample of sources presented here.
Alternatively, should this explanation not be the case, then there must be 
some yet unknown mechanism 
that determines these unique conditions by either fixing the pressure or 
perhaps the dynamical state of the absorbing material.
Future studies of a larger sample of objects are required to investigate 
these possibilities.

We have compared the parameters of the ionized gas with 
the derived continuum luminosity,
$L_X$, in the 0.1--10~keV band (after correcting for absorption).
We find no clear relation between $N_{H,z}$ and $L_X$, but 
some indication that $U_X \propto L_X$ (Fig.~17).
If true, from the definition of $U_X$ (Eqn.~\ref{eqn:U_X}), this 
implies the ionized gas has similar densities, $n_H$, and is at 
similar radii, $r$ in all sources.
However, the majority of the sources considered here
for which $U_X$ is well constrained lie within a restricted range of $L_X$
($10^{43}$--$10^{44}\ {\rm erg\ s^{-1}}$).
A recent observation of the 
quasar PG~1114+445 ($L_X \sim 6\times 10^{44}\ {\rm erg\ s^{-1}}$)
also exhibits strong absorption features due to ionized gas
with 
$N_{H,z} \simeq 2\times10^{22}\ {\rm cm^{-2}}$
and $U_X \simeq 0.1$
(George et al. 1997b). 
Furthermore, 
Nandra et al (1997e) have recently reported the results from {\it ASCA}
observations of the RQQ MR~2251-178
($z = 0.068$, $L_X \simeq 2\times 10^{45}\ {\rm erg\ s^{-1}}$),
finding 
$N_{H,z} \sim 2\times10^{21}\ {\rm cm^{-2}}$
and $U_X \sim 0.07$.
Inclusion of these objects on Fig.~17
argues against any obvious relation between 
$U_X$ and $L_X$.

We found no evidence for any relationship between any other pairs of 
parameters derived from the X-ray results
(e.g. $L_X$ vs $\Gamma$; $\Gamma$ vs $U_X$; $\Gamma$ vs $N_{H,z}$).
In most cases this is again primarily 
due to the limited range of $L_X$ and/or $\Gamma$
exhibited by most the objects in the sample.
Thus, 
further progress requires high signal-to-noise observations of 
similar sources covering the high and low $L_X$ regimes.

\subsection{Constraints on ionized emitters}
\label{Sec:disc-emis}

The dominant effects of the ionized gas on the observed spectrum from these
sources is absorption by the material along the line-of-sight.
However, 
ionized material out of the line-of-sight will give rise to emission lines
and recombination continua, along with the (absorbed) continuum
Compton-scattered back into the line-of-sight.
The inclusion of such an emission component leads to an improvement 
at $>$99\% confidence in 11 datasets (9 objects) in the full covering case
(model {\it C(i)}), 
and 8 datasets (7 objects) in the partial covering case
(model {\it C(ii)}).
However there is no case for which this component is 
{\it required} (i.e. a satisfactory fit only obtained for 
model {\it C}).
It should also be noted that the inclusion of the 
emission component generally does not have a significant effect on the 
values derived for $U_X$ and $N_{H,z}$.
For the value of $U_X$ and $N_{H,z}$ derived for the majority of sources,
the emitted spectrum contains significant line emission in the
0.6--2.0~keV band, but constitutes only a relatively
small fraction ($\lesssim 10$\%) of the total flux observed in this band,  
even when 
the ionized gas subtends a large solid angle 
at the central source.
The best-fitting values of $\Omega$ are ill-constrained, but 
in most cases consistent with $\Omega = 4\pi$, and
there is no case where $\Omega > 4\pi$ at $>95$\% confidence.
Formally we find $< \log \Omega/4\pi >$  $= 0.23^{+0.24}_{-0.23}$ for 
the datasets for which we obtained acceptable fits, 

In cases where the ionizing continuum source is constant, then
the normalization of the emission spectrum is a direct measure of 
the geometry of the emitting gas. However in cases where the ionizing continuum 
varies, then the normalization of 
the emission spectrum will follow the luminosity history of the 
continuum source, with a lag due to light travel-time effects. 
In the latter case we are therefore offered the opportunity to determine the 
location of the ionized gas via the delay in the intensity of the 
emission features following variations in the ionizing continuum.
Unfortunately however, the 
signal-to-noise ratio is too low for the few objects presented here 
for which there are multiple observations (\S\ref{Sec:disc-multi})
to place even crude 
constraints on whether the intensity of the emission 
changed in an appropriate manner.

\subsection{Evidence for an unattenuated component}
\label{Sec:disc-Df}

In \S\ref{Sec:basic_models} \& \S\ref{Sec:additional_models}, 
we find that allowing a fraction $D_f$ of the powerlaw continuum to be 
observed without suffering attenuation by the column density $N_{H,z}$ 
significantly improves the fit 
(compared to the corresponding model with $D_f=0$)
for a large number of dataset/model combinations.
Indeed, inspection of 
$F(\frac{Aii}{Ai})$ (Table~4) reveals 
all but 3 datasets are improved in the case of model {\it A(ii)},
and inspection of 
$F(\frac{Bii}{Bi})$ and $F(\frac{Cii}{Ci})$ 
(Tables~6 and 8)
reveals that 8 datasets are improved in the case of 
models {\it B(ii)} and {\it C(ii)}.
However, in many of these cases the best-fitting solutions have 
$D_f \sim 1$ (Fig.~5) and as described in 
\S\ref{Sec:complex_cont} are an artifact of spectral curvature.
Indeed, as shown in 
Table~12, only in the case of 2 datasets 
(1 object: NGC~4151) are solutions which satisfy our 
criteria for acceptability obtained {\bf only} if $D_f>0$.
Thus we conclude 
most of the sample show solutions close to the full-covering case 
and allowing $D_f$ as a free parameter
does not significantly change the properties of the sample.

As noted earlier, this model is appropriate to 
a partial covering of the cylinder--of--sight and to a geometry in 
which some fraction of the underlying continuum, initially emitted 
in other directions, is scattered back into the line-of-sight.
As discussed in \S\ref{Sec:disc-multi}, a possible 
explanation for the spectral variability observed in NGC~4151 consists of 
clumps of photoionized gas (of differing column density) moving through the
cylinder--of--sight on timescales $\lesssim 1$~day.
Under such an hypothesis, the fact that $D_f$ is similar for all the 
observations (at $\sim$5\%) suggests scattering as the most likely 
explanation. 
Circumstantial support is provided by the observation of a similar 
component in many type-2 AGN,
with implications for unification schemes (e.g. Turner et al 1997a,b).
Indeed the optical spectrum of NGC~4151 has been observed to exhibit 
both type-1 and type-2 characteristics at various times.
It should be noted that the parameterization used here
approximates the scattered 
continuum as a simple powerlaw as would be appropriate in the case of 
pure electron scattering in a very highly-ionized medium. In a more realistic
treatment, the ionization state of the scattering gas would be taken into
account, which under a wide range of parameter space would lead to 
emission and absorption features further complicating the spectrum 
$\lesssim 2$~keV.
It should be noted that the fact that such a component 
is unambiguously observed only in NGC~4151 amongst our sample is simply 
due to the suppression of the transmitted continuum by the high 
column density of gas within the cylinder--of--sight of this source
(with the gas having $U_X$ such that there's still significant opacity)

\subsection{Implications of the '1~keV deficit'}
\label{Sec:disc-2ion}

A number of sources have evidence for a deficit of counts 
at $\sim$1~keV compared to the predictions of our photoionization models. 
As noted above, such a deficit occurs in those sources exhibiting 
the strongest absorption features.
In \S\ref{Sec:2ion} we showed that this could be modeled with 
an additional edge (consistent with O{\sc viii}) in the rest frame of the 
source. 

There are several possible explanations of such a feature.
First, it could be indicative that the cylinder-of-sight contains at least
two absorption systems, each of different density, $n_H$, and column density,
$N_{H,z}$.
Such an hypothesis is supported in the case of MCG-6-30-15 
by the short-timescale variability observed in 
the depth of the O{\sc viii} edge whilst the depth of the 
O{\sc vii} edge remains constant (\S\ref{Sec:disc-multi}).
Circumstantial evidence for such a possibility also comes from the 
{\it Ginga} observations that a large fraction of Seyfert 1 galaxies 
(including some of those considered here) contain an Fe
$K$-shell absorption edge, far deeper than that predicted by
the models considered here (Nandra \& Pounds 1994; 
\S\ref{Sec:Fe-absorption}).
Unfortunately the geometry of the absorption systems cannot be 
constrained by the current data. 
One possible geometry is that both absorbing systems consist of complete 
screens, but at different radii from the central source.
A full and correct treatment of such a scenario 
requires the ionization-equilibrium of the second screen of gas to be 
calculated in the same way as the first screen, but with the
photoionizing continuum appropriate for that which has already 
passed through the first screen.
Such a treatment is beyond the scope of the current paper.
An alternative geometry is that the absorbing screen is clumpy 
on scale-sizes smaller than the emission region, resulting in the
observed spectrum being the sum of the spectra transmitted by 
regions of different density and $N_{H,z}$.
Yet another possibility is that there is indeed only a single 
absorbing cloud along the line-of-sight at any instant, 
but that different clouds (of different densities \& $N_{H,z}$) move 
across the line-of-sight on timescales much shorter than the typical 
{\it ASCA} observation. In this case our analysis reveals time-averaged 
parameters of the absorber. 
Finally, we note that 
some of our detailed assumptions regarding atomic processes 
may be inapplicable to AGN, such as other processes important in determining 
the ionization equilibrium for O{\sc vii}/O{\sc viii}.

Given the uncertainty in the form of the 'primary' continuum,
the spectral resolution and typical signal-to-noise ratios typically 
achieved, it appears to be extremely difficult to 
distinguish between these possibilities  using {\it ASCA} data.

\subsection{The location of the ionized gas}
\label{Sec:disc-location}

The location and geometry of this ionized gas is currently unclear, although
there are prospects for future progress.
Of course if one knows the density, $n_H$, of the ionized gas, $U_X$ and the 
intensity of the photoionizing source, one can use eqn.~\ref{eqn:U_X} to 
derive its radius as
\begin{equation}
r_{ld} \simeq 3\times10^{5} (\frac{L_{X44}}{U_X n_H})^{0.5}\ {\rm light-days}
\end{equation}
Assuming $n_H = 10^{10}\ {\rm cm^{-3}}$ (as used for the model calculations),
$U_X \simeq 0.1$ and $L_{X44}\sim 0.1$--1 (as appropriate for the majority of 
the objects presented here), one obtains $r_{ld} \sim$few light-days.
Such a location is comparable to that of the BELR
in these objects.
However as noted in \S\ref{Sec:ion_model}, our models are insensitive to
densities in the range 
$10^{4} \lesssim n_H \lesssim 10^{10}\ {\rm cm^{-3}}$, thus values of 
$r_{ld} \lesssim$few light-years cannot be excluded on this basis.

As described in \S\ref{Sec:disc-multi},
the apparent variations in the column density of the ionized gas seen in the 
X-ray band can be used to place upper limits on the radius of the gas 
if one assumes these variations are due to inhomogeneities in the 
matter passing through the cylinder--of--sight due to 
{\it only} Keplerian motion.
For the few cases observed to-date, such an assumption also implies a location 
close to the BELR
and hence is consistent with all/part of the same gas being 
responsible for the absorption features seen in the optical/UV.
However such estimates are currently extremely crude, and it should be 
stressed that the bulk motion of the ionized gas is highly uncertain and 
could easily be dominated by other kinematics components.
Constraints on the location could be placed by applying 
(under certain assumptions) photoionization models to both the UV/optical
and X-ray absorption features. As will be described in 
\S\ref{Sec:XUVabso-disc}, 
the results from current studies are mixed and future progress requires 
simultaneous, high-resolution observations in all three wavebands.
Reverberation mapping using the emission features produced within the ionized 
gas is another means whereby the location of the material could be 
determined. However as described in \S\ref{Sec:disc-emis}, current results 
are inconclusive. 

However we believe that in all likelihood there is ionized gas throughout the 
nuclear region in AGN. 
Multiple components, separated in at least velocity-space
are seen explicitly in the optical/UV absorption features of some objects.
At least two components are implied spectroscopically and/or from 
temporal variations in the X-ray absorption features of some objects.
In addition, highly ionized gas (occupying a different region of 
$U_X$, $N_{H,z}$ parameter-space than that responsible for the 
absorption on Seyfert 1 galaxies)
is commonly invoked to explain the 
'scattered' radiation in Seyfert 2 galaxies (e.g. see Turner et al 1997b,c).

%As described by Reynolds (1997), from a consideration of 
%the mass and the flow rate, 
%For the 'outer absorber' in MCG-6-30-15, Reynolds derives 
%a mass $M \sim 10^3\ {\rm M_{\odot}}$ and suggests this material is 
%most likely outflowing at a rate 
%$\dot{M} \sim 1\ {\rm M_{\odot}\ yr^{-1}}$
%(i.e. a factor $\sim 10^2$ greater than the putative accretion rate). Thus 
%it appears that the ionized material is a major component of the circumnuclear 
%environment of AGN

\subsubsection{The possible effect on UV--IR observations}
\label{Sec:disc-dust}

Should the volume of ionized gas
contain embedded dust, and is located outside the appropriate emission regions, then one
might expect a correlation between the column density $N_{H,z}$ derived from
X-ray observations and the various reddening indicators at longer wavebands.
Such studies therefore have the potential for further constraining the 
location of absorbing material and/or geometry of the nuclear regions.
Brandt, Fabian \& Pounds (1996) have recently suggested that the
infrared-bright quasar IRAS~13349+2438, which exhibits a large degree of 
reddening in the optical, might contain photoionized gas with internal dust.
Reynolds (1997) noted that MCG-6-30-15, whose X-ray spectrum is clearly 
attenuated by ionized gas, also exhibits a large degree of 
reddening at longer wavelengths.
The study of the creation, survival and implications of dust within the
nuclear regions of AGN has received much attention for several decades 
(e.g. see Laor \& Draine 1993 and references therein).

Unfortunately the use of the various reddening indicators in order to 
determine the column density of dust along the line--of--sight (and which can 
then be compared to that observed in the X-ray regime) rely upon making one 
or more assumptions.
The primary assumption is of course that one knows the intrinsic value of 
the quantity from which the observed reddening is determined.
If one knew the form of underlying continuum, multiwaveband continuum 
measurements could be applied in the optical--UV . However such a technique
often suffers from a lack simultaneous observations (as well as observational 
'contamination' of other spectral components).
Broad emission line ratios, in particular broad hydrogen lines, are
poor reddening indicators since the lines are significantly affected
by optical depth effects and there is no satisfactory theory to predict
their intrinsic ratios (Netzer 1990).
Clearly all techniques also require assumptions concerning the composition 
(chemical and grain-size distribution) of the dust itself, and well as 
the dust/gas mass ratio.
With these points in mind, we consider only a single reddening indicator here
(resonance-absorption line studies in the UV/optical are also considered
in \S\ref{Sec:XUVabso-disc}, but there do not provide a direct
diagnostic of the dust-phase material).

Fig.~18 shows $(f_{125}/f_{220})_{obs}$ (defined 
as the mean ratio of the observed flux at 125~nm to that at 220~nm, determined 
as described in \S\ref{Sec:sample} and listed in Table~1) is 
shown against $N_{H,z}$ for the datasets for which model {\it C(ii)} was 
considered acceptable (\S\ref{Sec:ion_pc_emis}). It can be seen that the 
majority of sources are consistent with
$2 \lesssim (f_{125}/f_{220})_{obs} \lesssim 5$, but that 3 datasets
(NGC~3227, MCG-6-30-15(1,2)) have values a factor $\sim 10$ lower. Thus it is 
tempting to make the assumption that the intrinsic flux 
ratio\footnote{It should be stressed that line emission may contribute to 
$f_{125}$ and (to a lesser extent) $f_{220}$, and thus 
$(f_{125}/f_{220})_{int}$ is not necessarily a good indication of the 
underlying UV continuum. However, the use of the color here is only based on 
noting $(f_{125}/f_{220})_{obs}$  appears to be crudely constant (within a 
factor $\sim 2$) for the majority of the objects considered.}
is in the range
$2 \lesssim (f_{125}/f_{220})_{int} \lesssim 5$, with the 
$(f_{125}/f_{220})_{obs}$ lower in the case of NGC~3227 \& MCG-6-30-15 due to 
reddening. 
Also shown in 
Fig.~18 is the predicted $(f_{125}/f_{220})_{obs}$
assuming $(f_{125}/f_{220})_{int} = 3.1$ for various values of the 
{\it difference} between the optical depths at 125~nm and 
220~nm (where $\Delta \tau =  \tau_{125} - \tau_{220}$). It can be seen that 
NGC~3227 and MCG-6-30-15(1,2) are consistent with 
$(f_{125}/f_{220})_{int}$ similar to the other objects if their fluxes at 
125~nm and 220~nm are absorbed by a column density similar to that observed 
in the X-ray band {\it and} $\Delta \tau = 0.5 N_{H,z}/10^{21}$ (dashed line).
Interestingly, the 'standard Galactic' extinction curve (for a 1:1.12
silicate:graphite mixture of dust grains with sizes $5 \leq a \leq 250$~nm,
number density $\propto a^{-3.5}$ and total dust/gas mass ratio of $10^{-2}$ 
--- see Laor \& Draine 1993 and references therein) gives $\Delta \tau$ 
close to this value ($\Delta \tau = 0.42 N_{H,z}/10^{21}$). Thus in the case 
of NGC~3227 and MCG-6-30-15(1,2), and assuming a Galactic dust/gas mass ratio, 
this UV color is consistent with reddening by a column density of dusty 
material similar to that inferred by the X-ray observations.
There are several implications of these results. 

First, we consider the implications on the location of the dust and gas 
in the case of NGC~3227 and MCG-6-30-15.
The consistency between the $N_{H,z}$ derived from the X-ray absorption
studies and the UV-reddening in the case of these two sources, assuming a 
'reasonable' dust/gas mass ratio, implies that the dust- and gas-phase 
material might occupy the same volume. If true, this allows us to 
place constraints on the location of this material based upon the survival of 
the dust.
For a typical AGN continuum of bolometric luminosity 
$L_{bol} = f_{bolX} L_{X44}$ (as defined in \S\ref{Sec:bolX}), and assuming  
maximum equilibrium temperature of $\sim 1750$~K, Laor \& Draine find 
sublimation radii of $r_{sub} \simeq 20 (f_{bolX} L_{X44})^{1/2}$~light-days 
and $\simeq 150 (f_{bolX} L_{X44})^{1/2}$~light-days for graphite grains with 
$a = 10^{-3}$~cm and $5\times10^{-7}$~cm respectively.
(Silcate grains, with a slightly lower maximum equilibrium temperature 
($\sim 1400$~K), have slightly larger values of $r_{sub}$.) Assuming 
$f_{bolX} \simeq 4$, $r_{sub} \sim 10$~light-days and $\sim 5$~light-days for 
large graphite grains in MCG-6-30-15 ($L_{X44} \simeq 0.3$) and NGC~3227 
($L_{X44} \simeq 2\times10^{-2}$) respectively. The corresponding value in 
the case of NGC~5548 ($L_{X44} \simeq 1$) is $r_{sub} \sim 40$~light-days 
for large graphite grains. The latest results from the intensive reverberation 
mapping of the broad emission line region (BELR)
in NGC~5548 have suggested lags of 
$\lesssim 2 \sim$14~days, with the lowest lags seen in the highest ionization 
lines (e.g. Korista et al 1995 and references therein).
As first pointed out by Netzer \& Laor (1993),
since both the radius of the BELR and $r_{sub}$ scale as $L_{bol}^{1/2}$
we conclude dust can only survive (i.e. in {\it equilibrium}) at radii just 
beyond the BELR in these objects.
If the dust covers a substantial fraction of the BELR, then it will 
attenuate the broad 
emission lines in these objects. For instance, assuming the same 
'standard Galactic' extinction curve used above,
$\tau({\rm C{\sc iv}}) = 9 N_{H,z}/10^{22}$ at the energy of
C{\sc iv} emission.
Thus from the values of $N_{H,z}$ found in MCG-6-30-15 and NGC~3227, 
a broad C{\sc vi} emission line is predicted to be totally absent in both 
sources, consistent with observations (see e.g. Courvoisier \& Paltani 1992).
The extinction in the optical is lower
(e.g. $\tau({\rm H}_{\beta}) = 3 N_{H,z}/10^{22}$,
$\tau({\rm H}_{\alpha}) = 2 N_{H,z}/10^{22}$) 
but sufficient to attenuate $\sim$80\% and $\sim$95\%
of the broad ${\rm H}_{\beta}$ line, and 
$\sim65$\% and $\sim$85\% of the 
broad ${\rm H}_{\alpha}$ line in 
NGC~3227 and MCG-6-30-15 respectively (again assuming 
the dusty absorber completely covers the entire BELR).
However it is clear from the value of $r_{sub}$ above and 
the discussion in \S\ref{Sec:multi-mcg6}
that dusty clouds cannot survive at a sufficiently 
small radius that Keplerian motion alone can offer an explanation of 
the short-timescale variability ($\sim 10^4$~s) in the O{\sc viii}~edge 
observed in MCG-6-30-15.
Thus either the dusty absorber is extremely non-uniform and/or its 
kinematics are extremely non-Keplerian, or (far more likely)
the gas giving rise to the bulk of the O{\sc viii}~edge is at much 
smaller radii than that with embedded dust.
Interestingly NGC~3227 and MCG-6-30-15 have the lowest axial ratios 
($a/b$, Table~1)
in our sample, with the exception of the well-known, edge-on galaxy
IC~4329A.

We now consider the majority of the objects on our sample, for which 
$(f_{125}/f_{220})_{obs}$ provides no evidence for reddening. 
There are two obvious explanations.
First, the column density of ionized gas within the cylinder--of--sight
may simply not contain any embedded dust in these objects.
This might arise from differences in the origin of the 
material, or as a result of the material in these objects being 
within $r_{sub}$.
Alternatively the dust may be present yet unobservable.
This will true if the dust is of a different composition in which
$\Delta \tau \sim 0$ (such as is the case if the grains are large), 
or if the dust clouds cover only a fraction of the UV emission region.

\subsubsection{Relationship to 'associated absorbers' in the UV/optical}
\label{Sec:XUVabso-disc}

The presence of 
highly-ionized gas, instrinsic to the Seyfert 1 nuclei is also 
obtained from observations carried out in the UV.
{\it IUE} observations revealed only a small fraction ($\sim$3\%) of
sources to exhibit narrow absorption lines close to the systemic velocity
(e.g. Ulrich 1988). However, recent results taking advantage of the higher
spectral resolution and signal-to-noise ratio available with the FOS and GHRS
instruments onboard {\it HST} have revealed that up to 50\% of Seyfert 1
galaxies  may
in fact contain such absorption features (e.g. Crenshaw 1997).
Such features have also been reported in {\it HUT} observations of
NGC~4151 (Kriss et al. 1992) and NGC~3516 (Kriss et al. 1996b).
Furthermore variability in the UV absorption features have been observed
on timescales of weeks--years in NGC~4151 (Bromage et al. 1985),
NGC~3783 (Maran et al 1996),
and NGC~3516 (Koratkar et al. 1996),
and multiple radial velocity components have been reported
in Mrk~509 (Crenshaw, Boggess \& Wu 1995)
and NGC~3516 (Crenshaw, Maran \& Mushotzky 1997).

As noted by Crenshaw (1997), there is a reassuring trend whereby 
Seyfert 1 galaxies
which have evidence for strong absorption features due to ionized gas in the
X-ray band also tend to have strong absorption lines in the UV/optical.
This raises the question as to whether the associated absorbers seen the 
UV/optical are 
caused by (part or all of) the same ionized material which gives rise to the 
absorption features observed in the X-ray band.
The UV absorption components are often blueshifted with respect to the 
emission lines, but with velocity shifts 
(typically $\sim$few$\times 10^{2}\ {\rm km\ s^{-1}}$) too small to be 
detectable with current X-ray observations. Thus direct 
tests for kinematic consistency are not yet possible.
However 
various attempts have been made to relate the UV/optical and X-ray 
absorption features using photoionization calculations
(e.g. Mathur 1994; Mathur et al. 1994; Mathur, Wilkes \& Elvis 1995;
Shields \& Hamann 1996; Kriss et al 1996a,b).
These have met with varying degrees of success, which is hardly 
surprizing given the uncertainties in the location and kinematics of the 
gas and the spectral form of the photoionizing continuum.
Nevertheless there have been a number of impressive successes 
linking the UV/optical and X-ray absorbers (e.g. see Mathur 1997).
The observed UV/optical absorption features are often superimposed on the 
wings of broad emission lines. This clearly places the gas responsible for these
'associated absorbers' at radii greater than that where the bulk (of the
blue-wing) of that emission line originates ($\lesssim 2 \sim$14~light-days in 
the case of NGC~5548).
However ionized gas could exist at a wide range of radii, 
with only part of the gas responsible for the X-ray absorption
within the cylinder--of--sight to the UV/optical 
emission line region(s). Indeed the presence of multiple components 
to the UV/optical absorption features (and the indication of 
additional absorption zones in the X-ray --- e.g. \S\ref{Sec:disc-2ion})
clearly indicates a single screen of gas to be an over-simplification.
Thus the column densities implied by the 
UV/optical absorption features may be small (and dominated by 
different ionization states) compared to those implied from 
the X-ray features.
Furthermore, the emission and absorption features predicted in the UV/optical 
from the ionized gas are somewhat dependent on the form of the 
photoionizing continuum in the UV--X-ray band. This is poorly determined 
for all objects, leading to further ambiguities.
In addition, in at least some objects the absorption features in 
both the UV and X-ray are known to vary with time. To date, most comparisons 
between the absorption features have been performed using non-simultaneous 
data, leading to obvious risk of deception.
Finally, there is the usual irritation of uncertainties in the abundances 
within the absorbing material.
The best hope for progress is provided by intense, 
simultaneous monitoring of the UV/optical and X-ray absorption features.

\section{SUMMARY \& OPEN ISSUES}
\label{Sec:open_issues}

We have presented the systematic analysis of the 0.6-10 keV
{\it ASCA} spectra of 18 Seyfert 1 galaxies in order to examine
the characteristics of X-ray absorption by ionized gas. 

\subsection{Summary of Results}

In summary, we find:
\begin{itemize}

\item
For the majority of sources, an underlying  
powerlaw with $\Gamma \sim 2$ dominates the emission within 
the {\it ASCA} bandpass (\S\ref{Sec:disc-2plaw}
and Fig.~2).

\item
Absorption by neutral material does not provide a
satisfactory model for the X-ray attenuation for most objects
(\S\ref{Sec:disc-freq-neut}). 
However, there is
evidence for a significant column density of ionized gas in the line-of-sight
to 13/18 objects
(\S\ref{Sec:disc-freq}).

\item
The X-ray ionization parameter for the ionized material is 
strongly peaked at $U_X \sim 0.1$ (Fig.~7), which may in 
part be the result of selection effects.
The column densities of ionized material
are typically in the range $N_{H,z} \sim 10^{21}$--$10^{23} {\rm cm}^{-2}$
(Fig.~3), 
although highly ionized (and hence psuedo-transparent) column densities 
up to $10^{24} {\rm cm}^{-2}$ cannot be excluded in some cases
(see also \S\ref{Sec:disc-hagai})

\item
Allowing a fraction of the continuum to be observed without
attenuation significantly improves the fit to
many sources (\S\ref{Sec:disc-Df} and Fig.~5).
Such a component is required only in the case of NGC~4151.

\item
Inclusion of the emission from the ionized material significantly 
improves the fit to many sources, at an intensity consistent 
with the material subtending $\sim 4 \pi$~sr at the central source.
However such a component is not required in any individual source 
(\S\ref{Sec:disc-emis} and Fig.~11).

\item
A deficit of counts is observed at $\sim 1$~keV in the sources 
exhibiting the strongest
absorption features (\S\ref{Sec:disc-2ion}).
This is consistent with the presence of a second zone
of (more highly) ionized gas, which might have been seen previously
in the deep Fe $K$-shell edges observed in
some {\it Ginga} observations.

\item
There is evidence that the ionized material in NGC~3227 and MCG-6-30-15 
contains embedded dust, whilst no such evidence in the other sources
(\S\ref{Sec:disc-dust} and Fig.~18).

\end{itemize}

\subsection{Open issues}

While X-ray absorption by ionized gas appears to play a very important role
in shaping the observed X-ray spectra of AGN, there remain a large
number of important questions which cannot be answered with the
available data, including:

\begin{itemize}

\item
What is the frequency of occurrence of warm absorbers at higher
luminosities. Is there a simple relationship between $U_X$ and
$L_X$ ?

\item
Are there several distinct warm absorber components
attenuating the X-ray bandpass ?

\item
What is the location, density and distribution of the
material comprising the X-ray warm absorber ?
Sensitive measurements
of the emission spectrum from the gas will help to constrain the
gas distribution.

\item
Is the warm absorber flowing into or out from the active nucleus ?
Do the UV associated-absorbers arise from the same X-ray absorbing 
gas ?

\item
Does the gas to dust ratio differ significantly between objects, and how
does this affect the observed properties of the source ?
How important is this component in the unification of AGN ?

\item
Is the absorption variability due to bulk motion of material across the
line-of-sight, variable ionization  or a combination of both effects ?

\item
How stable is the warm absorbing gas ?

\end{itemize}

\acknowledgements
We thank the {\it ASCA} team for their operation of the satellite, and 
Keith Gendreau, 
Mike Crenshaw and the members of the {\it ASCA} GOF at
NASA/GSFC for helpful discussions.
We would also like to thank an anonymous referee whose carefully reading
of the manuscript and thoughtful comments led to a great improvement in 
the final version.
We acknowledge the financial support of
the Universities Space Research Association (IMG, TJT), 
National Research Council (KN) and 
a special grant from the Israel Science Foundation (HN).
This research has made use of the Simbad database, operated by CDS, 
Strasbourg, France; of the NASA/IPAC Extragalactic database, which is 
operated by the Jet Propulsion Laboratory, Caltech, under contract with 
NASA; and
of data obtained through the HEASARC on-line service, provided by
NASA/GSFC.

\clearpage
\section*{APPENDIX: NOTES ON INDIVIDUAL OBJECTS}
\label{Sec:appendix}

The sources presented here are all well-known, bright Seyfert 1 galaxies, 
detected early in the history of X-ray astronomy.
Thus, most have been observed by all the major X-ray astronomy satellites.
The results from, and references to,
almost all X-ray observations published prior to the end of 1992 are 
included in the compilation of Malaguti, Bassini \& Caroli (1994).
The following abbreviations are used for papers cited 
numerous times in this section:
NP94 -- Nandra \& Pounds (1994);
R97 -- Reynolds (1997);
TP89 -- Turner \& Pounds (1989); 
T91 -- Turner et al (1991);
W95 -- Weaver, Arnaud \& Mushotzky (1995).
As stated in \S\ref{Sec:disc-freq}, R97 also fitted a single-zone 
photoionization model to a number of sources, and quotes the ionization 
parameter $\xi$ ($= L/n_{H} R^2$, where $n_{H}$ and $R$ are as 
defined in eqn.\ref{eqn:U_X}, but where $L$ is the integrated luminosity) 
calculated over the 0.0136--13.6~keV band assuming a single powerlaw.
For convience, below we give the corresponding $U_X^{R97}$
(defined as in eqn.\ref{eqn:U_X} for such a continuum over the 
0.1--10.0~keV band) using the conversion factors shown 
in Fig.~1b.
{\it Ginga} observations have shown evidence for a significant Fe absorption 
edge in the 7.1--8.9~keV band for a number of sources (NP94), in some 
cases confirming suggestions in earlier observations. These 
sources are 
noted, but as described in \S\ref{Sec:Fe-absorption}
a detailed analysis of such features has not been attempted here.
Unless otherwise stated, the Seyfert classification is from Whittle (1992)
and all limits on a parameter are quoted at 90\% 
confidence for 1 interesting parameter.

As a reminder, models {\it A(i)}--{\it C(ii)} all assume a 
single powerlaw continuum absorbed by a neutral column density at $z=0$
(constrained to be $\geq N_{H,0}^{gal}$, the Galactic H{\sc i} 
column density along that line-of-sight) and an additional column 
density, $N_{H,z}$, at the redshift of the source.
In models {\it A} (\S\ref{Sec:zwabspo}--\ref{Sec:zpcfpo})
$N_{H,z}$ is assumed to be neutral, whilst in 
models {\it B} (\S\ref{Sec:ion}--\ref{Sec:ion_pc})
\& {\it C} (\S\ref{Sec:ion_emis}--\ref{Sec:ion_pc_emis})
$N_{H,z}$ is assumed to be photoionized, with a corresponding ionization 
parameter $U_X$.
Models {\it C} also include the emission features expected from the ionized 
gas (assuming it subtends a solid angle $\Omega$ at the central source).
In the models denoted {\it (i)}, the absorbing gas covers the entire 
cylinder-of-sight to the source, whilst in models {\it (ii)} a fraction 
$D_f$ of the underlying continuum are observed without attenuation.
See \S\ref{Sec:models} for further details.

\underline{\bf Mrk 335:}\\
This Seyfert 1.0 galaxy was first detected in X-rays by 
{\it UHURU} (Tananbaum et al 1978).
From our analysis of the data from {\it ASCA} observation performed 
in 1993 December, we find 
model {\it A(i)} to provide an unsatisfactory 
description of the spectra (Table~3).
Fits satisfying our criteria for acceptability are obtained 
if $\sim$60\% of the continuum escapes without suffering attenuation by
neutral material (model {\it A(ii)}) and for all the
models in which the absorbing gas is assumed to be ionized
(models {\it B(i)}--{\it C(ii)}). 
However, in \S\ref{Sec:2plaw} we obtained a superior fit if the continuum 
is assumed to be the sum of a steep powerlaw 
($\Gamma_s = 3.3^{+2.0}_{-0.7}$) dominating at 
energies $\lesssim 1$~keV and a flatter powerlaw 
($\Gamma_h \sim 1.8^{+0.2}_{-0.5}$) 
dominating at energies $\gtrsim 3$~keV.
This model gave $\chi^2/dof = 621/642$, had $R_{s/h} \sim 1.6$
and required no absorption in addition to $N_{H,0}^{gal}$.
Such a double-powerlaw continuum is in good agreement with 
previous observations
(Turner et al 1993b and references therein).
The flux in the 2--10~keV band during the {\it ASCA} observation 
was $F_{2-10} = 9.5\times10^{-12}\ {\rm erg\ cm^{-2}\ s^{-1}}$, 
similar to that during the {\it BBXRT} era, 
and indeed the X-ray spectrum of Mrk~335 
appears relatively stable on long timescales, 
though the source exhibits 
variability on short timescales 
(Pounds et al 1987; Lee et al 1988; Paper~I).
In an independent analysis of the same {\it ASCA} dataset, 
R97 modeled the continuum in terms of single powerlaw
and a blackbody (with $kT \sim 0.14$~keV and 
luminosity $L_{bb} \sim 10^{43}\ {\rm erg\ s^{-1}}$) to account for the 
steepening to low energies.
We confirm these results, but find such a model provides an 
inferior description of the data ($\chi^2/dof = 629/641$)
compared to the double-powerlaw model. 
Explanations for the X-ray continuum in Mrk~335 have been 
presented which involve $\rm e^{\pm}$--pair cascades (Turner et al 1993b)
and the Comptonization of the spectrum emitted by an accretion disk
(Zheng et al 1995a).

We find any features imprinted by any 
ionized gas to be relatively weak during the 
epoch of the {\it ASCA} observation, resulting in poor constraints on 
$N_{H,z}$ and $U_X$ in models {\it B(i)}--{\it C(ii)}.
Turner, George \& Mushotzky (1993) found evidence for an absorption
edge ($\tau \sim 0.3\pm0.1$) in the 0.7--0.9~keV band during a
{\it ROSAT} PSPC observation of Mrk~335.
Furthermore, R97 found the addition of 
an O{\sc vii} edge (of optical depth $\tau_{O7}\sim0.3$)
significantly improved the fit 
to a single powerlaw model ($\Gamma \sim 2.1$) to these {\it ASCA} data.
However, 
we find only a marginal improvement ($\Delta \chi^2 = 4$) when such an edge 
($\tau_{O7} = 0.2\pm 0.2$) is added to our double-powerlaw model, and 
no improvement when an O{\sc viii} edge is also included
($\tau_{O8} \lesssim 0.1$).
Some evidence exists for a Fe $K$-shell edge in the 7.1--8.9~keV band
in two {\it Ginga} observations (NP94), and possibly in
an {\it EXOSAT} observation (Turner et al 1993b). 
The energy of the features could not be constrained in any of the datasets,
but NP94 inferred $N_{H,z} \sim 2\times10^{23}\ {\rm cm^{-2}}$.

There is no evidence for any UV absorption features due to ionized gas 
at the redshift of Mrk~335 (Zheng et al 1995a; Crenshaw 1997).

\underline{\bf Fairall 9:}\\
This Seyfert 1.0 galaxy was suggested to be an X-ray source 
based upon its location in {\it UHURU} and {\it Ariel-V}
error boxes
(Ricker 1978).
From our analysis of the data from {\it ASCA} observation performed 
in 1993 November, we find no model in \S\ref{Sec:basic_models} which
satisfies our formal criteria for acceptability. 
However, this is primarily due to scatter in 2--4 keV band, most likely due
to an underestimation of systematic errors for this dataset. 
In view of this, we consider model {\it A(i)} to provide 
an adequate description of the data. However model {\it A(ii)} 
provides a significant improvement in the goodness-of-fit 
if $\sim80$\% of the continuum escapes without attenuation by a 
large column density ($N_{H,z} \sim 3\times10^{23}\ {\rm cm^{-2}}$)
of neutral material which 
gives rise to a relatively subtle upturn in 
the spectrum $\gtrsim 5$~keV (Fig~4).
No significant improvement 
in the goodness-of-fit was obtained assuming the more complex models 
described in \S\ref{Sec:basic_models} or 
\S\ref{Sec:additional_models}.
For all models we find no evidence for absorption in addition to 
$N_{H,0}^{gal}$ and a derived slope ($\Gamma \simeq 2.0$) 
consistent with the 
{\it Einstein} (T91) and 
{\it EXOSAT} (TP89)
observations of the source, slightly steeper than that observed by {\it Ginga} 
($\Gamma = 1.84$, NP94) and only
marginally consistent with the {\it ROSAT} PSPC 
observations ($\Gamma = 2.2\pm0.2$, Walter \& Fink 1993).
X-ray variability has been reported on both short
and long timescales (Morini et al 1986; Paper~II).
Unfortunately these {\it ASCA} do not offer any insight into 
the possible spectral variability suggested by the 
{\it EXOSAT} observations (Morini et al 1986).
However, we note that the flux in the 
2--10~keV band observed during both the {\it Ginga} 
(NP94)
and this {\it ASCA} observation 
($F_{2-10} = 2\times10^{-11}\ {\rm erg\ cm^{-2}\ s^{-1}}$) 
indicates that the historical decline in the brightness of 
Fairall~9 in the X-ray band (Morini et al 1986) has ceased.

We find any features imprinted by any ionized gas to be weak 
during the epoch of the {\it ASCA} observation, resulting in poor 
constraints on 
$N_{H,z}$ and $U_X$ in models {\it B(i)}--{\it C(ii)}.
R97 also found a lack of evidence for
O{\sc vii} and O{\sc viii} absorption
edges in their analysis of the {\it ASCA} data from Fairall~9.
They find a best-fitting photon index consistent with our value, 
although unlike our findings, require a small column density 
($N_{H,z} \sim 10^{20}\ {\rm cm^{-2}}$) of intrinsic absorption.

There is no evidence for any UV absorption features due to ionized gas 
at the redshift of Fairall~9 (Zheng et al 1995b; Crenshaw 1997).

\underline{\bf 3C 120:}\\
This bright, flat-spectrum radio source, has Seyfert 1 properties in the
optical regime. The source was first detected in X-rays by {\it SAS-3} 
(Schnopper et al 1977), and has since been observed by all the 
major X-ray astronomy satellites except {\it Ginga}. 
From our analysis of the data from the {\it ASCA} observation performed 
in 1994 February, we find model {\it A(i)} to provide an adequate
description of the spectra, with some evidence for intrinsic absorption
($N_{H,z} \sim 5\times10^{20}\ {\rm cm^{-2}}$).
However, we find some evidence for a 'hard tail' in this source, with 
significantly superior fits obtained if $\sim30$\% of the continuum is 
attenuated by $N_{H,z} \sim 5\times10^{23}\ {\rm cm^{-2}}$ (model {\it A(ii)};
Fig.~4), 
or by a flattening of the underlying continuum $\gtrsim 6$~keV 
(\S\ref{Sec:2plaw}).
We find no compelling evidence for ionized gas in 3C~120 (intense emission 
from ionized gas is implied by the best-fitting solution assuming model 
{\it C(ii)}, but at a level incompatible with strength of the observed 
Fe $K$-shell line -- see Fig~12).
This confirms the finding of R97 who also found a lack of evidence for
O{\sc vii} and O{\sc viii} absorption edges in their analysis of these 
{\it ASCA} data.
They find a best-fitting photon index consistent with 
our value, but a slightly larger absorbing column density instrinsic to the 
source ($N_{H,z} \sim 8\times10^{20}\ {\rm cm^{-2}}$) then we found 
for model {\it A(i)}.

In all cases, the derived slope is $\Gamma \simeq 2.0$, and is 
consistent with a poorly constrained {\it ROSAT} PSPC 
spectrum ($\Gamma \sim 1.5^{+1.2}_{-1.8}$, Boller et al 1992), and
slightly steeper than than those derived from 
{\it HEAO-1}, {\it Einstein} and {\it EXOSAT} 
($\Gamma \sim 1.5$--1.9, W95; 
T91; TP89).
Spectral variability was apparent in the {\it Einstein} observations
(T91).
Unfortunately the {\it ASCA} data are unable to constrain the 
Compton-Reflection component seen in {\it HEAO-1} observations
(W95), although such a component clearly offers an an alternative 
explanation of the hard tail.
We also note that many of the fits to the 3C~120 dataset imply 
a local column density $> N_{H,0}^{gal}$ 
($\Delta N_{H,0} \simeq 5 \times 10^{20}\ {\rm cm^{-2}}$). Furthermore
this dataset seems particularly badly affected by a 
trough-like deficit of counts at energies $<0.6$~keV
compared to the extrapolated model. It is unclear how these facts 
may be related to the 
well-known, but poorly-understood calibration problems of the 
XRT/SIS at these energies (see also \S\ref{Sec:calib_uncert}).

\underline{\bf NGC 3227:}\\
This Seyfert 1.2 galaxy was first detected in X-rays by {\it Ariel-V} 
(Elvis et al 1978).
In our analysis of the data from {\it ASCA} observation performed 
in 1993 May, we find 
model {\it A(i)} to provide an unsatisfactory 
description of the spectra (Table~3).
Fits satisfying our criteria for acceptability are obtained 
if $\sim$35\% of the continuum escapes without suffering attenuation by
neutral material (model {\it A(ii)}; Fig.~4). However
yet superior fits are obtained if the absorbing gas is assumed to be ionized
(model {\it B(i)}; Fig.~6).
The parameters of the ionized gas are fairly well constrained, 
with $U_X \simeq 10^{-2}$ and 
$N_{H,z} \sim 4\times10^{21}\ {\rm cm^{-2}}$
consistent with the findings of Ptak et al (1994) and R97
($U_X^{R97} \simeq 0.02$, $N_{H,z} \sim 4\times10^{21}\ {\rm cm^{-2}}$)
in their independent analysis of this {\it ASCA} observation.

We find that statistically there is no requirement for any of the underlying 
continuum to escape without suffering attenuation by the ionized material, 
or for any significant emission from the ionized gas, 
(models {\it B(ii)}--{\it C(ii)}).
In all cases there is no requirement for additional absorption by 
neutral gas in excess of $N_{H,0}^{gal}$.
The derived photon index ($\Gamma \sim 1.6$) found for NGC~3227, 
is flatter than for the 
majority of the sources (Fig.~2), consistent with previous 
measurements by 
{\it HEAO-1} (W95),
{\it Einstein} (T91),
{\it EXOSAT} (TP89),
{\it Ginga} (NP94)
and 
{\it ROSAT} (Rush et al 1996).
However, it is important to note that {\it EXOSAT} observations of 
NGC~3227 revealed significant spectral variability on timescale as 
short as $10^4$~s, most likely as a result of changes in absorption
(TP89). Furthermore, such variability was also present during the 
{\it ASCA} observation presented here
(Ptak et al 1994; Paper~I). Thus it should be remembered that 
our results most likely represent the time-averaged values.
We note evidence was found for a Fe $K$-shell edge in the 7.1--8.9~keV in 
two {\it Ginga} observations of the source. NP94 found 
evidence for absorption by ionized gas with 
$N_{H,z} \sim 3\times10^{22}\ {\rm cm^{-2}}$, higher than that
observed here.

\underline{\bf NGC 3516:}\\
This source, one of the original galaxies classified by
Seyfert (1943) as a type 1.0 galaxy, was first detected in X-rays by 
{\it Einstein} (Maccacaro, Garilli \& Mereghetti 1987).
In our analysis of the data from an {\it ASCA} observation performed 
in 1994 April, we find no model in \S\ref{Sec:basic_models} which
satisfies our formal criteria for acceptability. 
Nevertheless, we find model {\it B(i)} to provide a significant improvement 
over models {\it A(i)} \& {\it A(ii)}, offering clear evidence for 
ionized gas and 
confirming the results of R97 and those from a second {\it ASCA} observation 
carried out in 1995 March (Kriss et al 1996a).
However, whilst a single-zone photoionization model clearly reveals a 
strong '$\sim$1~keV' deficit in this source, we find that 
despite providing a vast improvement, a statisfactory fit is still not 
obtained with the inclusion of an additional absorption edge (\S\ref{Sec:2ion}).
Interestingly, we note that neither R97 nor Kriss et al (1996a) were 
able to find a solution which satisfy our criteria for acceptability.
We find a derived index ($\Gamma \sim 1.9$), steeper than that which has been 
claimed from previous {\it Einstein} (Kruper, Canizares \& Urry 1990),
{\it EXOSAT} (Ghosh \& Soundararajaperumal 1991)
and {\it Ginga} (Kolman et al 1993) observations in the 2--10~keV band.
However all these observations required a neutral $N_{H,0} >> N_{H,0}^{gal}$, 
undoubtedly due to the incorrect modelling of the deep O{\sc vii} and 
O{\sc viii} edges in this source with neutral gas.
A steeper index ($\Gamma \gtrsim 2.4$) is implied when a single powerlaw model 
is applied to the {\it ROSAT} PSPC data (Boller et al 1992), again probably
the result of the deep O{\sc vii} and O{\sc viii} edges.
For models {\it B(i)}--{\it C(ii)}, 
we find a neutral $N_{H,0} \sim 5\times10^{20} \ {\rm cm^{-2}}$
and an ionized $N_{H,z} \sim 9\times10^{21} \ {\rm cm^{-2}}$, 
with $U_X \sim 0.1$.
For comparison, R97 find $N_{H,z} \sim 10^{22}\ {\rm cm^{-2}}$, 
$\Gamma \sim 1.7$ and $U_X^{R97} \sim 0.06$ from an independent analysis 
of the {\it ASCA} dataset obtained in 1994 April, and assuming a single-zone 
model.
Kriss et al (1996b) have performed a detailed analysis of the 1995 March 
observation (using {\sc XSTAR}), and for a single-zone model
find $U_X \sim 0.03$ (assuming their quoted spectrum),
$N_{H,z} \sim 10^{22}\ {\rm cm^{-2}}$ and $\Gamma \sim 2.0$.
These workers also find adding a second zone of ionized gas significantly
improves the fit, obtaining a best-fitting solution with
$U_X \sim 0.1$,  
$N_{H,z} \sim 1.4\times10^{22}\ {\rm cm^{-2}}$ 
and 
$U_X \sim 0.02$,  
$N_{H,z} \sim 7\times10^{21}\ {\rm cm^{-2}}$ 
for the two zones.
Nandra et al (1997c) have recently shown the Fe emission line varied 
significantly between the two {\it ASCA} observations.
Evidence for absorption by ionized gas was found 
in 3 {\it Ginga} observations
of NGC~3516, with 
$N_{H,z} \sim 2$--$3\times10^{23}\ {\rm cm^{-2}}$ (Kolman et al 1993).
We also note that {\it ROSAT} HRI observations show possible
evidence for elongation of some fraction of 
the soft X-ray emission in NGC~3516 in a direction consistent 
with the extended narrow-line region (Morse et al 1995).

NGC~3516 has long been known to exhibit deep, variable absorption lines in the 
UV (Koratkar et al. 1996; Kriss et al 1996a and references therein)
with at least 4 distinct radial components visible in C{\sc iv}
(Crenshaw, Maran \& Mushotzky 1997).
From the combined analysis of the simultaneous {\it ASCA} and {\it HUT},
Kriss et al (1996b) conclude that a range of column densities and 
ionization parameters are required to explain the 
depths of all the X-ray and UV absorption features seen in NGC~3516.

\underline{\bf NGC 3783:}\\
This Seyfert 1.2 galaxy was first detected in X-rays by 
{\it Ariel-V} (Cooke et al 1976).
Absorption due to ionized gas in NGC 3783 was first suggested by 
{\it ROSAT} PSPC observations (Turner et al 1993).
In our analysis of the data from 2 {\it ASCA} observations separated by 
4 days in 1993 December, we find models involving absorption by neutral
material (model {\it A(i)} \& {\it A(ii)}) to provide unsatisfactory 
descriptions of the spectra.
However fits satisfying our criteria for acceptability are obtained 
if the absorbing gas is assumed to be ionized
(models {\it B(i)} \& {\it B(ii)}), with a significant 
improvement in the goodness-of-fit with the inclusion of the ionized 
emitter (models {\it C(i)} \& {\it C(ii)}).
In all cases, the parameters of the ionized gas are fairly well constrained, 
with $U_X \sim 0.1$--0.15 and 
$N_{H,z} \sim 2\times10^{22}\ {\rm cm^{-2}}$
(e.g. Fig.16).
This confirms the findings of 
George, Turner \& Netzer (1995), and R97 who found the addition of 
O{\sc vii} and O{\sc viii} 
edges (of optical depths $\tau_{O7}\sim1.2$ \& $\tau_{O8}\sim1.4$)
significantly improved the fit 
to a single powerlaw model ($\Gamma \sim 1.4$) to these {\it ASCA} data.
R97 also fitted a single-zone photoionization model and found
$U_X^{R97} \sim 0.08$,
$N_{H,z} \sim 2\times10^{22}\ {\rm cm^{-2}}$, $\Gamma \sim 1.7$.
The index derived for these {\it ASCA} observations ($\sim 1.8$) is 
somewhat steeper than that found by George et al (1995), primarily due 
to these workers not adequately modeling the Fe $K$-band.
A flatter spectral index has also been suggested by some (but not all) 
previous observations in the 2--10~keV band (e.g. Ghosh et al 1992).
This may indicate spectral variability or (more likely) be an artifact 
of incorrectly modeling the deep absorption troughs due to the 
ionized gas in this source.
As described in \S\ref{Sec:multi-3783}, the behaviour of the ionized 
gas in response to the change in flux seen between the two 
{\it ASCA} observations is consistent with expectations.
NP94 found 
evidence for absorption by ionized gas with 
$N_{H,z} \sim 10^{23}\ {\rm cm^{-2}}$, higher than that
observed here.

UV observations of NGC~3783 have revealed absorption due to Ly$\alpha$, 
N{\sc v} and C{\sc iv}, with the latter exhibiting variability on a 
timescale of $\lesssim 6$ months
(Reichert et al 1994; Lu et al 1994; Maran et al 1996).
Shields \& Hamann (1997) have compared the implied column densities 
with the photoionization modeling of the X-ray absorption presented 
in George et al (1995). Whilst alternative scenarios could not be excluded, 
they found the UV and X-ray results to be consistent with a single-phase,
photoionized plasma. By comparing the observed variability timescale in 
C{\sc iv} with the radiative recombination, they derived a density
$n \gtrsim 50\ {\rm cm^{-3}}$ for the ionized gas and hence,
from the ionization parameter, a location $r \lesssim 30$~pc from 
the ionizing source.
Since the parameters associated with the X-ray absorption are 
are generally consistent with those in George et al (1995), we 
agree with the conclusions of Shields \& Hamann (1997).

\underline{\bf NGC 4051:}\\
This Seyfert 1.5 galaxy was first detected in X-rays by 
{\it Einstein} (Marshall et al 1983) and has been seen to exhibit rapid 
variable on timescales as short as few$\times10^2$~s (e.g. Lawrence et al 1987).
Indeed, in Paper~I we showed NGC~4051 the source to exhibit the most pronounced 
variability of any of the sources considered.
Absorption due to ionized gas in this object was first suggested by 
{\it ROSAT} PSPC observations (Pounds et al 1994).
In our analysis of the data from the {\it ASCA} observation performed 
in 1993 April, we find no model in \S\ref{Sec:basic_models} which
satisfies our formal criteria for acceptability. 
A satisfactory fit is found however, if a strong (${\cal F} \sim 4$)
Compton-reflection component is included in the model
(\S\ref{Sec:reflection}). 

However, such a curvature in the observed spectrum may be due in part 
to the spectral variability observed during the observation (Paper~I).
Nevertheless, we do find a significant improvement in the fits for models
including ionized gas, with 
$U_X \sim 0.7$,
$N_{H,z} \sim 2$--$10\times10^{22}\ {\rm cm^{-2}}$, 
$\Gamma \sim 2.3$ for all models {\it B(i)}--{\it C(ii)}, 
and 
$U_X \sim 0.4$,
$N_{H,z} \sim 10^{22}\ {\rm cm^{-2}}$,
$\Gamma \sim 2.4$ for the model including Compton-reflection
(Table~10).
Mihara et al (1994) and R97 have both performed independent analyses of this 
{\it ASCA} dataset. Both find the addition of two edges to a single powerlaw 
model significantly improved the goodness-of-fit over the 
0.6--1.0~keV band, confirming the results from the {\it ROSAT} PSPC
and from a later {\it ASCA} observation carried out in 1994 July 
(Guainazzi et al 1996).
R97 fixed the edge energies to those appropriate for O{\sc vii} and 
O{\sc viii} and found $\tau_{O7}\sim \tau_{O8}\sim0.2$. 

Photoionization models were also applied to these {\it ASCA} data
by Mihara et al and R97, as well as to the earlier {\it ROSAT} PSPC
data (Pounds et al 1994; McHardy et al 1995) and data from the later {\it ASCA}
observation (Guainazzi et al 1996).
Unfortunately detailed comparisons are difficult due to differing assumptions 
regarding the form of the observed and photoionized spectra.
Mihara et al find $\xi \sim 50$ (but for an unspecified XUV continuum)
and $N_{H,z} \sim 2\times10^{21}\ {\rm cm^{-2}}$ when the observed continuum
is modeled by a (relately) flat powerlaw ($\Gamma \sim 1.9$) plus a 
low temperature blackbody ($kT \sim 0.1$~keV and
luminosity $L_{bb} \sim 3\times10^{41}\ {\rm erg\ s^{-1}}$).
R97 find $U_X^{R97} \sim 0.02$, 
$N_{H,z} \sim 10^{21}\ {\rm cm^{-2}}$ and $\Gamma \sim 1.9$, but also 
find evidence for a similar blackbody component.
From the analysis of combined {\it ROSAT}--{\it Ginga} data
Pounds et al (1994) also prefer the inclusion of 
a blackbody ($kT \sim 0.2$~keV and
$L_{bb} \sim 5\times10^{41}\ {\rm erg\ s^{-1}}$) 
and powerlaw ($\Gamma \sim 2.1$)
finding 
$N_{H,z} \sim 7\times10^{22}\ {\rm cm^{-2}}$ and 
$U \sim 0.8$ (but for an unspecified XUV continuum).
From subsequent {\it ROSAT} data
McHardy et al (1995) find
$U_X \sim 0.5$ (given their continuum),
$N_{H,z} \sim 8\times10^{22}\ {\rm cm^{-2}}$ and $\Gamma \sim 2.2$, 
with no requirement for a cool thermal component.
Finally, Guainazzi et al find
$N_{H,z} \sim 2\times10^{22}\ {\rm cm^{-2}}$, $\Gamma \sim 2.1$
and $U_X \sim 0.2$ (given their continuum). These workers found evidence for 
additional emission $\lesssim 1$~keV, which could be modeled by either 
a blackbody or a series of emission lines. However given the 
differential variability between the depths of the O{\sc vii} and
O{\sc viii} edges seen on a timescale $\sim 10^4$~s during the 1994 
observation, Guainazzi et al considered neither explanation to be satisfactory.
We find no strong requirement for a low temperature blackbody component in 
our analysis of NGC~4051 (with our models for having 
$1.0 \lesssim \overline{R_{0.6}} \lesssim 1.2$). 
For our model including Compton-reflection we find 
$L_{bb} \lesssim 3\times10^{41}\ {\rm erg\ s^{-1}}$
for $kT=0.1$~keV.
Instead our findings do imply 
significant emission from the photoionized gas in NGC~4051 at energies 
$\lesssim 1$~keV (with a luminosity of $\sim 10^{41}\ {\rm erg\ s^{-1}}$
in the 0.1-1~keV band), emission which was ascribed to the 
blackbody component by previous workers.
However we do note that the derived intensity of this emission is 
up to a factor 3 times that expected given the intensity of the 
continuum during the observations. 
It is currently unclear whether these results indicate
that the source was brighter in the past, that there is indeed an 
additional component at the softest energies, or the physical conditions 
with NGC~4051 are much more complex.
We consider it unlikely that significant progress in decoupling the complex 
spectral variability exhibited by NGC~4051 $\lesssim 1$~keV will not be
made until high signal-to-noise observations are performed with high spectral 
and temporal resolution. Nevertheless, the time-averaged results presented 
here are useful for comparison with those from those sources which appear 
to exhibit less dramatic variability.

NP94 found 
evidence for absorption by ionized gas in {\it Ginga} observations, 
with $N_{H,z} \sim 6\times10^{22}\ {\rm cm^{-2}}$.

\underline{\bf NGC 4151:}\\
This Seyfert 1.5 galaxy was first detected in X-rays by 
{\it UHURU} (Giacconi et al 1974), and has been extensively 
studied ever since at at all wavebands. Interestingly, despite being
one of the brightest Seyfert 1 galaxies in the X-ray band, the source appears to be
somewhat atypical (e.g see Warwick et al 1996 and references therein).
Specifically, it has long been known that NGC 4151 appears to have a 
flat powerlaw ($\Gamma \sim 1.3$--1.7), a substantial 
column density of absorbing gas instrinsic to the source
($N_{H,z} \sim 10^{22}$--$10^{23}\ {\rm cm^{-2}}$) and 
a strong 'soft-excess' at energies $\lesssim 2$~keV. 
At present, the consensus is that the usually hard underlying continuum is the 
result of different physical conditions associated with the emission mechanism 
compared to most other objects in its class (but see 
Zdziarski, Johnson \& Magdziarz 1996). The soft-excess has been interpreted
as either due to separate emission component 
(e.g. Warwick, Done \& Smith 1995) and/or the result of the absorbing gas
only partially covering the source (e.g. Holt et al 1980).
The suggestion that the absorbing gas in NGC~4151 might be photoionized was 
first made by Yaqoob, Warwick \& Pounds (1989). However subsequent 
observations, in particular the different variability characteristics observed 
below and above $\sim1$~keV, have suggested the situation may be more complex
(e.g. Weaver et al 1994a,b; Warwick et al 1996).
{\it Einstein} HRI (Elvis, Briel \& Henry 1983)
and {\it ROSAT} HRI (Morse et al 1995) observations have 
shown spatially resolved X-ray emission, probably associated with 
the extended narrow-line region, at an intensity likely to also contribute to 
the observed spectrum $\lesssim1$~keV.

In our analysis of the data from 6 {\it ASCA} observations carried out 
prior to 1994 May, we found 3 datasets fulfilling the criteria defining this 
'sample' (see also Paper~I). As summarized in Table~12,
we find that a fit satisfying our criteria for acceptability is achieved for 
the first two observations (NGC~4151(2,4)) only when the absorbing material 
is allowed to be ionized {\it and} $\sim$5\% of the underlying continuum is 
allowed to escape without suffering any attenuation (i.e. models {\it B(ii)} 
\& {\it C(ii)}). Such an hypothesis is strongly supported by the third 
observation (NGC~4151(5)), even though none of the models in 
\S\ref{Sec:basic_models} provide a solution which formally satisfies our 
criteria for acceptability (see Fig.~9).
In \S\ref{Sec:thermal} we found evidence in all 3 datasets for
additional emission due to an optically-thin plasma at a temperature 
and intensity consistent with the extended emission.
In all cases we find best-fitting values of
$\Gamma \sim 1.5$, 
$U_X \sim 0.07$,
$N_{H,z} \sim 4$--$7\times10^{22}\ {\rm cm^{-2}}$,
and $L_X \sim 1.4$--$2.0\times10^{43}\ {\rm erg\ s^{-1}}$.
However, 
as described in \S\ref{Sec:multi-4151}, the ionized gas appears to 
respond in an interesting way in response to changes in the intensity of the 
illuminating continuum between observations.
A possible explanation of such behaviour was proposed in which 
the gas responsible for the absorption is 
clumped into regions of differing
$N_{H,z}$ with a  characteristic scale-size greater than that for
the central source. It was shown that Keplarian motion would allow 
such $N_{H,z}$ perturbations to traverse the cylinder--of--sight
on a timescale consistent with observations for reasonable range of
radii, giving rise to an apparent decoupling between 
variations in $U_X$, $N_{H,z}$ and the intensity of the underlying source. 

These NGC~4151 datasets have been independently analysed by other workers.
Weaver et al (1994b) considered the 4151(2) dataset 
and found a satisfactory fits with $\Gamma \sim 1.6$, 
$U \sim 0.3$ (for an unspecified XUV spectrum),
$N_{H,z} \sim 5\times10^{22}\ {\rm cm^{-2}}$,
$D_f \sim 0.03$. 
However when a similar model was applied to the 
4151(1) dataset, these workers found 
$U$ (and $N_{H,z}$) to be {\it higher} when the source was fainter.
(We confirm this general behaviour, finding $\Gamma \sim 1.4$, 
$U_X \sim 0.13$, $N_{H,z} \sim 10^{23}\ {\rm cm^{-2}}$, 
$D_f \sim 0.09$, $\Omega/4\pi \sim 4$ and
$L_X \sim 0.9\times10^{43}\ {\rm erg\ s^{-1}}$
when model {\it C(ii)} is applied to the 4151(1) dataset.)
Weaver et al also suggested an explanation is terms of 
inhomgeneities in the ionized material 
trasversing the cylinder-of-sight.

Zdziarski, Johnson \& Magdziarz (1996) 
have considered the 4151(1) {\it ASCA} dataset along with contemporaneous 
{\it CGRO} OSSE data.
They find the spectra $\gtrsim 3$~keV to be better described by thermal 
Comptonization, and dual, cold absorber (with twice cosmic abundance of Fe), 
rather than by an ionized absorber.
Furthermore, they show that if a Compton-reflection
component is added, the underlying spectral index of this source is steepened 
to $\Gamma \sim 1.8$ and hence similar to that observed in other 
Seyfert 1 galaxies.
However, since we have not included 4151(1) in our general analysis 
we feel unable to comment further on this result here
(although see above, and also note that in \S\ref{Sec:reflection}
we did not find a significant improvement when Compton-reflection was 
included in our analysis of the NGC~4151(2,4,5) datasets).
Nevertheless, we note Zdziarski, Johnson \& Magdziarz also consider
several archival X- and $\gamma$-ray datasets, and make a 
a number of points concerning the emission mechanism in this source.

Warwick et al (1996) considered the 4151(4,5) datasets
(as well as 4151(3,6), and psuedo-simultaneous {\it ROSAT} PSPC
and {\it CGRO} OSSE observations). 
They confirmed previous findings that the time-variability patterns differ 
in the soft and hard X-ray bands, with the {\it ROSAT} PSPC count 
rate in the 0.1--1~keV band remaining constant at 
$\sim 0.46\ {\rm count\ s^{-1}}$ whilst the 1--2~keV count rate increased 
a factor $\sim$2 (to $\sim 0.32\ {\rm count\ s^{-1}}$)
on a timescale $\sim2$~days (around the time of the 
NGC~4151(3) {\it ASCA} observation).
Unfortunately the {\it ROSAT} observations ended $\sim 2$~days later
(and at the time of the 4151(4) {\it ASCA} observation), but the 
subsequent {\it ASCA} observations showed that 1--10~keV count rate
remained relatively constant ($\lesssim30$\% variability) for a period of 
$\sim 6$~days thereafter.
We do not attempt a detailed comparison of our results with these 
findings here. However we note that our best-fitting model to 
the 4151(4) dataset does predict a count rate in the 1.0--2.0~keV band 
of the {\it ROSAT} PSPC of 
$0.30\ {\rm count\ s^{-1}}$, consistent with that observed.
The corresponding count rate prediction for the 0.1--1.0~keV is 
clearly a strong function of the spectrum assumed $< 0.6$~keV
especially the value assumed for $N_{H,0}$. 
This band is clearly poorly constrained with {\it ASCA} which 
results in a range of predicted PSPC count rates 
$\sim 0.2$--$0.8\ {\rm count\ s^{-1}}$ at this epoch, but which encompasses 
that observed.

In \S\ref{Sec:thermal} we suggested that the spectrum in $\lesssim 1$~keV 
was composed of a fraction 
$D_f \sim$5\% of the underlying continuum which is
allowed to escape without suffering any attenuation, the emission 
spectrum of the photoionized ionized gas 
and thermal emission from an optically-thin plasma.
At energies $\gtrsim 1.5$~keV, the observed spectrum becomes increasingly 
dominated by the underlying continuum transmitted by the ionized absorber
(e.g. Fig.~9).
The emission from the optically-thin plasma, identified with the 
extended emission observed in the source, is assumed to be constant 
with time and under the above assumptions is of relatively minor
importance to the observed flux (corresponding to a
count rate in the 0.1--1~keV band of the
{\it ROSAT} PSPC of $10^{-3}$--$0.15\ {\rm count\ s^{-1}}$).
The variability timescales for remaining spectral components differ, 
with that for the transmitted continuum determinded by the 
the variability timescale of the central source and/or the timescale 
over which significant changes in the absorbing column density 
occur within the cylinder--of--sight (above). 
The timescales on which variations in the intensity of the scattered continuum 
and emission from the photoionized gas occur are determined by the 
light travel-times to the corresponding regions.
If we assume that the rise in the 1--2~keV count rate observed 
by {\it ROSAT} is due to a brightening in the central source, then 
the lack of a corresponding increase $<1$~keV seen in NGC~4151(4,5)
implies these regions lie $\gtrsim 4$~light-days.
However if the rise seen by {\it ROSAT} is simply the result of a change in
the absorbing column, then no such constraints can be placed on the 
location of these regions.

We note that Weaver et al (1994a) presented the results of 
a {\it BBXRT} observation of NGC~4151, and 
considered a model similar to our model {\it C(ii)}
(but with $\Omega = 4\pi$).
They found (see Weaver et al \S4.3)
$\Gamma \sim 1.5$, 
$N_{H,z} \sim 10^{23}\ {\rm cm^{-2}}$, 
$D_f \sim 0.04$ (from their Fig.14), 
and $U_X \sim 10^{-2}$ for their 'high-state' dataset, 
decreasing by a factor $\sim 2$ during the 'low-state'
1~day later.

NGC~4151 has long been known to exhibit variable absorption features
in UV (e.g. Penston et al 1981). However the wide range of ionization 
states observed, including both low (Lyman series, C{\sc iii})
and high (N{\sc v}, O{\sc vi}) ionization UV lines,
exclude a single-zone photoionization model accounting for the absorption 
in both the UV and X-ray bands (Kriss et al 1995).

\underline{\bf Mrk 766:}\\
This Seyfert 1.5 galaxy was first detected in X-rays by {\it Einstein} 
(Kriss, Canizares \& Ricker 1980) 
and has since been revealed to be one of the most highly 
variable AGN (e.g. Ghosh \& Soundararajaperumal 1992b).
In our analysis of the data from {\it ASCA} observation perfomed 
in 1993 December, we found Mrk~766 to be second most variable 
source in our sample, with evidence that the amplitude was larger 
at $\lesssim 2$~keV (Paper~I). Leighly et al (1996b) have presented 
a more detailed analysis of this dataset and came to the same conclusion,
finding the amplitude of the variability to be highest at $\sim 1$~keV.
They suggest this is due to the intensity of a soft component decreasing 
slightly whilst the intensity and steepness of the underlying continuum 
$\gtrsim 1$~keV  intensity increased 
(from $\Gamma \sim 1.6$ to $\Gamma \sim 2.0$).
A similar change in the form of the continuum with flux state
was previously suggested on the basis of {\it ROSAT} PSPC results for this 
source (Molendi \& Maccacaro 1994; Netzer, Turner \& George 1994).

From the analysis of the mean spectrum presented here, we find 
model {\it A(i)} to provide an unsatisfactory 
description of the spectra.
Fits satisfying our criteria for acceptability are obtained 
if $\sim$50\% of the continuum escapes without suffering attenuation by
neutral material (model {\it A(ii)}). However
yet superior fits are obtained if the absorbing gas is assumed to be ionized
(model {\it B(ii)}) giving 
$\Gamma \sim 2$, $U_X \simeq 0.1$, 
$N_{H,z} \sim 1.6\times10^{22}\ {\rm cm^{-2}}$ and 
$D_f \sim 0.5$. 
We note that such a model requires no additional soft component
(eg. Fig~9).
Indeed, we find models {\it C(i)} \& {\it C(ii)} to offer only 
small improvements in the goodness-of-fit and thus find no compelling 
evidence for significant emission from the ionized gas.
In no case is there a requirement for any addition absorption by 
neutral gas in excess of $N_{H,0}^{gal}$.
It should be noted however, that 
that models {\it B(i)}--{\it C(ii)} all have 
$P(\chi^2 \mid dof) < 0.05$.

Leighly et al also report the presence of 
an absorption feature, identified with 
O{\sc vii}, and presented the results of photoionization calculations
assuming a model similar to our model {\it B(i)}.
They find a best-fitting column density 
$N_{H,z} \sim 6\times10^{21}\ {\rm cm^{-2}}$, in agreement with our findings
(Table~5). Unfortunately Leighly et al do not specify 
the limits on their assumed continuum, thus we are unable to compare their 
values to $U_X$ directly, however they do find an increase in
ionization parameter with intensity.

\underline{\bf NGC 4593:}\\
This Seyfert 1.0 galaxy was first detected in X-rays by {\it UHURU}
(Forman et al 1978).
In our analysis of the data from {\it ASCA} observation performed 
in 1994 January, we find model {\it A(i)} 
provides a satisfactory description of the data analysed, but fails 
our criteria for extrapolation $<0.6$~keV.
Fits satisfying our criteria for acceptability are obtained 
if $\sim$70\% of the continuum escapes without suffering attenuation by
neutral material (model {\it A(ii)}). However
yet superior fits are obtained if the absorbing gas is assumed to 
cover the entire source and is ionized
(model {\it B(i)}), giving $\Gamma \sim 2$, $U_X \simeq 0.1$, 
$N_{H,z} \sim 2\times10^{21}\ {\rm cm^{-2}}$.
This confirms the findings of 
R97 who found the addition of 
O{\sc vii} and O{\sc viii} 
edges ($\tau_{O7}\sim0.3$ \& $\tau_{O8}\sim0.1$)
significantly improved the fit 
to a single powerlaw model ($\Gamma \sim 2$) to these {\it ASCA} data.
R97 also fitted a single-zone photoionization model and found
$U_X^{R97} \sim 0.05$,
$N_{H,z} \sim 3\times10^{21}\ {\rm cm^{-2}}$, $\Gamma \sim 1.9$.

We find that statistically there is no strong requirement for any of the 
underlying  continuum to escape without suffering attenuation by the ionized 
material, or for any significant emission from the ionized gas, 
(models {\it B(ii)}--{\it C(ii)}).
In all cases there is no required for any addition absorption by 
neutral gas in excess of $N_{H,0}^{gal}$.
We note that a substantially flatter spectral index has been suggested 
in the past (eg. $\Gamma \sim 1.1$, T91).

\underline{\bf MCG-6-30-15:}\\
This Seyfert 1.0 galaxy was first detected in X-rays by {\it HEAO-1} A-2
(Marshall et al. 1979).
Absorption due to ionized gas first suggested by {\it Ginga}
observations, with $N_{H,z} \sim 10^{23}\ {\rm cm^{-2}}$
(Nandra, Pounds \& Stewart 1990).
Our analysis of the data from two {\it ASCA} observations separated 
by $\sim3$ weeks in 1993 July, we confirm the presence of ionized gas.
However whilst none of the models presented in \S\ref{Sec:basic_models} 
provide a fit which formally satisfies our criteria for 
MGC-6-30-15(1), this is primarily due 
to the presence of a deficit at $\sim 1$~keV (Fig.6).
An acceptable solution is obtained for MGC-6-30-15(2) for model {\it C(i)},
but again some evidence for a relatively small deficit at  $\sim 1$~keV 
(Fig.10).

However, as further discussed in \S\ref{Sec:multi-mcg6} there is a 
significant increase in column density 
(from $N_{H,z} \sim 6\times10^{21}\ {\rm cm^{-2}}$ 
to $\sim 9\times10^{21}\ {\rm cm^{-2}}$),
along with a slight {\it increase} in ionization parameter 
(from $U_X \sim 0.11$ to $\sim 0.15$)
despite a factor $\sim 2$ {\it decrease} in the intensity
of the illuminating continuum between the 2 observations
(e.g. Fig~16).
This, the presence of the $\sim 1$~keV deficit, and 
a decrease in the depth of the O{\sc viii} edge
on a timescale $\sim 10^4$~s (but not in the depth of the 
O{\sc vii} edge) observed in an {\it ASCA} observation performed in
1994 July (Otani et al 1996) are
impossible to reconcile with that expected from
a uniform shell of gas reacting to
changes in the intensity of the photoionizing continuum.

A single-zone photoionization models have also been applied to 
these data by Fabian et al (1994) and Reynolds et al (1995), assuming an 
ionizing continuum consisting of a single powerlaw of $\Gamma = 2$. Using 
Fig~1b to convert from their quoted values of $\xi$
we obtain results in good agreement with 
those found here ($N_{H,z} \sim 6\times10^{21}\ {\rm cm^{-2}}$, 
$U_X \sim 0.09$ and 
$N_{H,z} \sim 13\times10^{21}\ {\rm cm^{-2}}$, 
$U_X \sim 0.11$
for the MCG-6-30-15(1,2) datasets respectively).
Photoionization models for the extended 
(4 day) {\it ASCA} observation carried out in 1994 July have been 
presented by Otani et al (1996) and R97.
The source exhibited a factor $\sim 4$ variability in the 
observed flux during this observation (covering the two flux states 
represented by MCG-6-30-15(1,2))
leading Otani et al to divide the observation into 2 parts. 
They found 
$N_{H,z} \sim 5\times10^{21}\ {\rm cm^{-2}}$, 
$U_X \sim 0.04$ for the first 300~ks of the observation, but 
$N_{H,z} \sim 10\times10^{21}\ {\rm cm^{-2}}$, 
$U_X \sim 0.18$ for the last 60~ks.
For reference, 
R97 analysed the time-averaged spectrum and obtained
$U_X^{R97} \simeq 0.05$ and
$N_{H,z} \sim 6\times10^{21}\ {\rm cm^{-2}}$)
The physical implications of these results have been further explored
by Reynolds \& Fabian (1995).

Curiously (but see Reynolds \& Fabian 1995) MCG-6-30-15 appears to be 
heavily reddened in the optical/UV thus little is known regarding the 
presence of UV absorption features.

\underline{\bf IC 4329A:}\\
This Seyfert 1.0 galaxy was first detected in X-rays by {\it Ariel-V} 
(Elvis et al 1978).
In our analysis of the data from {\it ASCA} observation performed 
in 1993 August, we find no model in \S\ref{Sec:basic_models} which
satisfies our formal criteria for acceptability. 
However we do find a significant improvement in the fits for models
including ionized gas, with 
$U_X \sim 0.02$--0.04, $N_{H,z} \sim 2\times10^{21}\ {\rm cm^{-2}}$
and $\Gamma \sim 1.9$ for all models {\it B(i)}--{\it C(ii)}.
A substantial column density of neutral gas is also required for this source
$N_{H,0}\sim 4\times10^{21}\ {\rm cm^{-2}}$, far in excess of 
$N_{H,0}^{gal}$, but consistent with previous findings and the 
edge-on orientation of the host galaxy.
Cappi et al (1996) and R97 have both performed independent 
analyses of this {\it ASCA} dataset.
Both find the addition of two edges to a single powerlaw model 
($\Gamma \sim 1.9$) significantly improved the goodness-of-fit over the 
0.6--1.0~keV band. R97 fixed the edge energies to those appropriate 
for O{\sc vii} and O{\sc viii} and found $\tau_{O7}\sim0.6$ and
$\tau_{O8}\sim0.1$. Cappi et al allowed the line energies to be free
and obtained a different ratio of optical depths, but otherwise 
similar results. 
Both Cappi et al and R97 also fitted the data with a similar
photoionization model and obtained
$\xi \sim 10$ ($\equiv U_X^{R97} \sim 0.02$), 
$N_{H,z} \sim 3\times10^{21}\ {\rm cm^{-2}}$ and
$N_{H,0}\sim 3\times10^{21}\ {\rm cm^{-2}}$.
These results are consistent with those found in a {\it ROSAT}
PSPC observation (Madejski et al 1995).

We also found evidence for a strong Compton-reflection 
component in this source with ${\cal F} \sim 4$
(\S\ref{Sec:reflection}), similar to that found by Cappi et al.
This hard tail was also found in 
{\it HEAO-1} (W95) and 
{\it Ginga} (Miyoshi et al 1988; Piro, Yamauchi \& Matsuoka 1990)
observations, and 
offers an explanation the {\it CGRO} OSSE spectrum of this source
(Madejski et al 1995).

None of the models quoted by Cappi et al (1996) and R97
formally satisfy our criteria for acceptability (though 
some of the fits described by  Cappi et al
are considered acceptable to those workers).
It consider it likely that the
poverty of the various fits to the {\it ASCA} data
are a combination of unaccounted for systematics errors in the 
calibration and additional spectral complexity in the soft X-ray band.
As can be seen from Table~10, 
even we applied the Compton-reflection fit, an excess of 
counts remains $<0.6$~keV. A similar effect was reported by 
Cappi et al.
This may be the result of a real steepening 
of the intrinsic emission from IC~4329A, or the result of contamination 
of the {\it ASCA} data by emission 
from the nearby elliptical galaxy, IC~4329
(see Madejski et al 1995).

NP94 found 
evidence for absorption by ionized gas in {\it Ginga} observations, 
with $N_{H,z} \sim 10^{22}\ {\rm cm^{-2}}$.

\underline{\bf NGC 5548:}\\
This Seyfert 1.2 galaxy was first detected in X-rays by {\it OSO-7} 
(Hayes et al 1980).
Absorption by ionized gas in NGC~5548 was first suggested by 
{\it Ginga} observations (Nandra et al 1991) and confirmed by 
the {\it ROSAT} PSPC (Nandra et al 1993).
In our analysis of the data from {\it ASCA} observation perfomed 
in 1993 July, 
we find acceptable solutions with
$U_X \sim 0.1$, $N_{H,z} \sim 3\times10^{21}\ {\rm cm^{-2}}$
and $\Gamma \sim 1.9$ for all models {\it B(i)} \& {\it C(i)}.
Consistent results are also obtained for model {\it B(ii)}, with no requirement
for an unattenuated ($D_f = 0$).
This confirms the results of R97 who found the addition of 
O{\sc vii} and O{\sc viii} edges ($\tau_{O7}\sim0.25$ \& $\tau_{O8}\sim0.16$)
significantly improved their fit to a single powerlaw model ($\Gamma \sim 1.9$) 
to these {\it ASCA} data.
R97 also fitted a single-zone photoionization model and found
$U_X^{R97} \sim 0.08$, $N_{H,z} \sim 5\times10^{21}\ {\rm cm^{-2}}$, 
assuming a single optical--X-ray continuum $\Gamma \sim 1.9$.
Mathur, Elvis \& Wilkes (1995) have also presented photoionization 
calculations based on previously published {\it ASCA} results
and obtain $U_X \sim 0.06$ (for their quoted optical--X-ray continuum)
assuming $N_{H,z} \sim 4\times10^{21}\ {\rm cm^{-2}}$.

We find a significant improvement in the goodness--of--fit is achieved assuming 
model {\it C(ii)}. However the best-fitting solution occupies the region of 
high-$N_{H,z}$, high-$U_X$, high-$D_f$ parameter-space and does not extrapolate 
well $<0.6$~keV (Fig~12).
A yet superior improvement was found in 
\S\ref{Sec:2plaw} consisting of the combination of a steep powerlaw
($\Gamma_s \sim 3.2$) dominating the lowest energies and a flatter 
powerlaw ($\Gamma_h \sim 1.5$) dominating at highest energies.
The ionized gas in this case has 
$U_X \sim 0.2$, $N_{H,z} \sim 7\times10^{21}\ {\rm cm^{-2}}$
but also requires a neutral column density 
($N_{H,0} \sim 1.7\times10^{21}\ {\rm cm^{-2}}$) far in excess of
$N_{H,0}^{gal}$.

A spectral flattening towards higher energies also seen in {\it HEAO-1} (W95) 
and {\it Ginga} (NP94) along with 
evidence for absorption by ionized gas, 
with $N_{H,z} \sim 5\times10^{22}\ {\rm cm^{-2}}$.
A 'soft excess' has been proposed in NGC~5548 from the analysis of
{\it EXOSAT} (TP89)
and 
{\it ROSAT} (Nandra et al 1993; Done et al 1995)
datasets, but we find any such component to be insignificant 
$> 0.6$~keV from the analysis presented here.

NGC~5548 has been observed to have variable absorption lines of C{\sc iv},
N{\sc v} and Ly$\alpha$ in the UV (Shull \& Sachs 1993; 
Mathur, Elvis \& Wilkes 1995).
A detailed comparison between the UV and X-ray absorption systems 
indicates the same gas is probably responsible
(Mathur, Elvis \& Wilkes 1995).

\underline{\bf Mrk 841:}\\
This Seyfert 1.5 galaxy was first detected in X-rays by {\it Einstein}
(Tananbaum et al 1986) and has since been observed to exhibit 
dramatic spectral variability (eg.  George et al 1993 and references therein).
In our analysis of the data from two {\it ASCA} observations 
performed in 1993 August and 1994 February, 
we find model evidence for ionized gas in Mrk~841(1), with model {\it B(i)}
offering an acceptable description of the data
with $\Gamma \sim 1.9$, $U_X \sim 0.16$,
$N_{H,z} \sim 3\times10^{21}\ {\rm cm^{-2}}$.
However the features are relatively weak
(Fig~6) and not required by the Mrk~841(2)
dataset, presumably due to its lower signal-to-noise ratio.
A superior fit is obtained for Mrk~841(2) assuming model 
{\it B(ii)}. However the solution occupies the region of 
high-$N_{H,z}$, high-$U_X$, high-$D_f$ parameter-space and the 
parameters are poorly constrained, primarily since any derivations
from a single powerlaw are rather subtle 
(e.g. Fig~9).
R97 also found a lack of evidence for
O{\sc vii} and O{\sc viii} absorption
edges in their analysis of the {\it ASCA} data from Mrk~841(2).
Interestingly, all models suggest the underlying continuum was 
flatter during Mrk~841(2) (typically $\Gamma \sim 1.6$--1.7)
although the observed fluxes in the 2--10~keV band were similar
($F_{2-10} \sim 1.2$ and $1.1\times10^{-11}\ {\rm erg\ cm^{-2}\ s^{-1}}$
respectively).

No significant improvement in the goodness--of--fit are obtained for
models {\it C(i)} \& {\it C(ii)} with the high values 
for the intensity of the ionized emitter derived in these cases 
($\Omega/4\pi \sim 50$) most likely being an artifact 
of a separate soft component.
The existence of a separate soft component has been suggested 
by some (but not all) previous observations (e.g. see George et al 1993).
Most recently, the spectrum from an extended {\it ROSAT} PSPC observation of 
Mrk~841, performed over a $\sim 6$~day period in 1992 
can be well described by a powerlaw with $\Gamma \sim 2.2$ (i.e. steeper than 
that found here) along with a $kT \sim 0.08$~keV blackbody
(Nandra et al 1995).
However the data were equally well described by a yet steeper powerlaw 
($\Gamma \sim 2.4$) and a photoionized absorber with 
$N_{H,z} \sim 2 \times 10^{21}\ {\rm cm^{-2}}$
$U_X \sim 0.04$ (assuming a powerlaw optical--X-ray continuum).
Unfortunately, 
as described in \S\ref{Sec:multi-841},
given the signal-to-noise ratio of all the datasets and the dramatic spectral 
variability exhibited by the source, a further interpretation of these results 
is somewhat problematic.
However we do note that the extrapolation of the underlying powerlaw implied
by the PSPC spectrum does agree remarkably well with the UV data, implying 
the luminosity of the component dominating the 13.6--100~eV band 
may have been grossly overestimated in the past
(e.g. see fig 3 of Nandra et al 1995).

\underline{\bf NGC 6814:}\\
In the early 1990s, NGC~6814 had its 15 months of fame due to the 
apparent detection of a periodicity in its X-ray lightcurve 
(Mittaz \& Branduardi-Raymont 1989; Done et al 1992)
spawning a frenzy of theoretical activity.
However, subsequent observations by the {\it ROSAT} PSPC revealed the 
periodicity to be the result of strong contamination by a Galactic source 
within the field-of-view of previous instruments (Madejski et al 1993).
It now appears that the X-ray emission from this Seyfert 1.2 galaxy 
is relatively weak.  
{\it ASCA} observed  NGC 6814 on 3 occasions 
prior to 1994 May, but only one dataset (NGC 6814(1)) met the criteria 
for inclusion in our sample (Paper~I).
The count rate for this observation, 
performed in 1993 May, is a factor ten lower than any of the other sources 
in our sample.
Given the signal-to-noise ratio, it is perhaps not surprizing that we find 
the dataset to be consistent with a simple powerlaw ($\Gamma \sim 1.7$) 
absorbed by $N_{H,0}^{gal}$. R97 also found a lack of evidence for
O{\sc vii} and O{\sc viii} absorption edges in their analysis of this 
dataset.

\underline{\bf Mrk 509:}\\
This Seyfert 1.2 galaxy was first detected in X-rays by {\it Ariel-V} 
(Cooke et al 1978).
In our analysis of the data from an {\it ASCA} observation performed in 1994 April,
we find a simple powerlaw (model {\it A(i)} with $\Gamma \sim 1.9$) 
provides a reasonable description of the data (\S\ref{Sec:zwabspo}). 
Whilst a significant improvement in the goodness--of--fit 
is achieved assuming model {\it A(ii)}, $\sim 80$\% of the continuum is 
unattenuated giving rise to a relatively subtle change in 
the spectrum $\lesssim 1$~keV (Fig~4).
A further, significant improvement in the goodness-of--fit is obtained 
for model {\it B(i)}, with $\Gamma \sim 2.0$, $U_X \sim 10^{-2}$,
$N_{H,z} \sim 8\times10^{20}\ {\rm cm^{-2}}$, but again 
the features imprinted on the continuum are relatively weak
(Fig~6). 
This confirms the results of R97 who found the addition of shallow
O{\sc vii} and O{\sc viii} edges ($\tau_{O7}\sim0.11$ \& $\tau_{O8}\sim0.04$)
significantly improved their fit to a single powerlaw model ($\Gamma \sim 2$) 
to these {\it ASCA} data.
R97 also fitted a single-zone photoionization model and found
$U_X^{R97} \sim 0.1$, $N_{H,z} \sim 3\times10^{21}\ {\rm cm^{-2}}$, 
the discrepancy with our results possibly arising as a result of their 
flatter continuum ($\Gamma \sim 1.9$).

No significant improvement in the goodness--of--fit are obtained assuming the
subsequent models in \S\ref{Sec:basic_models}, despite 
the suggestion of line-like soft-excess by {\it Einstein} (T91) and 
{\it EXOSAT} (Morini, Lipani, Molteni 1987) observations.
However a significant 
improvement is achieved if a moderate ($1 \lesssim {\cal F} \lesssim 5$)
Compton-reflection component is included (\S\ref{Sec:reflection}).
Such a component is also suggested 
by {\it HEAO-1} (W95) and {\it Ginga} (NP94) observations of the 
source and its inclusion steepens slightly the underlying powerlaw 
($\Delta\Gamma \sim 0.2$) and gives rise to a larger ionized column
density 
($N_{H,z} \sim 2\times10^{21}\ {\rm cm^{-2}}$) whilst leaving 
$U_X$ as for model {\it B(i)}.

\underline{\bf NGC 7469:}\\
This Seyfert 1.2 galaxy was first detected in X-rays by {\it UHURU} 
(Forman et al 1978), and observed by 
{\it ASCA} on 3 occasions over a period of $\sim 9$~days
in 1993 November. However, only one observation (NGC~7469(2)) met the 
criteria for inclusion in our sample (Paper~I).
We find a simple powerlaw (model {\it A(i)} with $\Gamma \sim 2.0$) 
provides an acceptable description of the spectrum.
However there is evidence for some spectral curvature, leading to 
a significant improvement in the goodness--of--fit 
for models {\it A(ii)}, {\it B(ii)} and {\it C(ii)} if
the bulk of continuum is unattenuated ($D_f \gtrsim 0.6$)
whilst the remainder is absorbed by a large column density
($N_{H,z} \gtrsim $few$\times10^{23}\ {\rm cm^{-2}}$), 
or if there is a strong (${\cal F} \sim 7$)
Compton-reflection component
(\S\ref{Sec:reflection}).

Guainazzi et al (1994) have presented a combined analysis of the
NGC~7469(1,2) datasets. They found a $\sim40$\% decrease
in intensity between the two epochs, with some evidence for
a steeper spectrum in the soft X-ray band ($\lesssim 0.8$~keV) 
when the source is brighter. No such spectral variability was seen 
$\gtrsim 2$~keV, with both epochs consistent with a powerlaw with 
$\Gamma \sim 1.9$.
We see no reliable evidence for a soft-excess in our analysis
of the NGC~7469(2) dataset, however we note that the results of
Guainazzi et al (their Fig.~3) indicate such a component 
would be almost indistinguishable at that epoch.
Spectral complexity in NGC~7469 $\lesssim 1$~keV has been 
suggested by previous {\it Einstein}, {\it EXOSAT} and {\it ROSAT}
observations although its form remains illusive
(e.g. Brandt et al 1993; Leighly et al 1996a and references therein).
We note some soft flux is provided in our models which include emission
but generally this is small (except in the case of our model including 
a Compton-reflection component).

In their analysis of the mean spectra from the NGC~7469(1,2) datasets,
Guainazzi et al find the putative soft-excess to be consistent with 
a blackbody with $kT\sim 0.08$~keV. 
Guainazzi et al find no evidence for O{\sc vii} and/or O{\sc viii}
absorption edges, whilst R97 find the addition of 
an O{\sc vii} edge (of optical depth $\tau_{O7}\sim0.2$)
significantly improved the fit 
to a single powerlaw model ($\Gamma \sim 2.1$) to the NGC~7469(1) dataset.

In \S\ref{Sec:reflection} we found that a strong Compton-reflection component
is allowed by the data. Such a result is also obtained from 
previous {\it HEAO-1} (W95) and {\it Ginga} (NP94; Leighly et al 1996a) 
observations. 
%As mentioned in \S\ref{Sec:reflection}, such a result requires
%the Fe abundance to be substantially lower than solar in order to suppress 
%the strength of the predicted fluorescence line.

Finally we note that NGC~7469 has recently been the subject of an intense 
monitoring campaign in the optical, UV and X-ray bands. Most interestingly,
preliminary results from the {\it RossiXTE} observations 
reveal the X-rays to have exhibited large-amplitude variability 
apparently uncorrelated with the variations seen in the UV and optical
(Nandra et al 1997d).

\underline{\bf MCG-2-58-22:}\\
This Seyfert 1.2 galaxy was first detected in X-rays by {\it Ariel-V}
(Cooke et al 1978). In our analysis of the data from an {\it ASCA} observation 
perfomed in 1993 May, 
we find a simple powerlaw (model {\it A(i)} with $\Gamma \sim 1.7$) 
provides a reasonable description of the spectrum (\S\ref{Sec:zwabspo}). 
A similar result was obtained by Weaver et al (1995) from an independent 
analysis of these data, who suggested the source to be a 'bare' Seyfert 1 
continuum with weak (if any) additional absorption or emission features.
These workers also noted that the continuum was relatively flat compared 
to the canonical $\Gamma \simeq 1.9$--2.0 for most Seyfert 1 galaxies 
popularized by 
the {\it Ginga} results of NP94.
However, as shown in \S\ref{Sec:basic_models}, we do find significantly
better fits when the more complex models are assumed, and in some cases 
obtain a derived index $\Gamma \sim 1.9$ 
(e.g. models {\it B(ii)} \& {\it C(ii)}). Some of these models obviously 
contain ionized gas, but the constraints of $U_X$ and $N_{H,z}$ are poor.
We note that previous {\it Einstein} (T91), {\it EXOSAT} 
(Ghosh \& Soundararajaperumal 1992a)
and {\it ROSAT} PSPC 
(Turner, George \& Mushotzky 1993) observations of MCG-2-58-22 have 
shown evidence for additional emission (features) $\lesssim 1$~keV.

In \S\ref{Sec:reflection} we found a solution including Compton-reflection,
but its strength is poorly constrained (${\cal F} \sim 2$--40), and all
but the lower values difficult to reconcile with the 
observed equivalent width of the Fe florescent line. 
However, we note that NP94 found no evidence for a 
Compton-reflection component
in the {\it Ginga} data (${\cal F} \sim 0.1$),
but did find 
evidence for absorption by ionized gas in {\it Ginga} observations, 
with $N_{H,z} \sim 10^{23}\ {\rm cm^{-2}}$.

\clearpage

\begin{figure*}[h]
\plotone{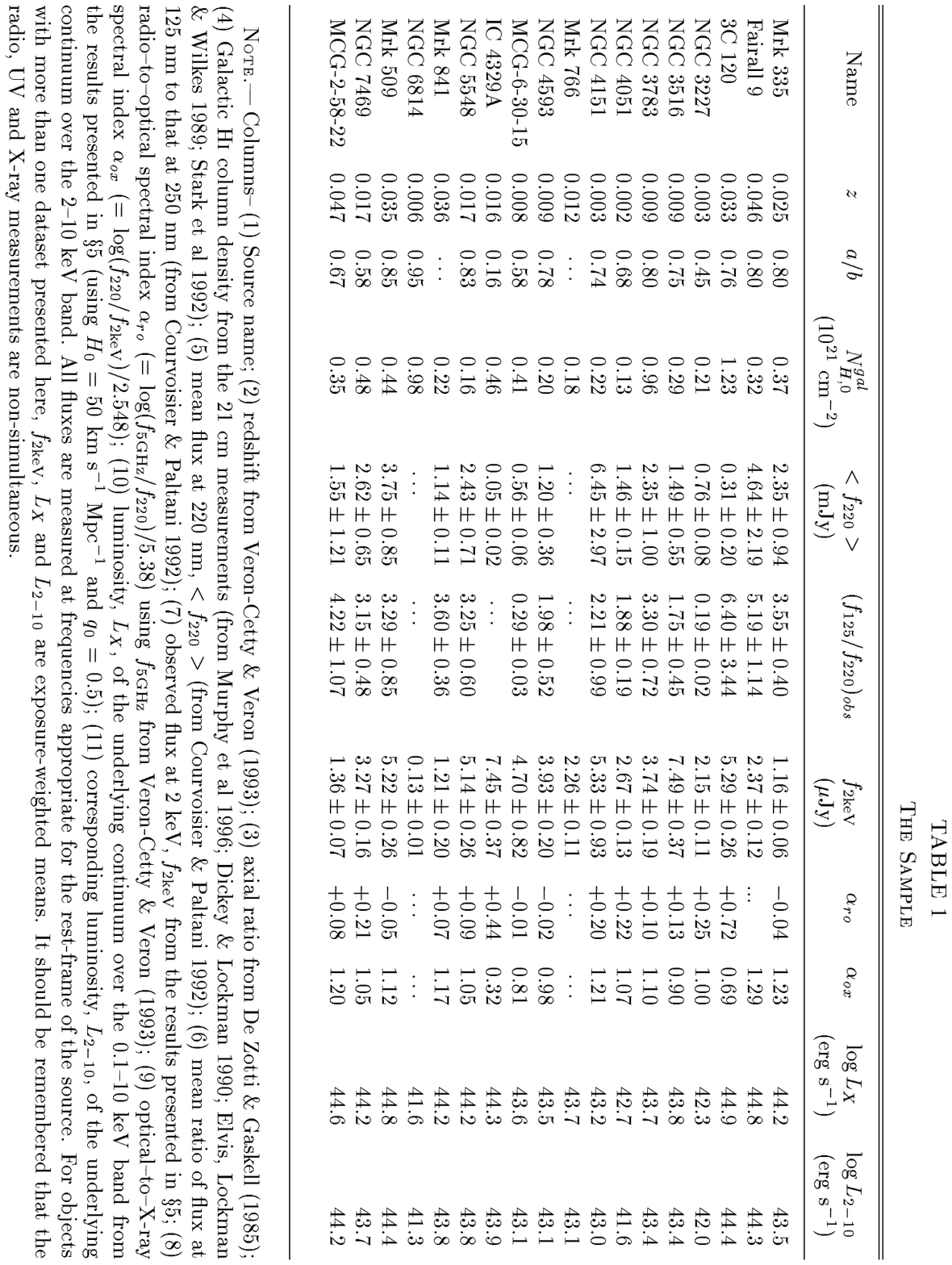}
\end{figure*}
\clearpage

\begin{figure*}[h]
\plotone{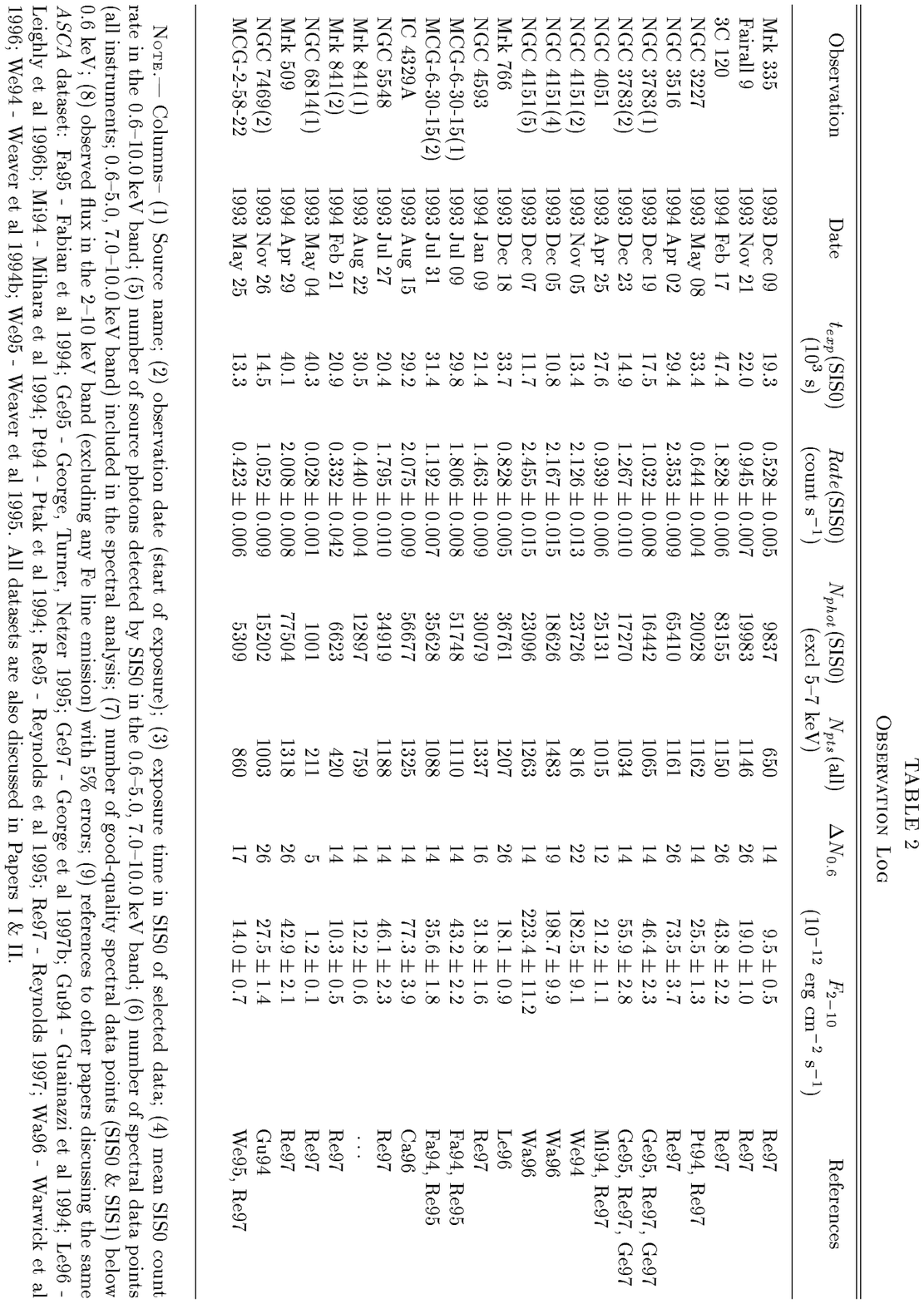}
\end{figure*}
\clearpage

\begin{figure*}[h]
\plotone{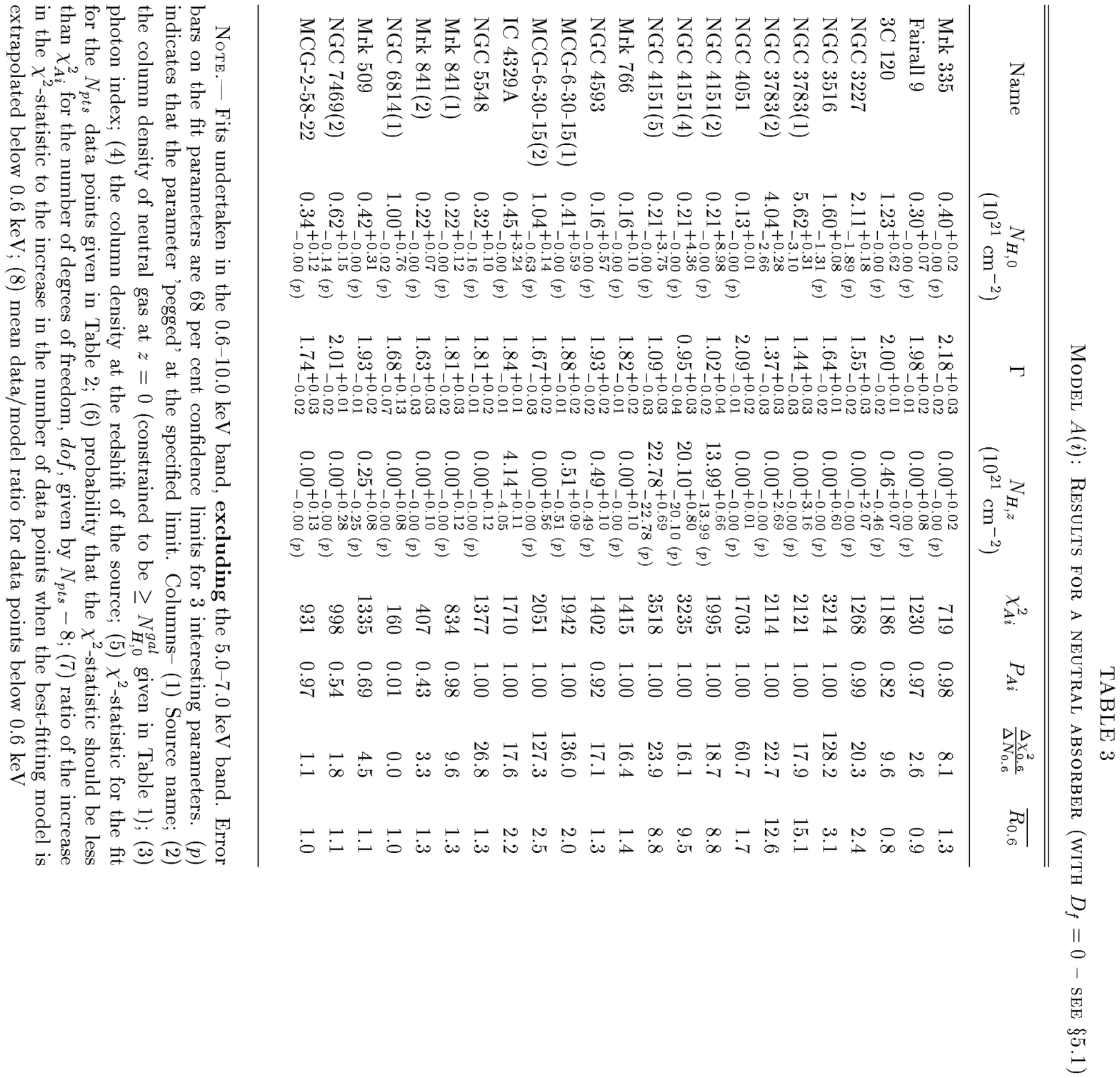}
\end{figure*}
\clearpage

\begin{figure*}[h]
\plotone{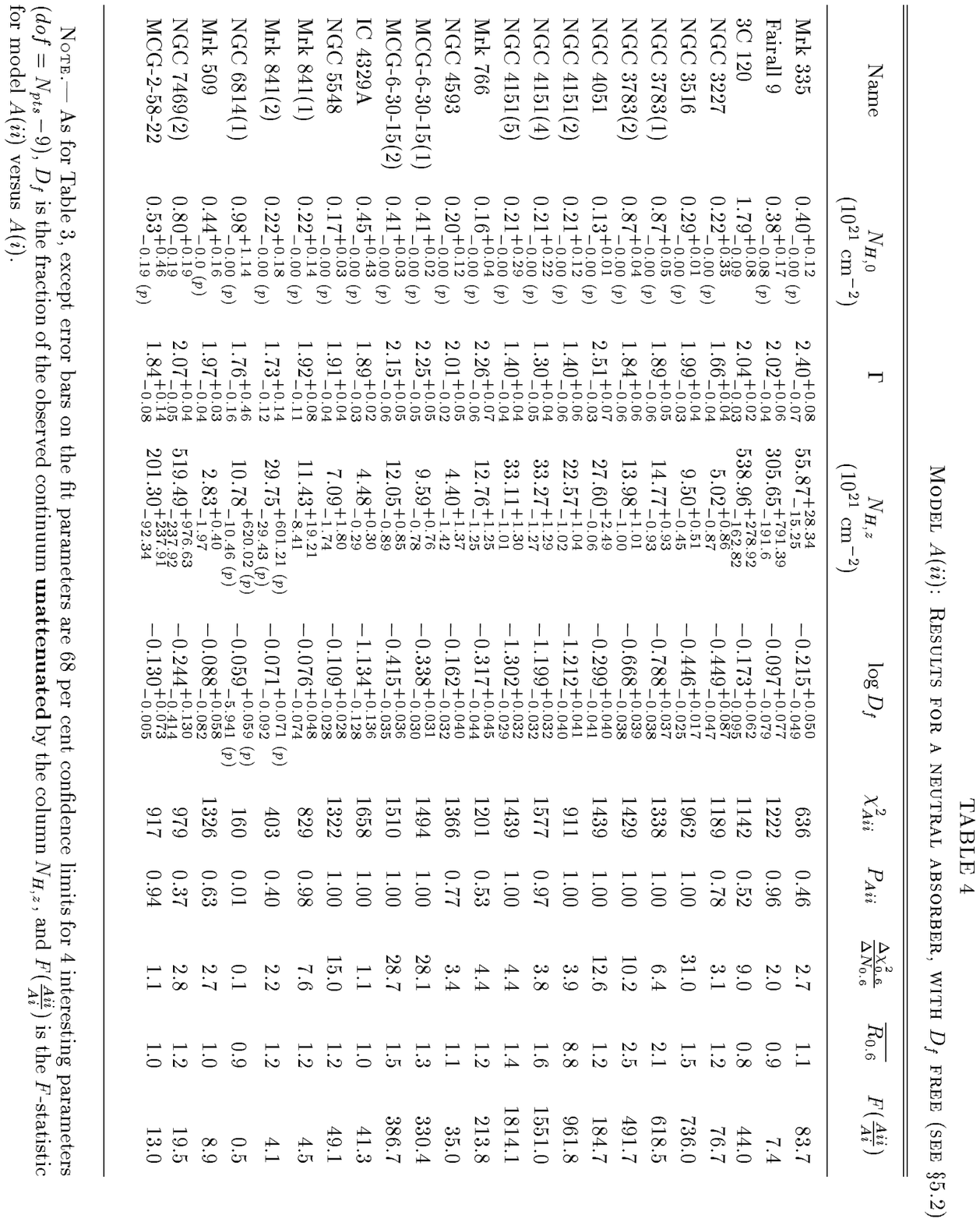}
\end{figure*}
\clearpage

\begin{figure*}[h]
\plotone{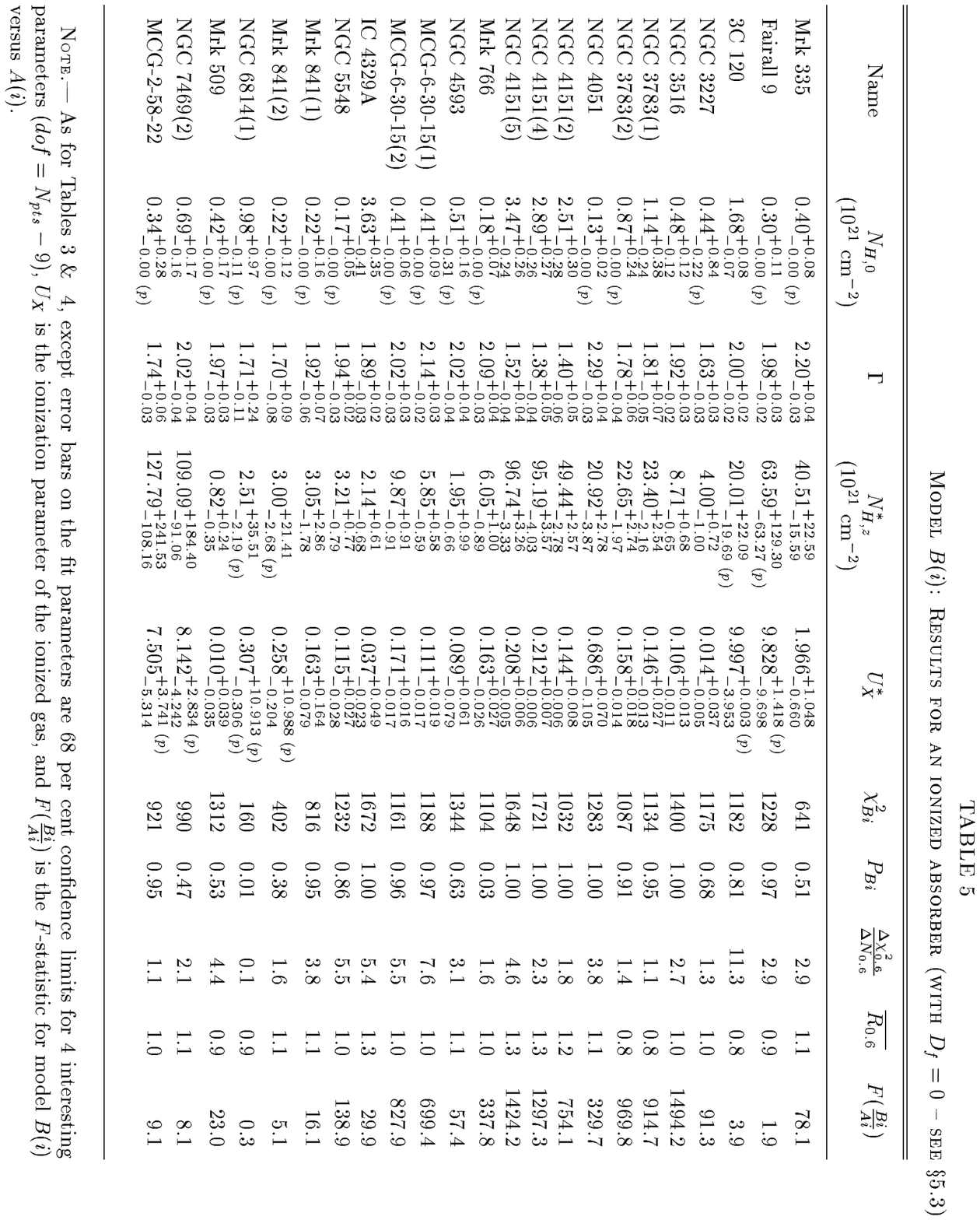}
\end{figure*}
\clearpage

\begin{figure*}[h]
\plotone{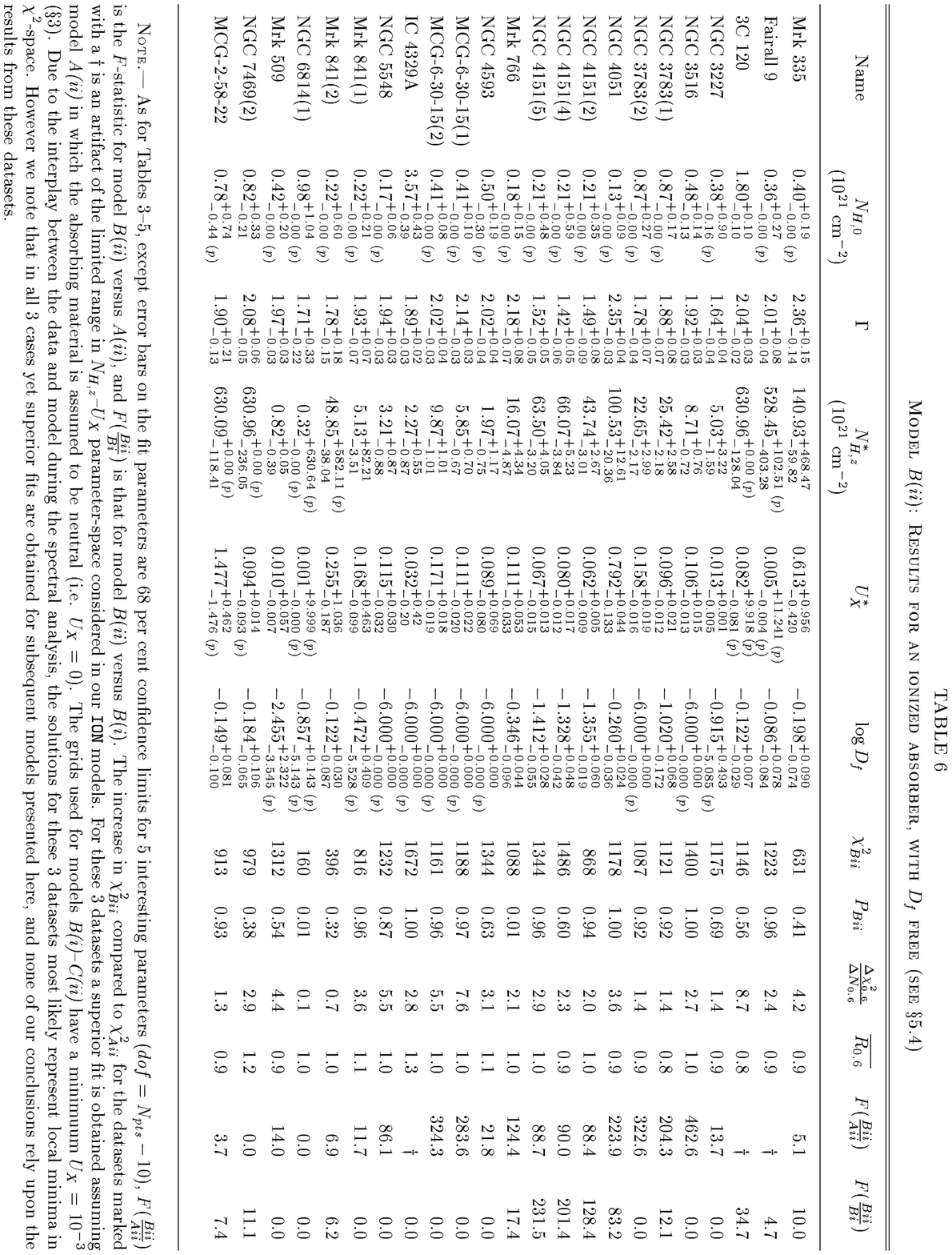}
\end{figure*}
\clearpage

\begin{figure*}[h]
\plotone{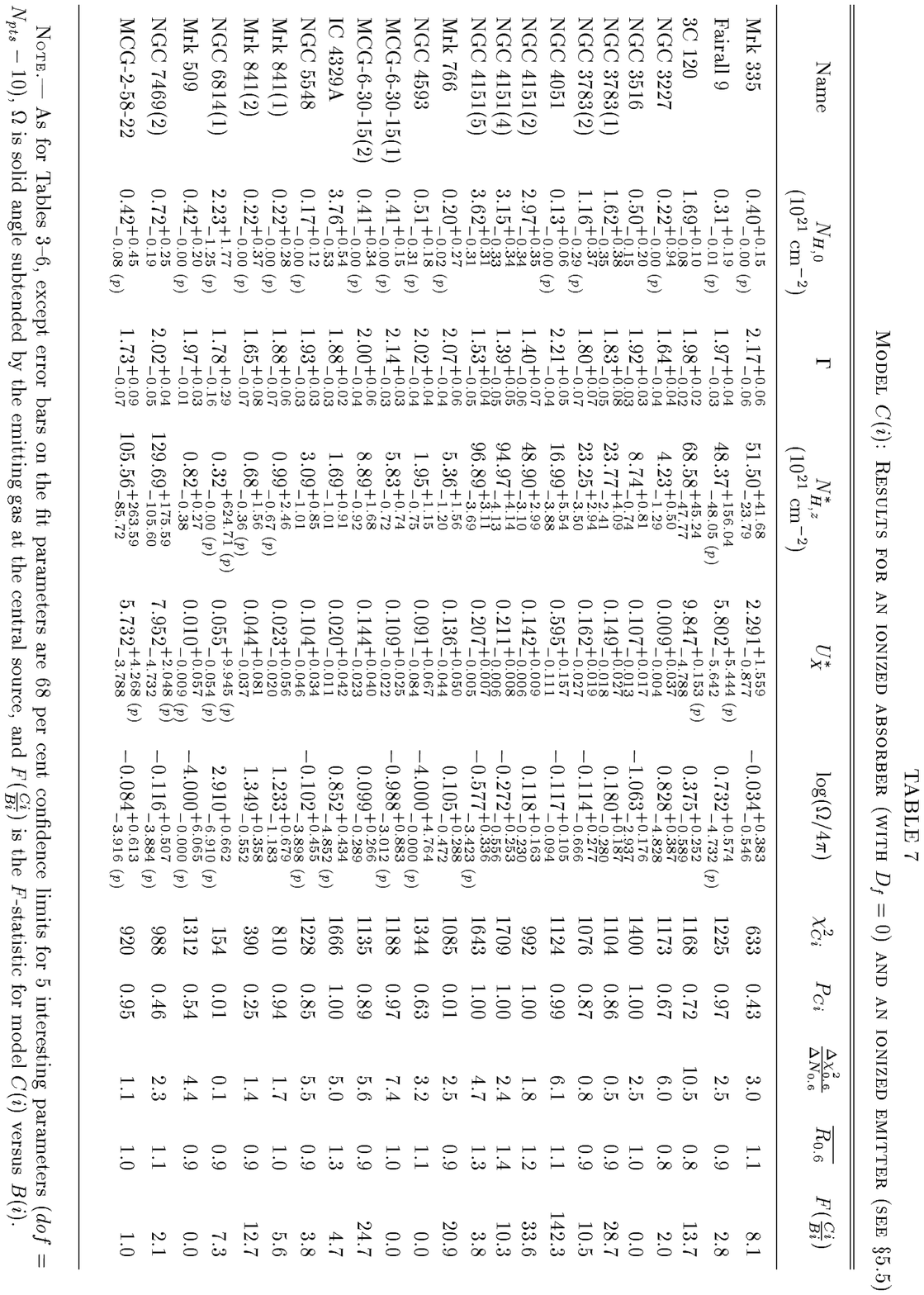}
\end{figure*}
\clearpage

\begin{figure*}[h]
\plotone{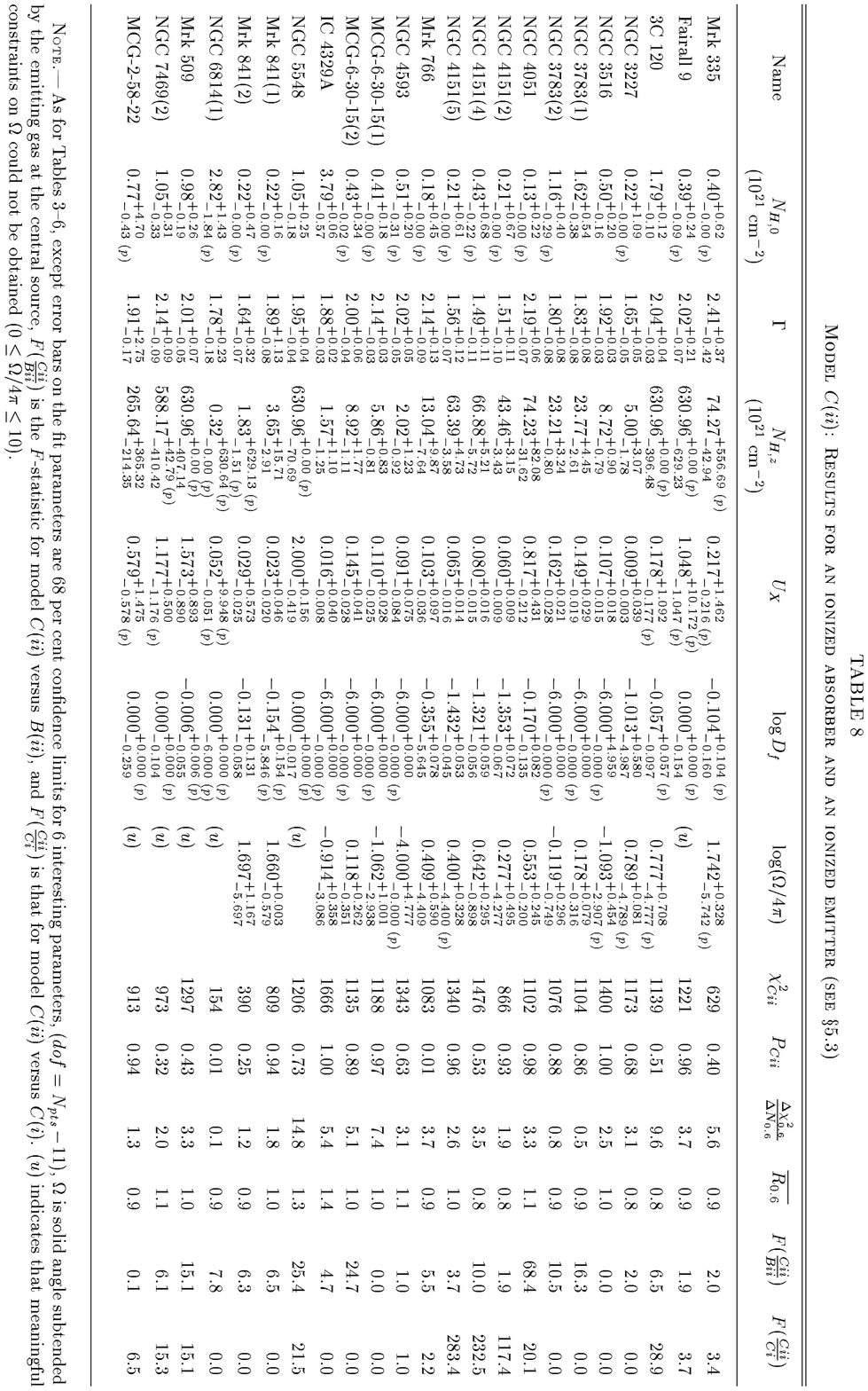}
\end{figure*}
\clearpage

\begin{figure*}[h]
\plotone{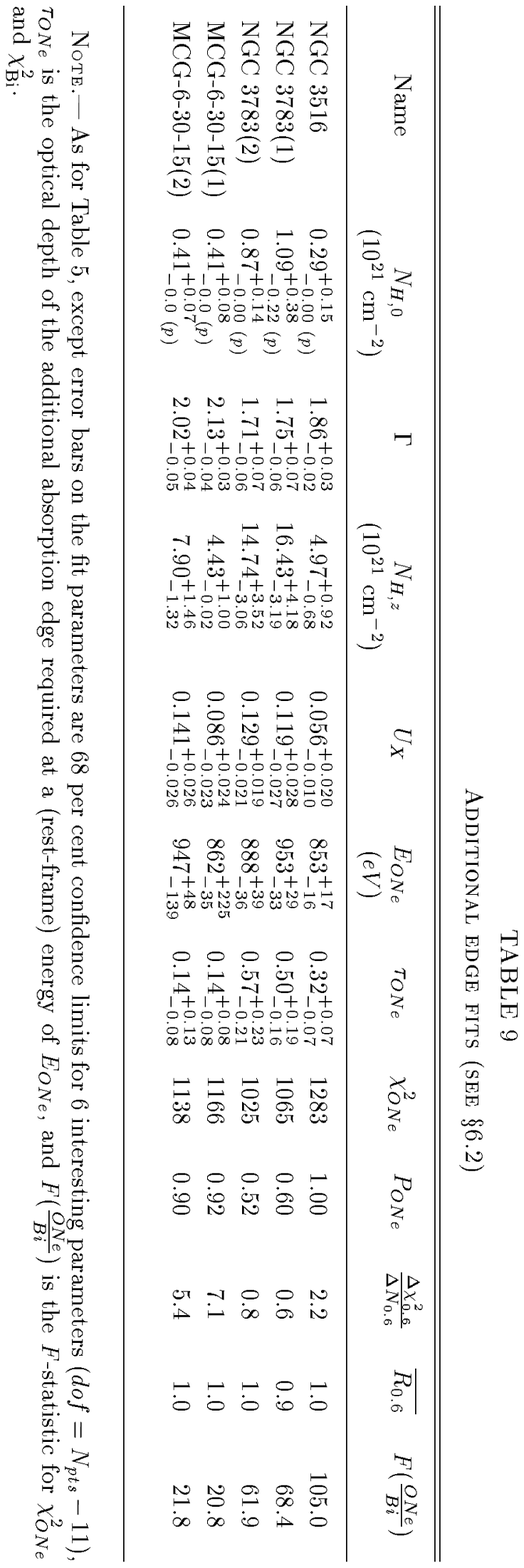}
\end{figure*}
\clearpage

\begin{figure*}[h]
\plotone{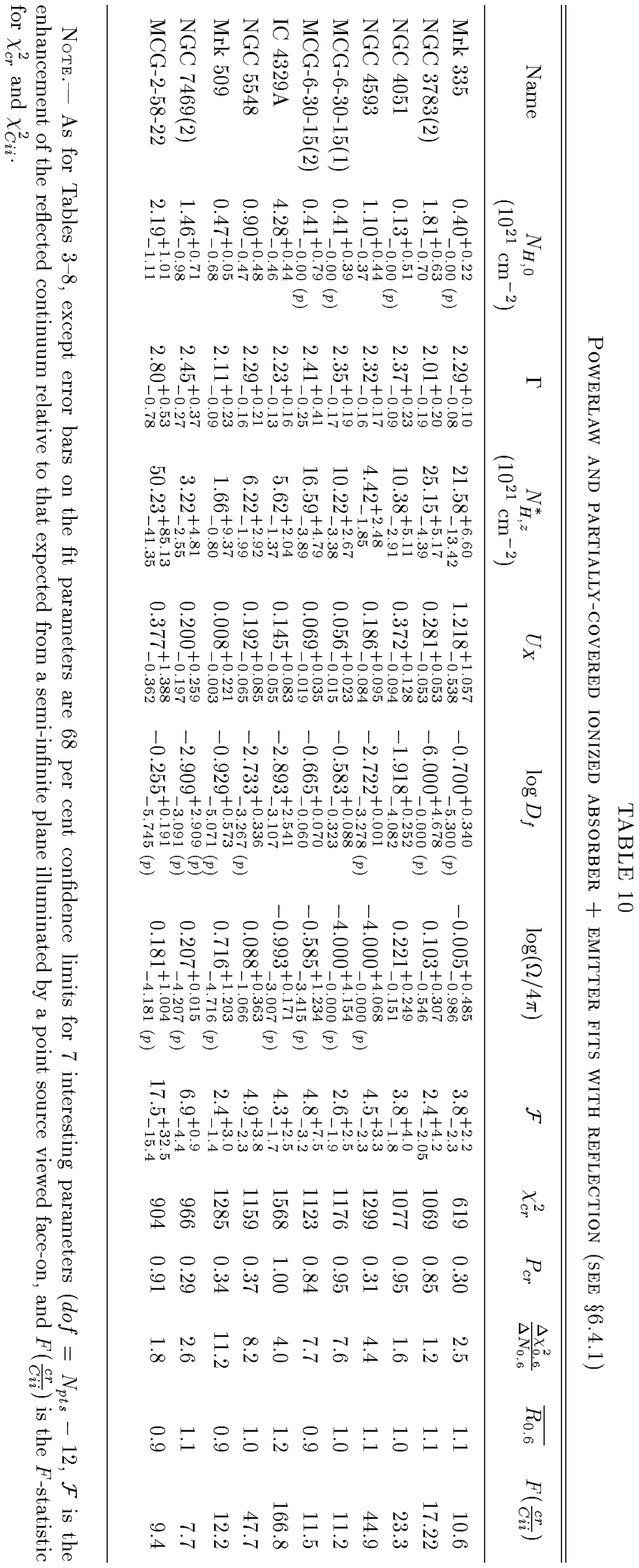}
\end{figure*}
\clearpage

\begin{figure*}[h]
\plotone{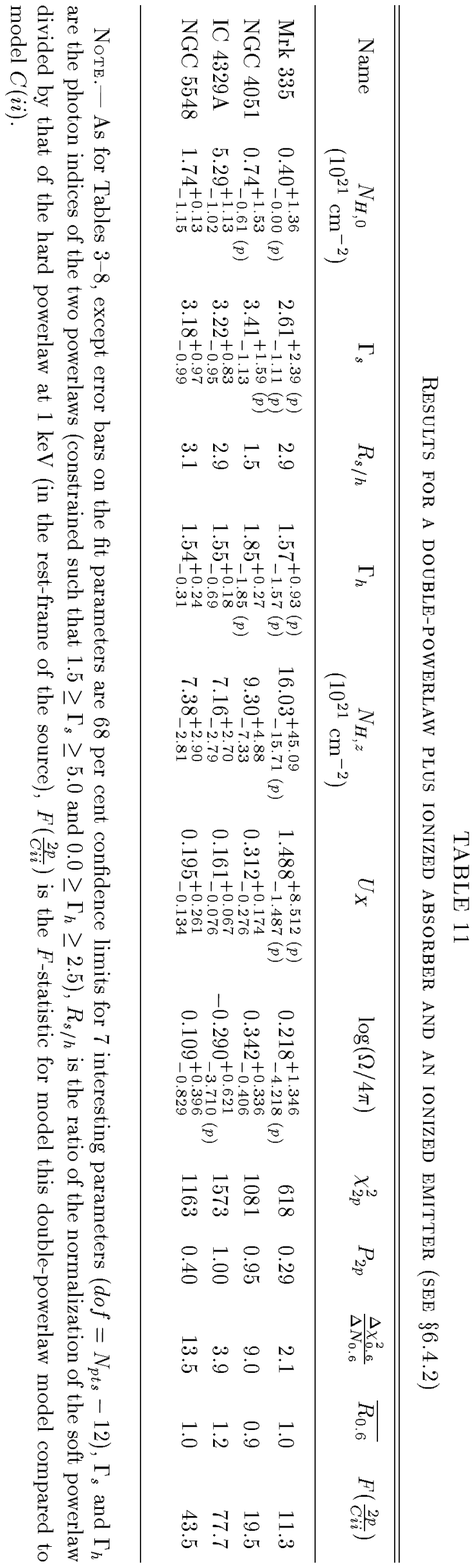}
\end{figure*}
\clearpage

\begin{figure*}[h]
\plotone{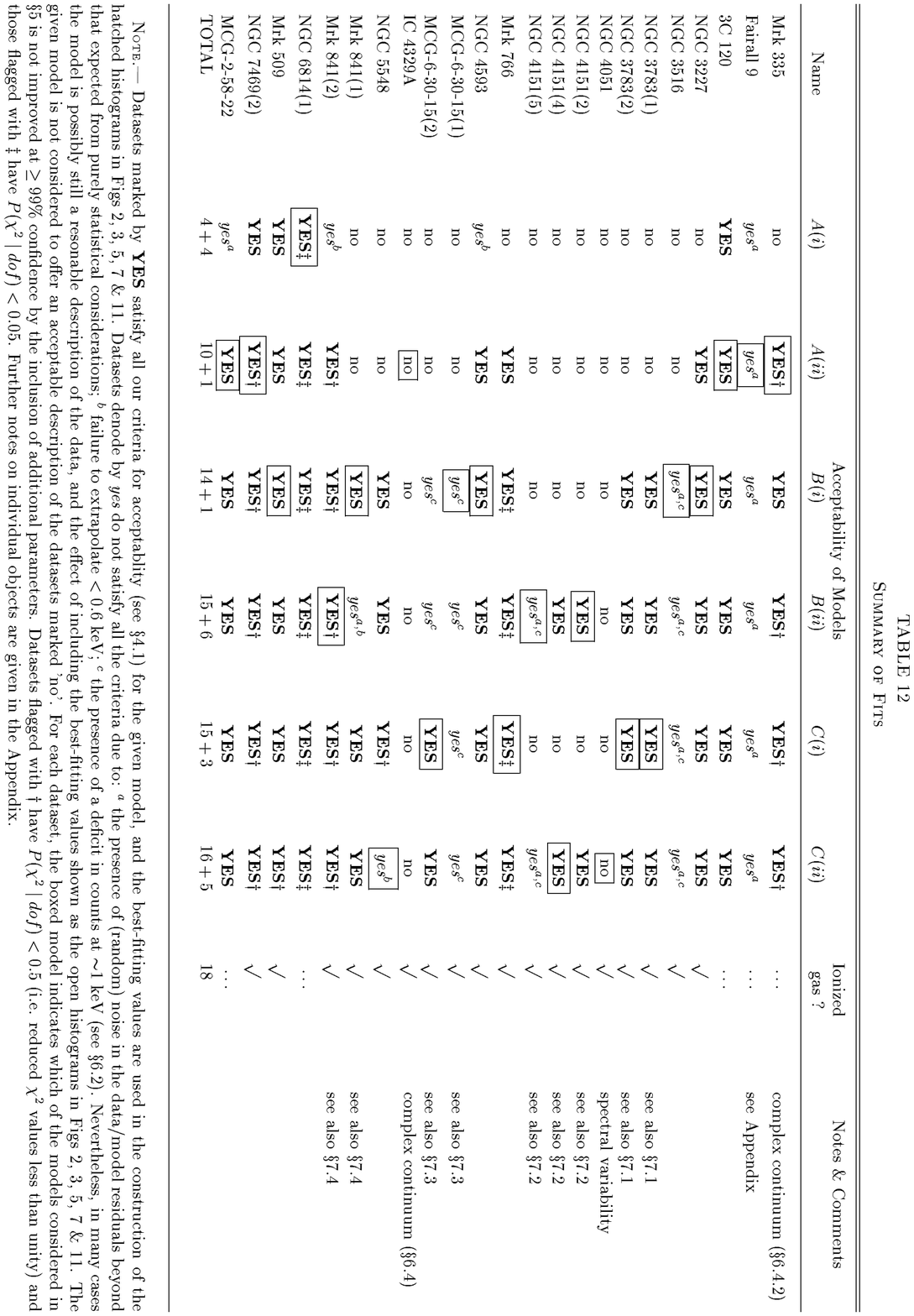}
\end{figure*}
\clearpage

\section*{Figure Captions}

% Figure contained in PS file f1.ps
{\bf Figure~1:}
(a) The solid curves give the ratio of the 'X-ray ionization parameter' 
    $U_X$ (eqn.~\ref{eqn:U_X}) to the more traditional ionization parameter 
    $U$ (defined over the entire Lyman continuum) for the form of the 
    optical--X-ray continua used here (see \S\ref{Sec:ion_model})
    assuming $\alpha_{ox} = 1.0$ (upper curve) and 
	$\alpha_{ox} = 1.5$ (lower curve). 
    The dashed curves give $U_X/U$ assuming the continuum 
    is a single powerlaw of photon index $\Gamma$ from 13.6~eV to 
    13.6~keV (upper curve) and to infinite energies (lower curve).
(b) The solid curves give the ratio of $U_X$ to $\xi$ 
    (in units of ${\rm erg\ cm\ s^{-1}}$), defined as $\xi = L/n_{H} R^2$, 
    where $n_{H}$ and $R$ are as in eqn.\ref{eqn:U_X}, but where $L$ is the 
    integrated luminosity over the 13.6~eV to 13.6~keV band for the form 
    of the optical--X-ray continua used here,
    assuming $\alpha_{ox} = 1.0$ (upper curve) and 
	$\alpha_{ox} = 1.5$ (lower curve). 
    The dashed curve gives $U_X/\xi$ assuming the continuum is a single 
    powerlaw of photon index $\Gamma$ from 13.6~eV to 13.6~keV
	(as used by Reynolds 1997 amongst others).
In both panels, the open and filled symbols indicate the corresponding ratios 
for the optical--X-ray continua given in Mathews \& Ferland (1987)
and Krolik \& Kriss (1995) respectively.

% Figure contained in PS file f2.ps
{\bf Figure~2:}
Histograms showing the range of photon index ($\Gamma$) of the underlying
continua derived assuming the various models described in
\S\ref{Sec:basic_models}. The histograms are generated by assigning a
rectangle of equal area to each dataset, of width determined from the
$1\sigma$ error bars for that parameter as listed in the relevant table.
In all cases the hatched region shows the histogram which formally
satisfy all the criteria given in \S\ref{Sec:acceptability}. The open
histogram shows the effect of also including the datasets which just fail to
meet our formal criteria but for which the model is considered likely to be
applicable (as described in the text and summarized in
Table~12).

% Figure contained in PS file f3.ps
{\bf Figure~3:}
As for Fig.~2, but for the column density ($N_{H,z}$)
of the material at the redshift of the source.
It should be remembered that in models {\it A(i)} \& {\it A(ii)}, the 
material is assumed to be neutral, whilst in models {\it B(i)}--{\it C(ii)}
it is photoionized (see \S\ref{Sec:models}).

% Figure contained in PS files f4a.ps, f4b.ps, f4c.ps
{\bf Figure~4:}
The derived spectrum and the corresponding rebinned, mean 
data/model ratios for the 8 datasets for which
model {\it A(ii)} (\S\ref{Sec:zpcfpo}; Table~4)
is acceptable and offers an improvement at $>$99\% confidence over
model {\it A(i)}. 
Also shown are the corresponding plots for Fairall~9 and IC~4329A as, 
although the fits to these 2 datasets fail to satisfy our formal criteria for 
acceptability, in neither case is the goodness--of--fit significantly 
improved by the inclusion of the additional parameters in subsequent models
presented in \S\ref{Sec:basic_models} (see text). 
The objects are presented in order of increasing Right Ascension.
The observed spectrum, corrected for the effects of Galactic absorption, is 
shown in the upper panel of each plot as the solid, bold curve. 
The individual spectral components which contribute to the observed spectrum
are also shown, as is the powerlaw (dashed) that would be visible if 
there was no attenuation along the line--of--sight at the redshift of 
the source (i.e. if $N_{H,z} = 0$).
As described in \S\ref{sec:anal}, for each observation 
the spectral analysis was performed 
by simultaneously fitting the mean spectra from all
4 instruments over the 0.6--5.0, 7.0--10.0~keV band.
The data/model ratios plotted as filled triangles in the lower panel 
are the (error-weighted) means of the 
ratios from the individual detectors within this energy range, rebinned in 
energy-space for clarity.
The open triangles show the corresponding rebinned, mean ratios when the 
best-fitting model is extrapolated $<0.6$~keV and into the 5--7~keV band.
It should be noted that significantly 
superior fits are obtained for more complex models 
in the case of Mrk~355 (\S\ref{Sec:2plaw}),
NGC~3227 \& NGC~4593 (\S\ref{Sec:ion}),
Mrk~766 (\S\ref{Sec:ion_pc})
and Mrk~509 (\S\ref{Sec:ion_pc_emis}).

% Figure contained in PS file f5.ps
{\bf Figure~5:}
As for Fig.~2, but for the fraction of the underlying
continuum ($D_f$) which does not suffer attenuation by 
material at the redshift of the source.

% Figure contained in PS file f6a.ps, f6b.ps, f6c.ps
{\bf Figure~6:}
As for Fig~4, but 
for the 8 datasets for which
model {\it B(i)}
(\S\ref{Sec:ion}; Table~5)
is acceptable and offers an improvement at $>$99\% confidence over
both models {\it A(i)} and {\it A(ii)}.
Also shown are the corresponding plots for NGC~3516 and MCG-6-30-15(1) as, 
although the fits to these 2 datasets fail to satisfy our formal criteria for 
acceptability, in neither case is the goodness--of--fit significantly 
improved by the inclusion of the additional parameters in subsequent models
presented in \S\ref{Sec:basic_models} (see text). 
It should be noted that significantly 
superior fits are obtained for more complex models 
in the case of 
NGC~3516 (\S\ref{Sec:2ion}),
NGC~3783(1,2) (\S\ref{Sec:ion_emis} \& \S\ref{Sec:2ion}),
Mrk~766 \& Mrk~841(2) (\S\ref{Sec:ion_pc})
and Mrk~509 (\S\ref{Sec:ion_pc_emis}).

% Figure contained in PS file f7.ps
{\bf Figure~7:}
As for Fig.~2, but for the ionization parameter ($U_X$) 
of the photoionized material, as defined by eqn.~\ref{eqn:U_X}.

% Figure contained in PS file f8.ps
{\bf Figure~8:}
The best-fitting values for 
$N_{H,z}$ and $U_X$ for each of the models including ionized gas
considered in \S\ref{Sec:basic_models}.
Shown are only those datasets for which the model offers an adequate 
description of the data {\it and}
leads to an improvement at $>99$\% over the corresponding 
model with neutral gas (i.e. model {\it A(i)} or {\it A(ii)}).
The solid symbols represent the 
datasets for which the model satisfies all our criteria for acceptability.
The open symbols represent the datasets which do not satisfy 
our formal criteria, but for which we consider the measurements of 
$N_{H,z}$ and $U_X$ to be reasonable estimates.
Overlaid on each panel are curves showing the $\tau=0.1$ (solid)
and $\tau=0.3$ (dotted) for O{\sc vii} and O{\sc viii}.
As futher discussed in \S\ref{Sec:disc-hagai}, ionized gas 
lying in the region of $U_X$--$N_{H,z}$ plane where 
$U_X > 10^{-22} N_{H,z}$ has an optical depth $\tau \lesssim 0.1$
for all ions with edges within the {\it ASCA} bandpass.
For the signal--to--noise ratio typical of {\it ASCA} observations
the ionized gas will thus be almost transparent.

% Figure contained in PS files f9a.ps, f9b.ps
{\bf Figure~9:}
As for Fig~4, but 
for the 5 datasets for which
model {\it B(ii)}
(\S\ref{Sec:ion_pc}; Table~6)
is acceptable and offers an improvement at $>$99\% confidence over
both models {\it A(ii)} and {\it B(i)}.
Also shown are the corresponding plots for  NGC~4151(5) as, 
although the fit to this dataset fails to satisfy our formal criteria for 
acceptability, the goodness--of--fit is not significantly 
improved by the inclusion of the additional parameters in subsequent models
presented in \S\ref{Sec:basic_models} (see text). 
It should be noted that significantly 
superior fits are obtained for more complex models 
in the case of 
NGC~3783(1) (\S\ref{Sec:ion_emis} \& \S\ref{Sec:2ion}),
NGC~4151(4) (\S\ref{Sec:ion_pc_emis})
and Mrk~509 (\S\ref{Sec:ion_pc_emis}).

% Figure contained in PS file f10.ps
{\bf Figure~10:}
As for Fig~4, but 
for the 4 datasets for which
model {\it C(i)}
(\S\ref{Sec:ion_emis}; Table~7)
is acceptable and offers an improvement at $>$99\% confidence over
both models {\it B(i)} and {\it B(ii)}.

% Figure contained in PS file f11.ps
{\bf Figure~11:}
As for Fig.~2, but for the solid angle ($\Omega$) 
subtended by the ionized gas at the origin of the illuminating continuum.

% Figure contained in PS file f12.ps
{\bf Figure~12:}
As for Fig~4, but
for the 4 datasets for which
model {\it C(ii)}
(\S\ref{Sec:ion_pc_emis}; Table~8)
is acceptable and offers an improvement at $>$99\% confidence over
both models {\it B(ii)} and {\it C(i)}.
It should be noted that 
NGC~4051 and NGC~5548 
fail to satisfy all our formal criteria for acceptability
assuming model {\it C(ii)}. Furthermore it should be noted that 
in the case of NGC~5548 the model predicts a stronger Fe $K\alpha$ 
emission than observed.
Caution is urged in when interpreting these results 
for NGC~4051, NGC~5548 and Mrk~509 as the best-fitting solutions 
are really only indicative of relatively
subtle curvature in the observed spectrum. Such curvature can arise 
as a result of curvature in the underlying continuum, spectral 
variability during the observation, and/or additional spectral 
components (see text and \S\ref{Sec:complex_cont}).

% Figure contained in PS file f13.ps
{\bf Figure~13:} 
The mean data/model ratios (in the {\it observer's} frame)
in the SIS and GIS for the 16 datasets satisfying our criteria
for acceptability assuming model {\it C(ii)}.
This plot was constructed by taking the error-weighted average of
the individual data/model ratios for each detector from
each of the appropriate observations.
These averages were then rebinned for display purposes.
Caution is urged when intrepreting such plots.
They potentally contain artificial features introduced by time-dependent
errors in the calibration of the instruments (such as slight offsets
in the gain of the detectors at the time of each observation),
along with real features due to errors in the calibration of the
instrument, as well as of astrophysical origin.
Furthermore such plots are dominated by the datasets with the
highest signal-to-noise ratio.

% Figure contained in PS file f14.ps
{\bf Figure~14:}
The mean data/model ratios in the 3--10~keV band (source frame)
obtained in Paper~II, and for models {\it B(ii)},
{\it C(i)} and {\it C(ii)} presented here.
The corresponding profile assuming model {\it B(i)} is 
almost identical to that assuming model {\it B(ii)} and hence is 
not shown. The $y$-axis for each panel covers the same dynamic 
range.
As described in the text, the 5--7~keV band is excluded from the 
spectral analysis presented here. The profiles were constructed 
from those datasets which satisfy our criteria for acceptability 
(\S\ref{Sec:acceptability}) assuming the best-fitting model and 
using the SIS data (only).
The vertical, dotted line shows the location of the K$\alpha$ 
flourescent line from neutral Fe (6.4~keV).
There are naturally slight differences between the profiles, in 
particular in the case of models {\it C(i)} \& {\it C(ii)}
(which include Fe line-emission for the photoionized-emitter).
However
it can be seen that the profiles obtained from the analysis presented 
here have the same characteristics of the profile found in Paper-II
Specifically, the peak of the line is consistent with that expected 
from Fe{\sc I}, the wings of the profile show a significant breath
(the FWHM of the instrumentatal resolution is $\sim 0.12$~keV at these 
energies), the profile appears assymmetric with more flux in the 
low-energy wing.
The implications are of such a profile in the context of models 
in which it is due to emission from the innermost regions of an accretion disk 
are discussed in detail in Paper-II.

% Figure contained in PS file f15.ps
{\bf Figure~15:}
As for Fig~4, but
for the 4 datasets for which the double-powerlaw
model 
(\S\ref{Sec:2plaw}; Table~11)
provides a significant improvement over the fit obtained 
assuming model {\it C(ii)} {\it and} in which both powerlaws of similar 
importance over a sizable fraction of the {\it ASCA} bandpass.
As described in the text, the goodness--of--fit of 
6 other datasets are also significantly improved assuming such a 
model. However in these cases
the second powerlaw component makes a significant
contribution only at either the very lowest 
or very highest energies.
It should be noted that best-fitting solution to IC~4329A fails to satisfy 
our formal criteria for acceptability.

% Figure contained in PS file f16.ps
{\bf Figure~16:}
The solid contours show the 90\% confidence limits (for 6 interesting 
parameters) in the $N_{H,z}$--$U_X$ plane for the individual observations of 
NGC~3783, NGC~4151 and MCG-6-30-15 assuming model {\it C(ii)}.
The dashed contours show the corresponding limits for NGC~4151
when the optically thin plasma with $kT \sim 0.7$~keV is 
also included in the spectral model (\S\ref{Sec:thermal}).
The dotted curves show the optical depths of the 
O{\sc vii} and O{\sc viii} edges in this region of parameter-space.
The implications of the changes in the state of the ionized gas in 
each source, particularly with regard to the corresponding behaviour of
the illuminating continuum, is discused in \S\ref{Sec:disc-multi}.

% Figure contained in PS file f17.ps
{\bf Figure~17:}
The derived continuum luminosity,
$L_X$, in the 0.1--10~keV band (after correcting for absorption)
versus the parameters of the ionized gas assuming model {\it C(ii)}.
The solid symbols represent the 
datasets for which the model satisfies all our criteria for acceptability.
The open symbols represent the datasets which do not satisfy 
our formal criteria, but for which we consider the measurements of 
$N_{H,z}$ and $U_X$ to be reasonable estimates.
As can be seen from the upper panel, there is no clear relation 
between $N_{H,z}$ and $L_X$. However from the lower panel it can be seen that
there is some 
indication that $U_X \propto L_X$, which would 
have implications for the distribution of gas in these sources.
However it should be noted that 
the majority of the sources considered here 
in which $U_X$ is well constrained cover only a decade in luminosity
($10^{43}$--$10^{44}\ {\rm erg\ s^{-1}}$).
The addition of the two quasars PG~1114+445 and MR~2251-178
to the plot argues against any obvious relation between
$U_X$ and $L_X$ (see \S\ref{Sec:disc-hagai}).

% Figure contained in PS file f18.ps
{\bf Figure~18:}
The column density, $N_{H,z}$, derived assuming model {\it C(ii)} versus 
the mean ratio of the flux observed at 125~nm to that at 220~nm, 
$(f_{125}/f_{220})_{obs}$ (from Table~1).
It can be seen that the majority of sources are consistent with 
$2 \lesssim (f_{125}/f_{220})_{obs} \lesssim 5$, whilst 3 datasets 
(NGC~3227, MCG-6-30-15(1,2)) have values a factor $\sim 10$ lower.
Clearly 
$(f_{125}/f_{220})_{obs} = (f_{125}/f_{220})_{int}\ e^{-\Delta \tau}$,
where $(f_{125}/f_{220})_{int}$ is the intrinsic flux ratio, and 
$\Delta \tau$ ($= \tau_{125} - \tau_{220}$) the difference in
optical depths at 125~nm and 220~nm.
We stress that $(f_{125}/f_{220})_{int}$ is not necessarily a good 
indication of the underlying UV continuum (see text). 
The curves show $(f_{125}/f_{220})_{obs}$ assuming 
$(f_{125}/f_{220})_{int} = 3.1$ and (from top to bottom)
$\Delta \tau = 0.0$, $0.005N_{21}$, $0.05N_{21}$ (dotted) 
and $\Delta \tau = 0.5N_{21}$ (dashed), where $N_{21} = N_{H,z}/10^{21}$.
It can be seen that NGC~3227 and MCG-6-30-15(1,2) are 
consistent with $(f_{125}/f_{220})_{int}$ similar to the other sources
if their fluxes at 125~nm and 220~nm are absorbed by a column density
similar to that observed in the X-ray band and $\Delta \tau = 0.5N_{21}$.
A comparison of such a value of $\Delta \tau$ with that expected from
'standard' dust models is provided in the text.

\clearpage
\begin{figure*}[h]
\plotone{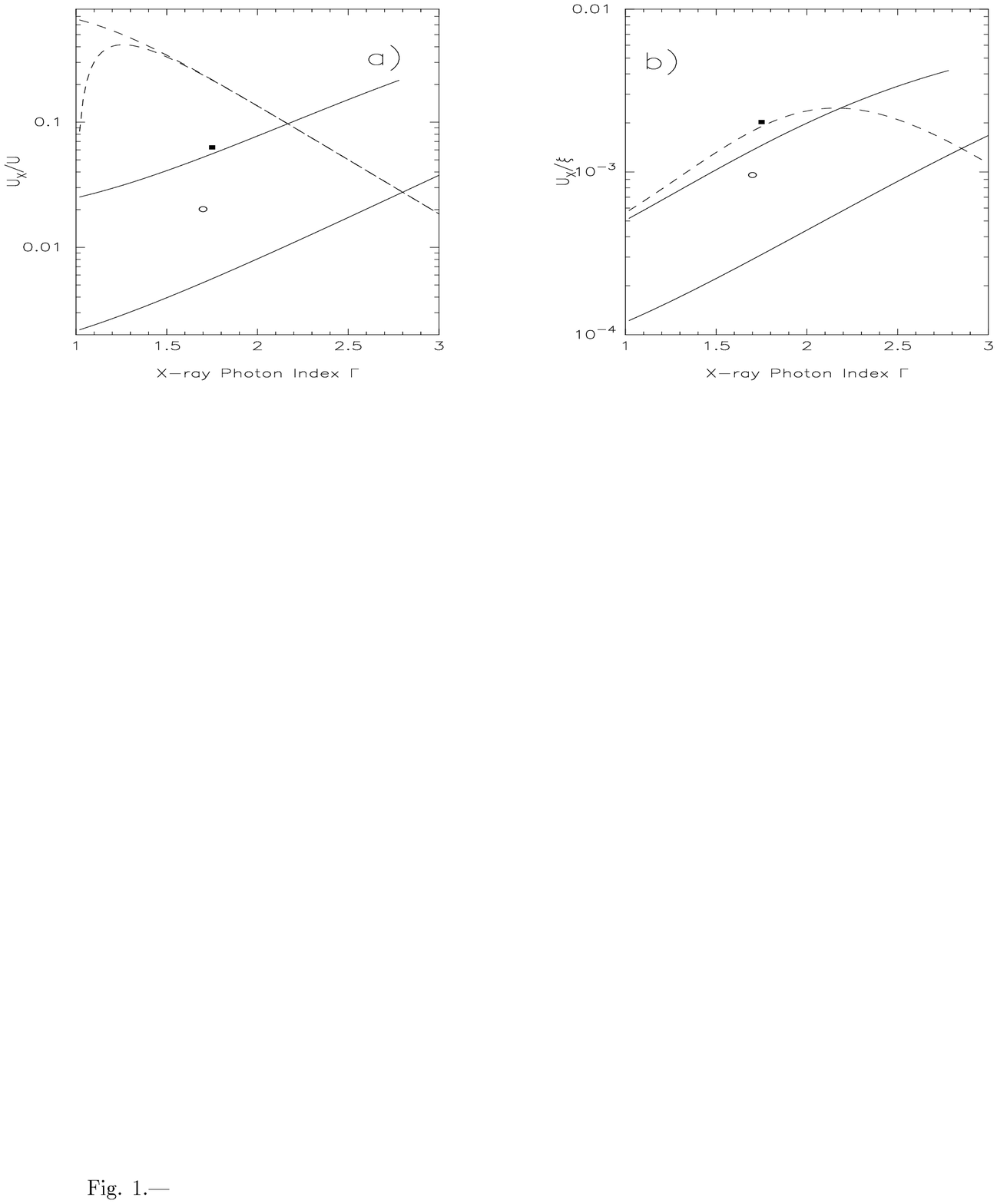}
\end{figure*}
\clearpage

\begin{figure*}[h]
\plotone{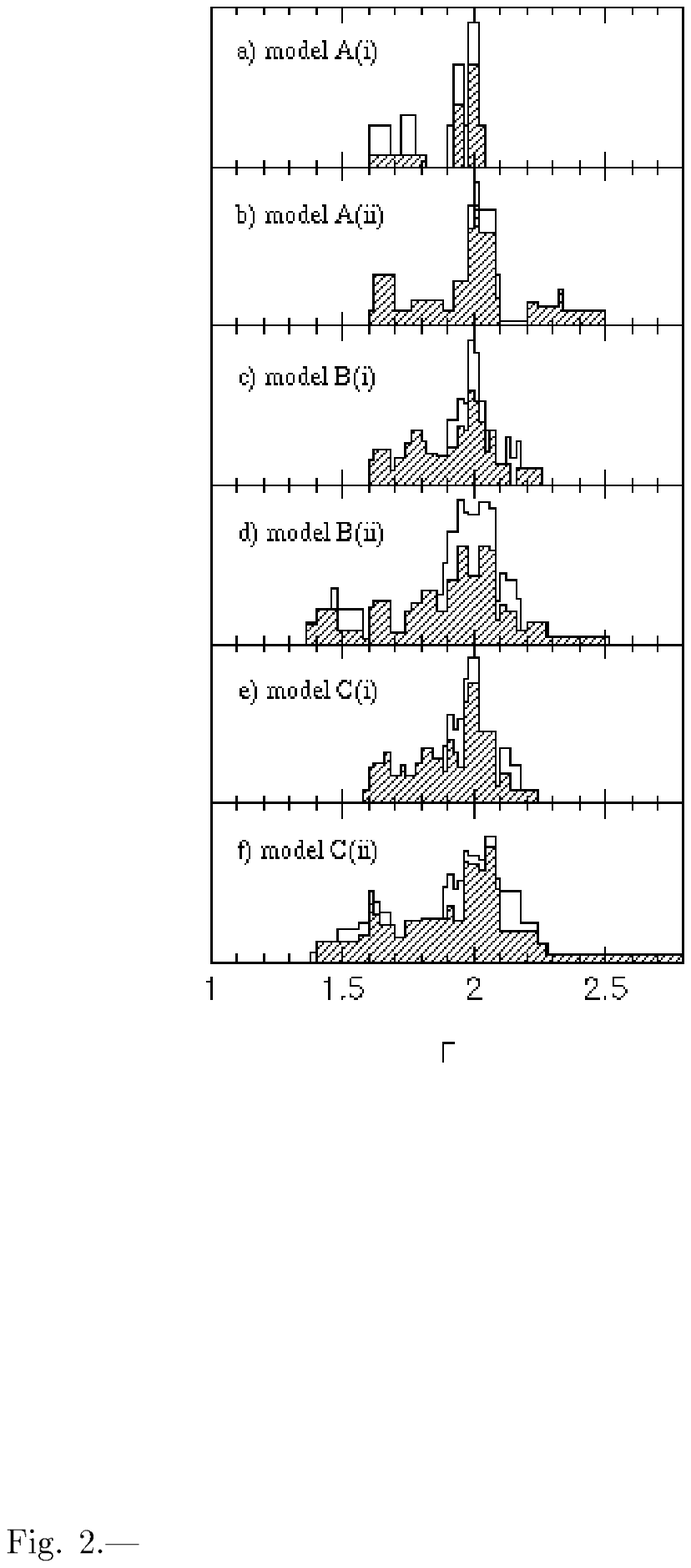}
\end{figure*}
\clearpage

\begin{figure*}[h]
\plotone{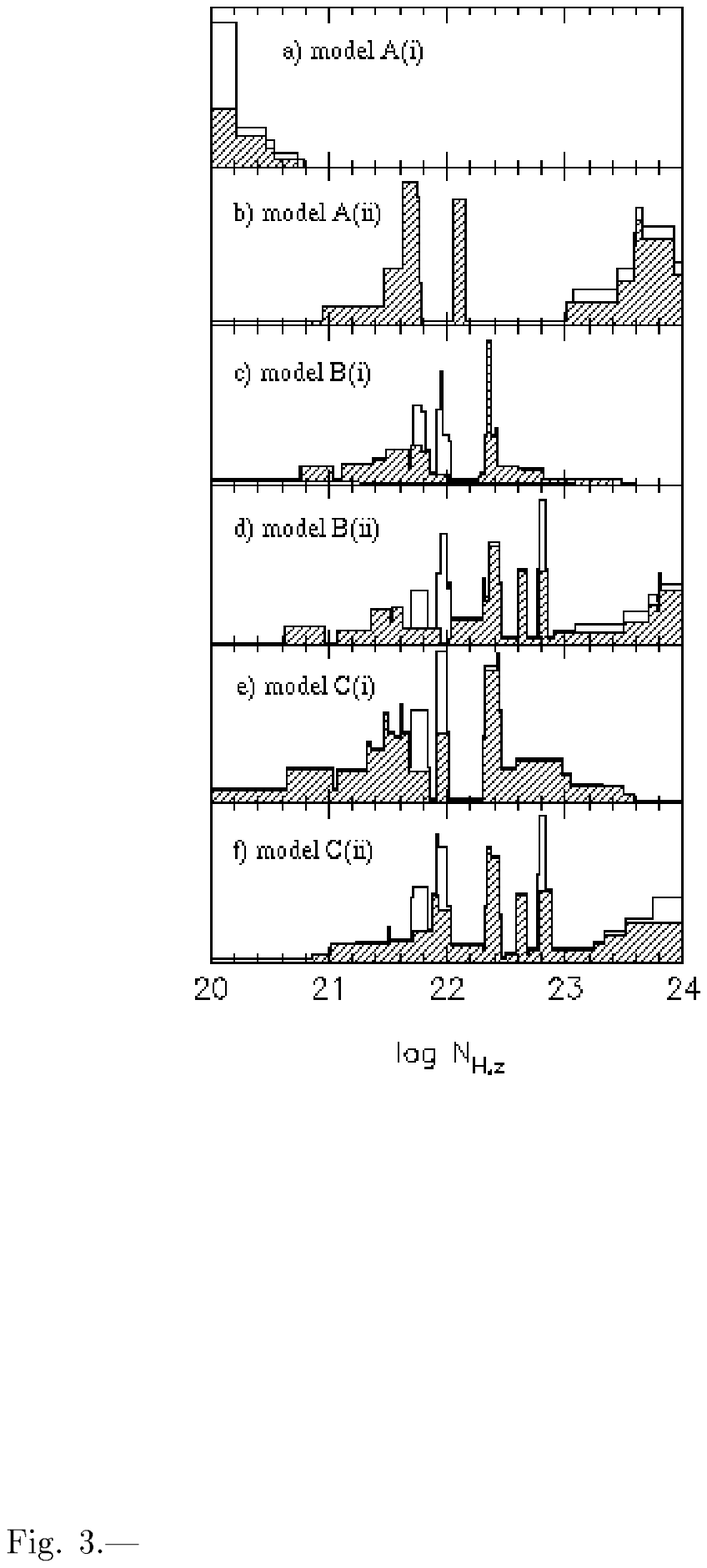}
\end{figure*}
\clearpage

\begin{figure*}[h]
\plotone{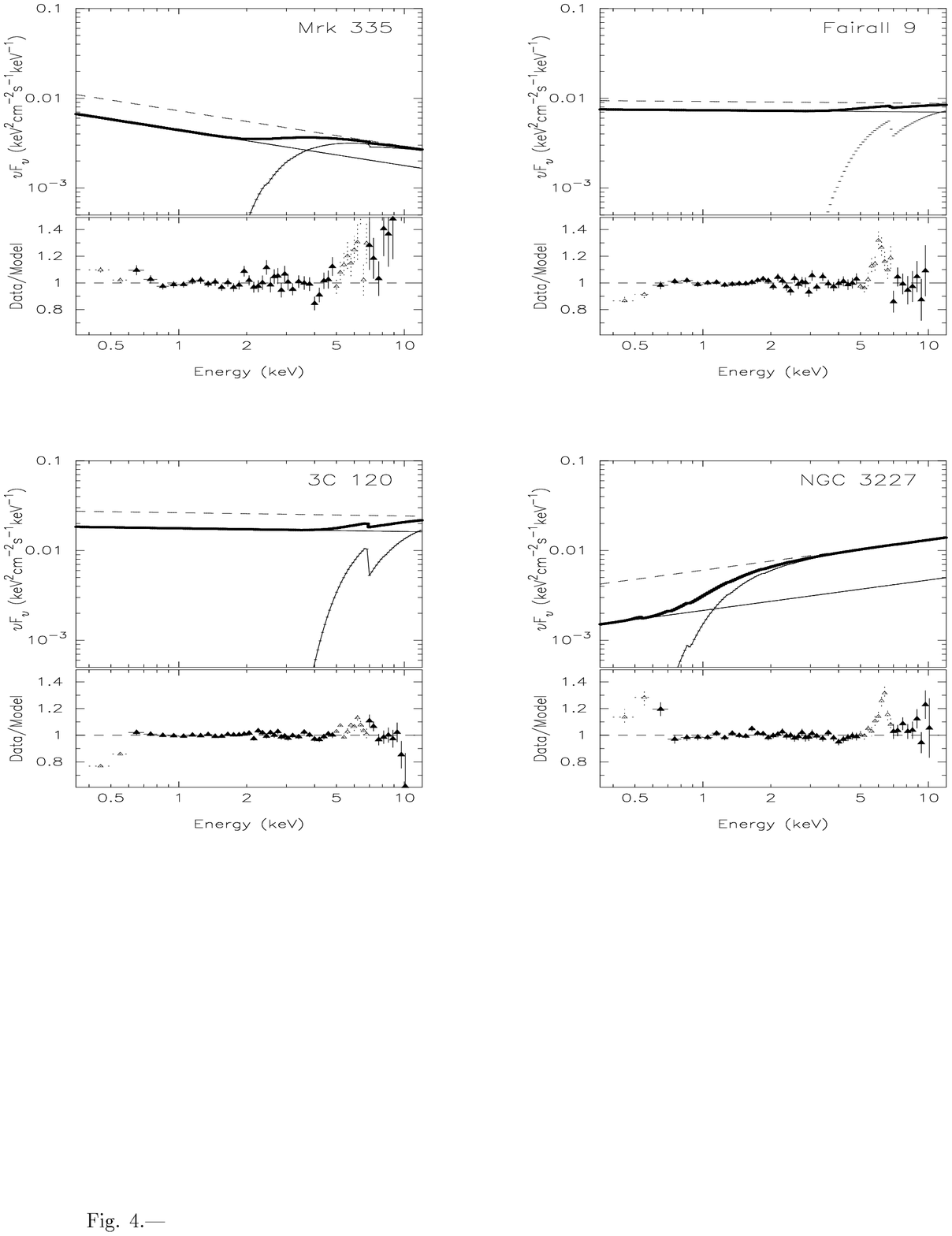}
\end{figure*}
\clearpage

\begin{figure*}[h]
\plotone{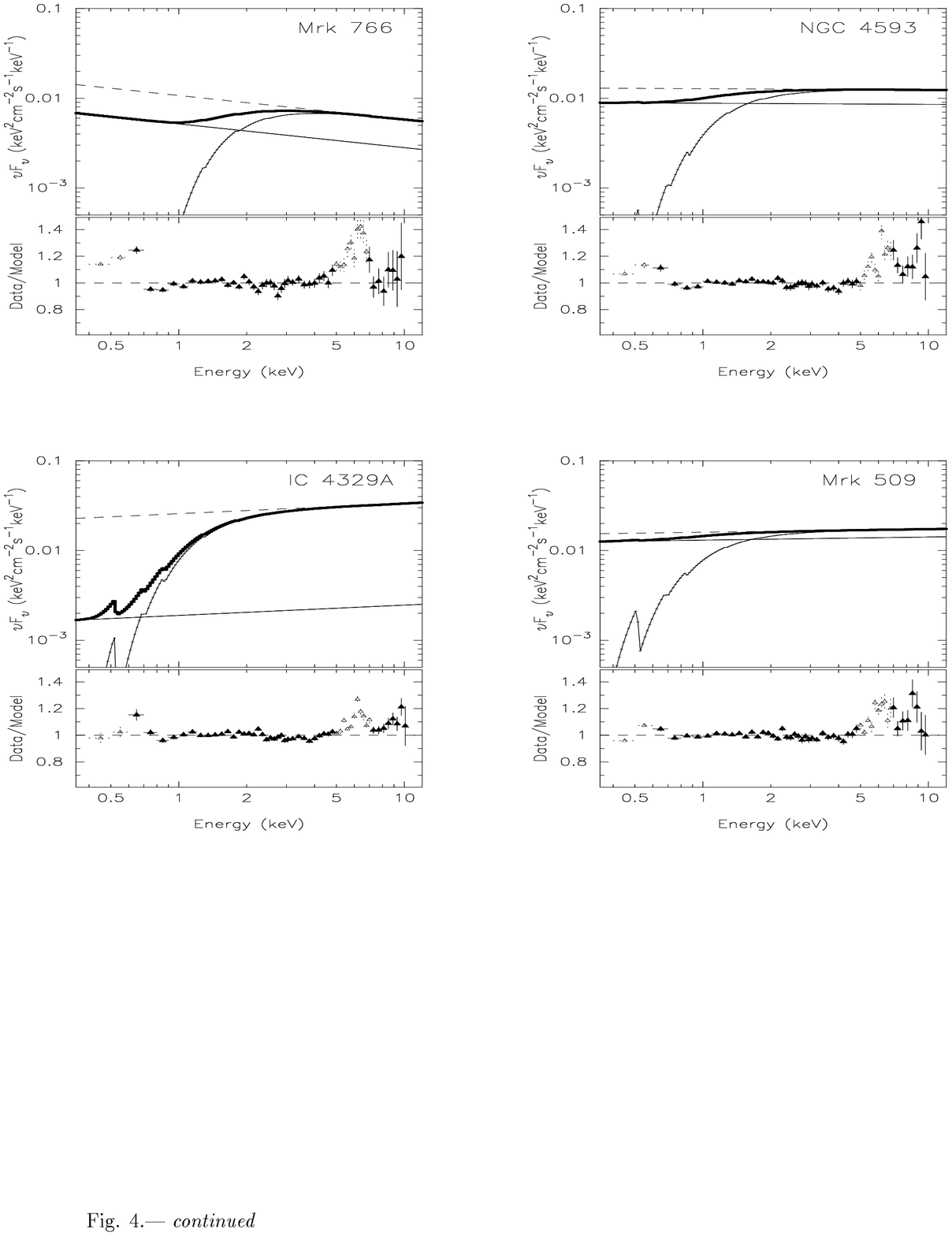}
\end{figure*}
\clearpage

\begin{figure*}[h]
\plotone{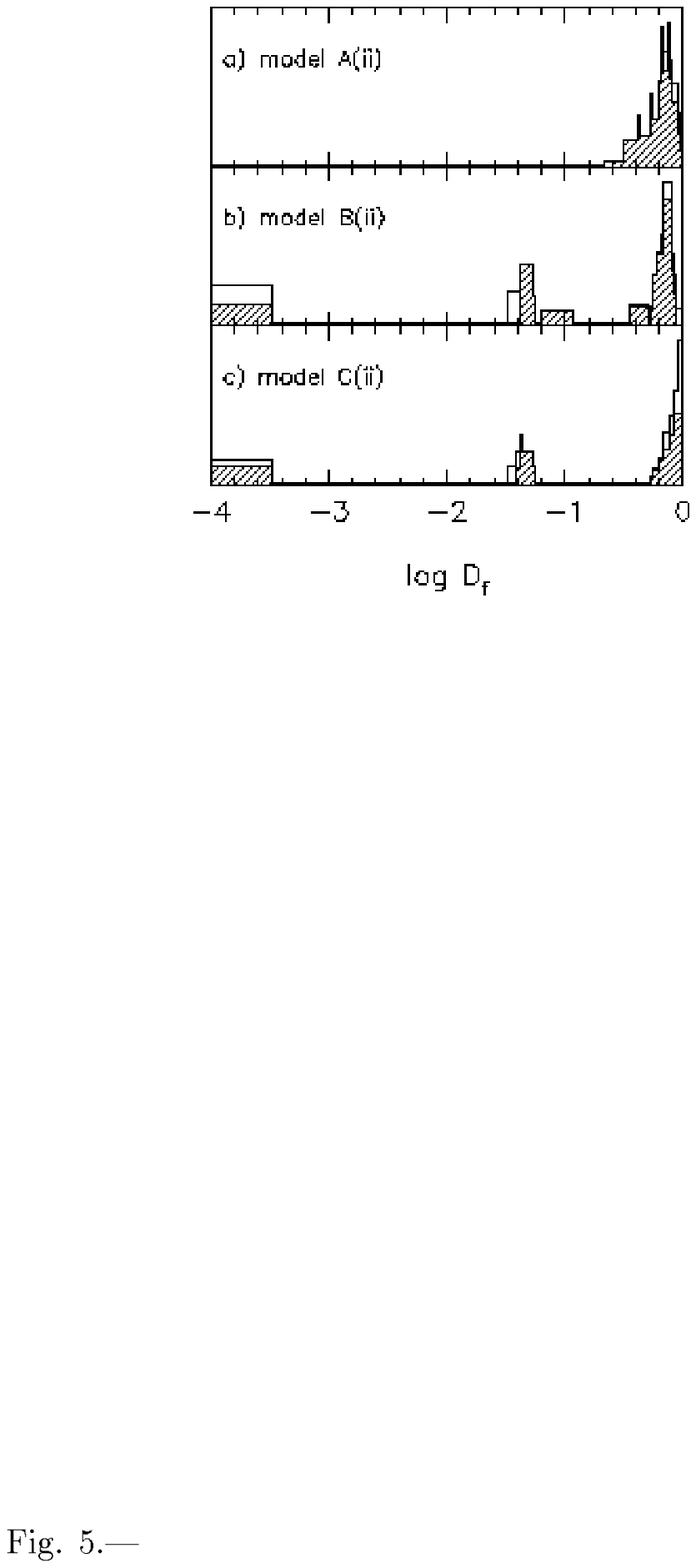}
\end{figure*}
\clearpage

\begin{figure*}[h]
\plotone{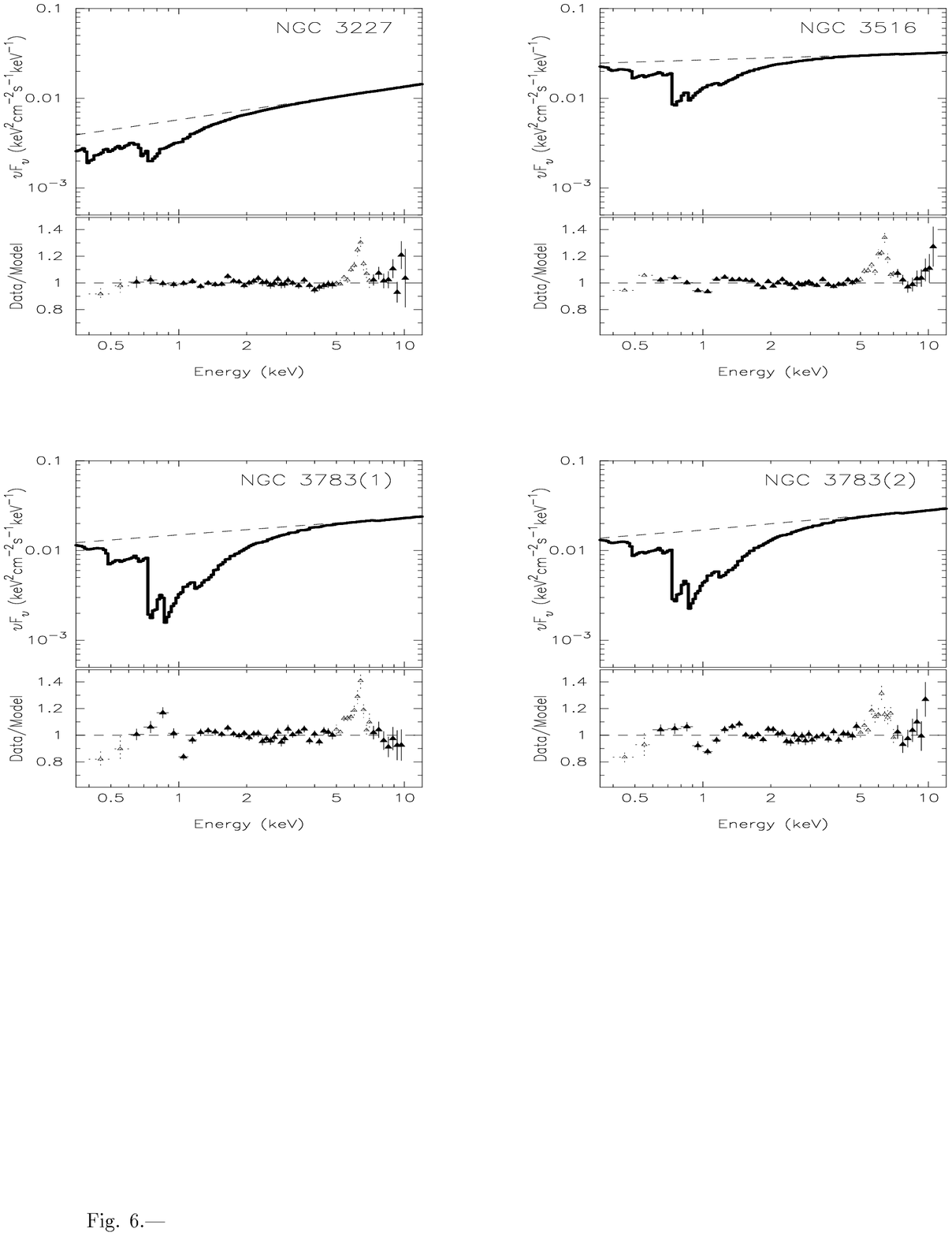}
\end{figure*}
\clearpage

\begin{figure*}[h]
\plotone{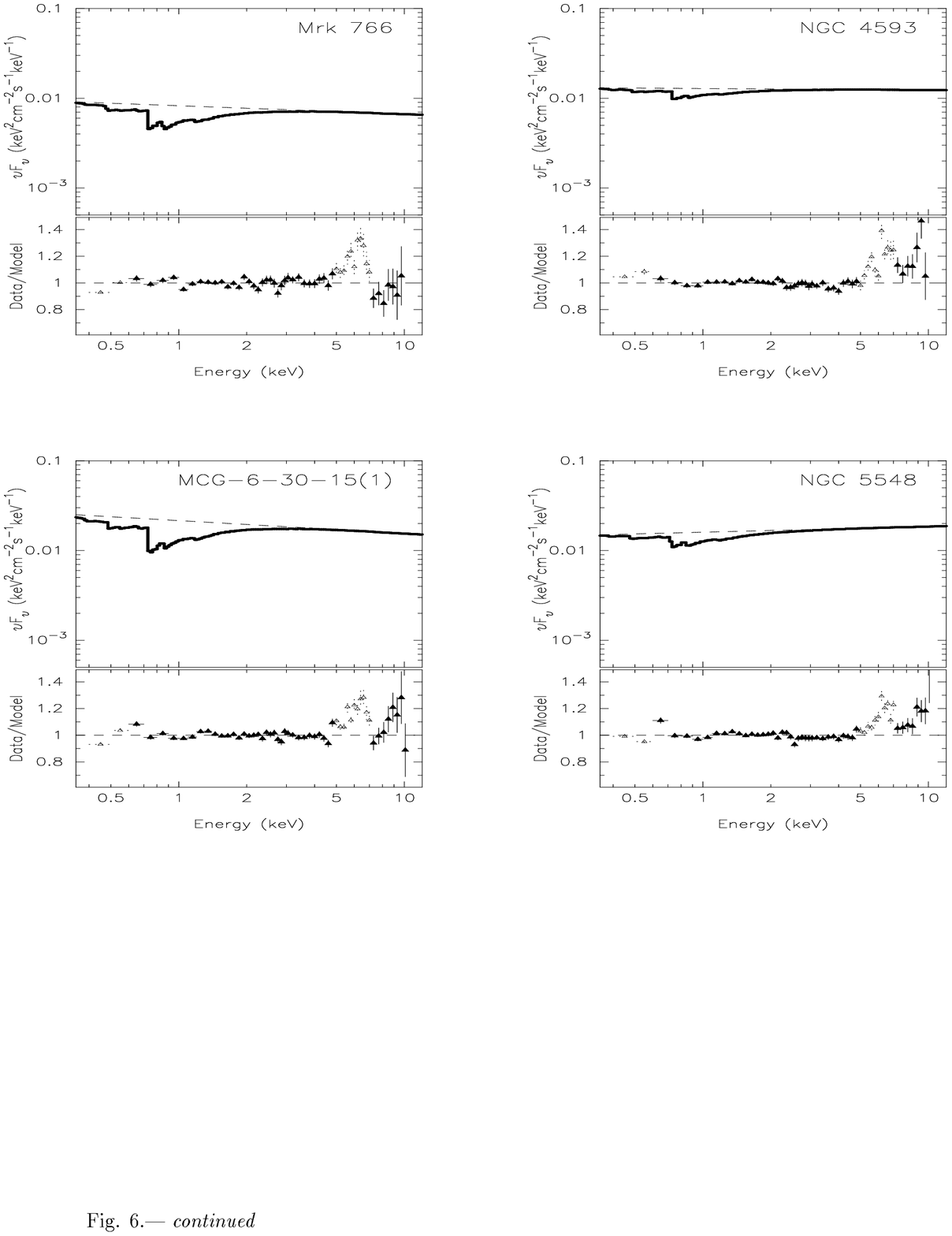}
\end{figure*}
\clearpage

\begin{figure*}[h]
\plotone{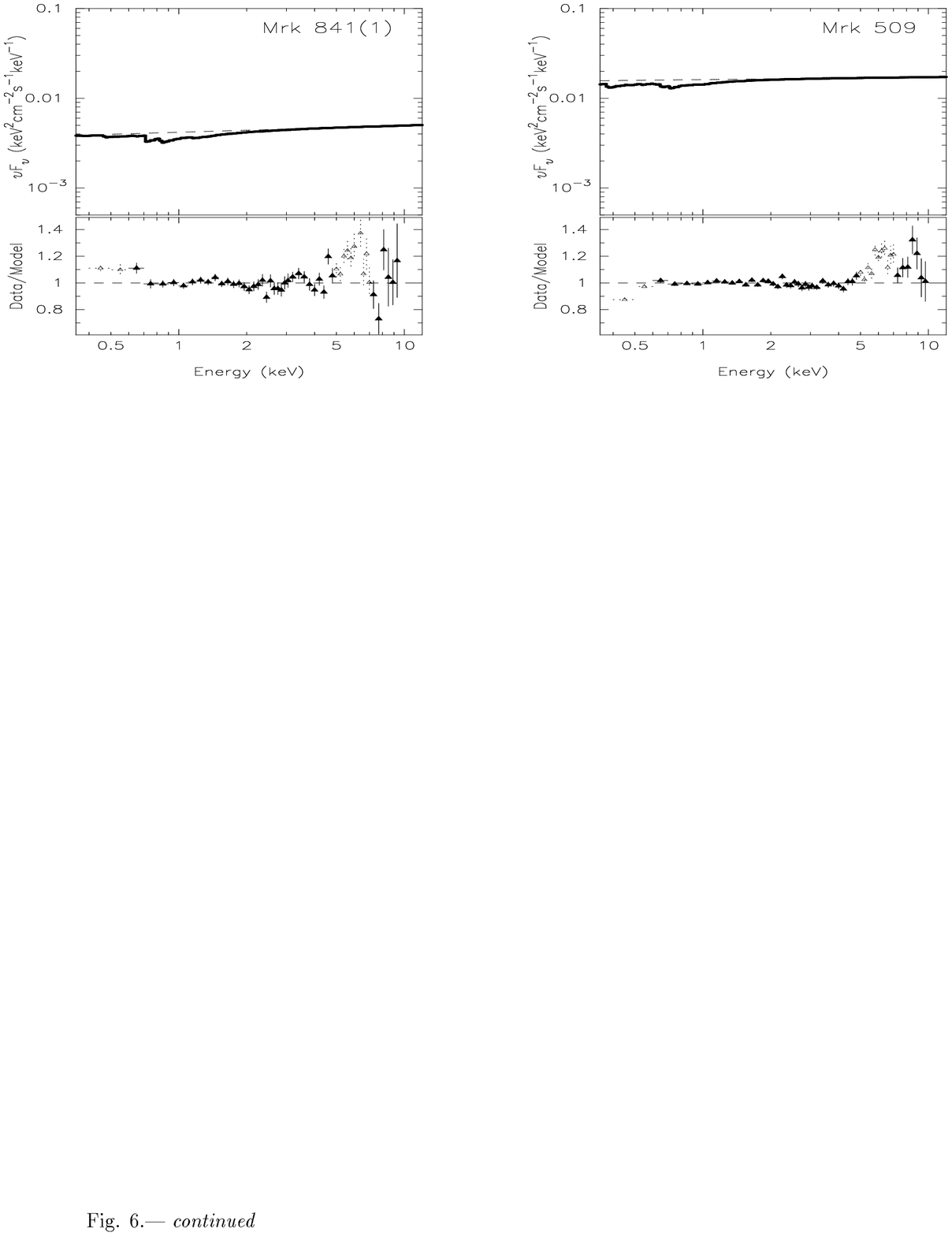}
\end{figure*}
\clearpage

\begin{figure*}[h]
\plotone{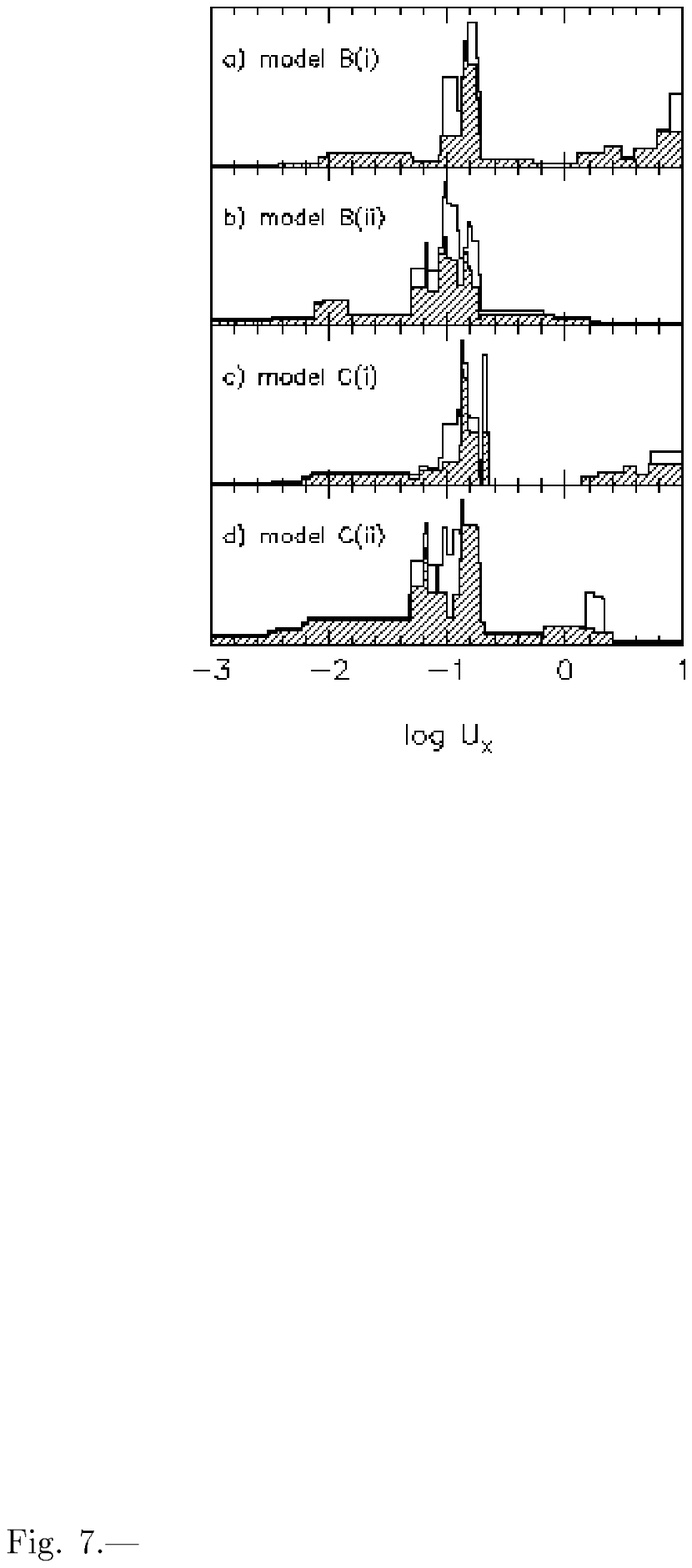}
\end{figure*}
\clearpage

\begin{figure*}[h]
\plotone{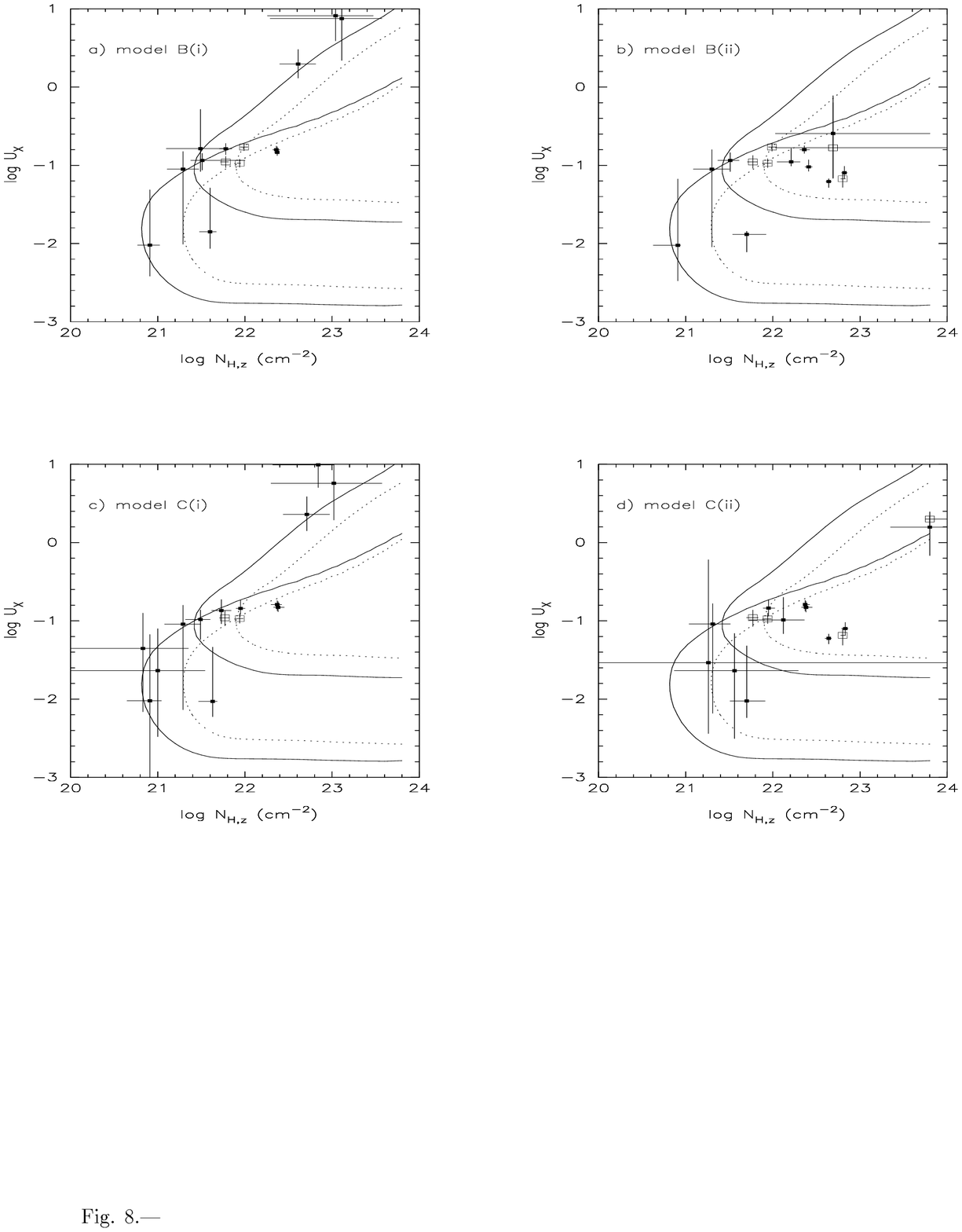}
\end{figure*}
\clearpage

\begin{figure*}[h]
\plotone{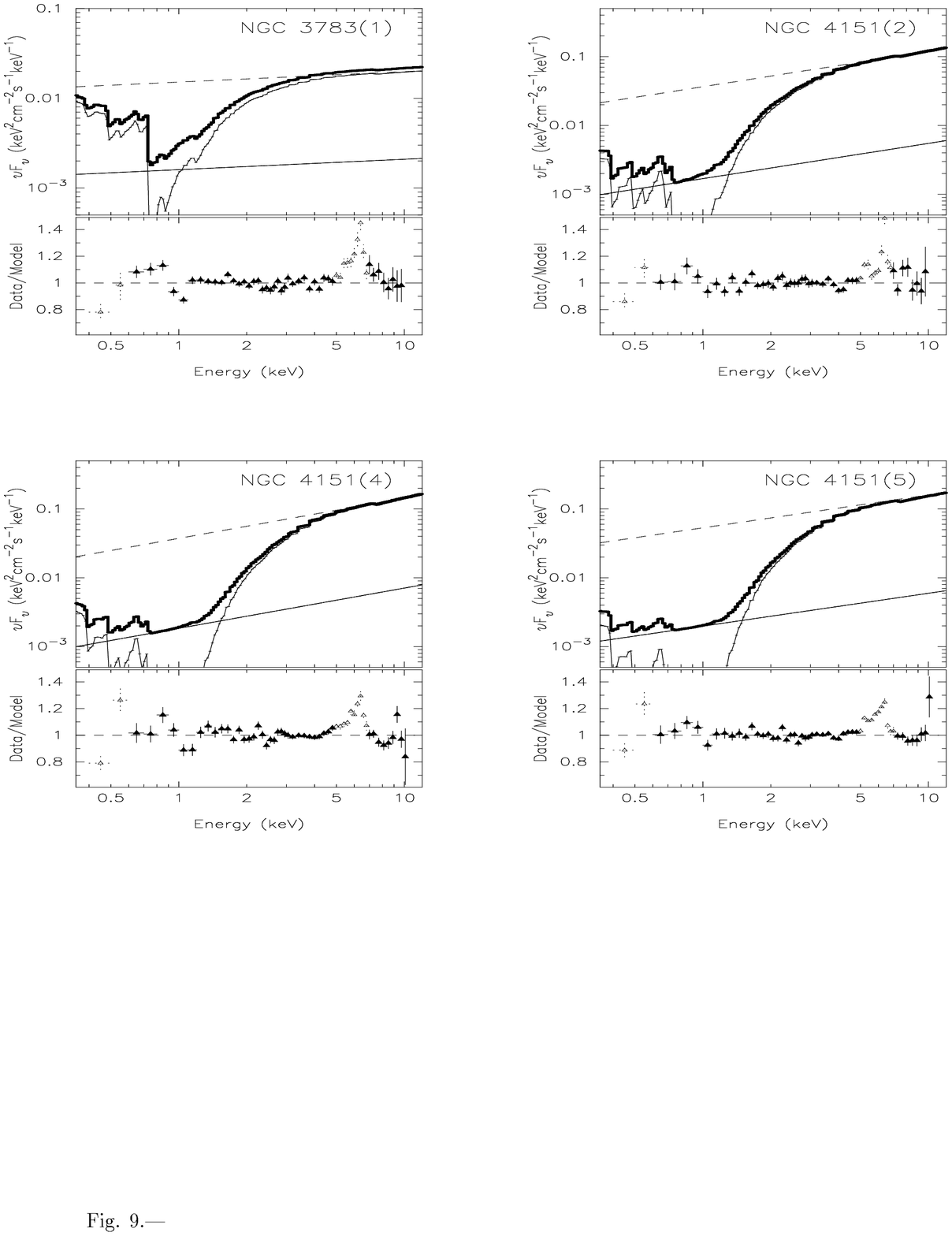}
\end{figure*}
\clearpage

\begin{figure*}[h]
\plotone{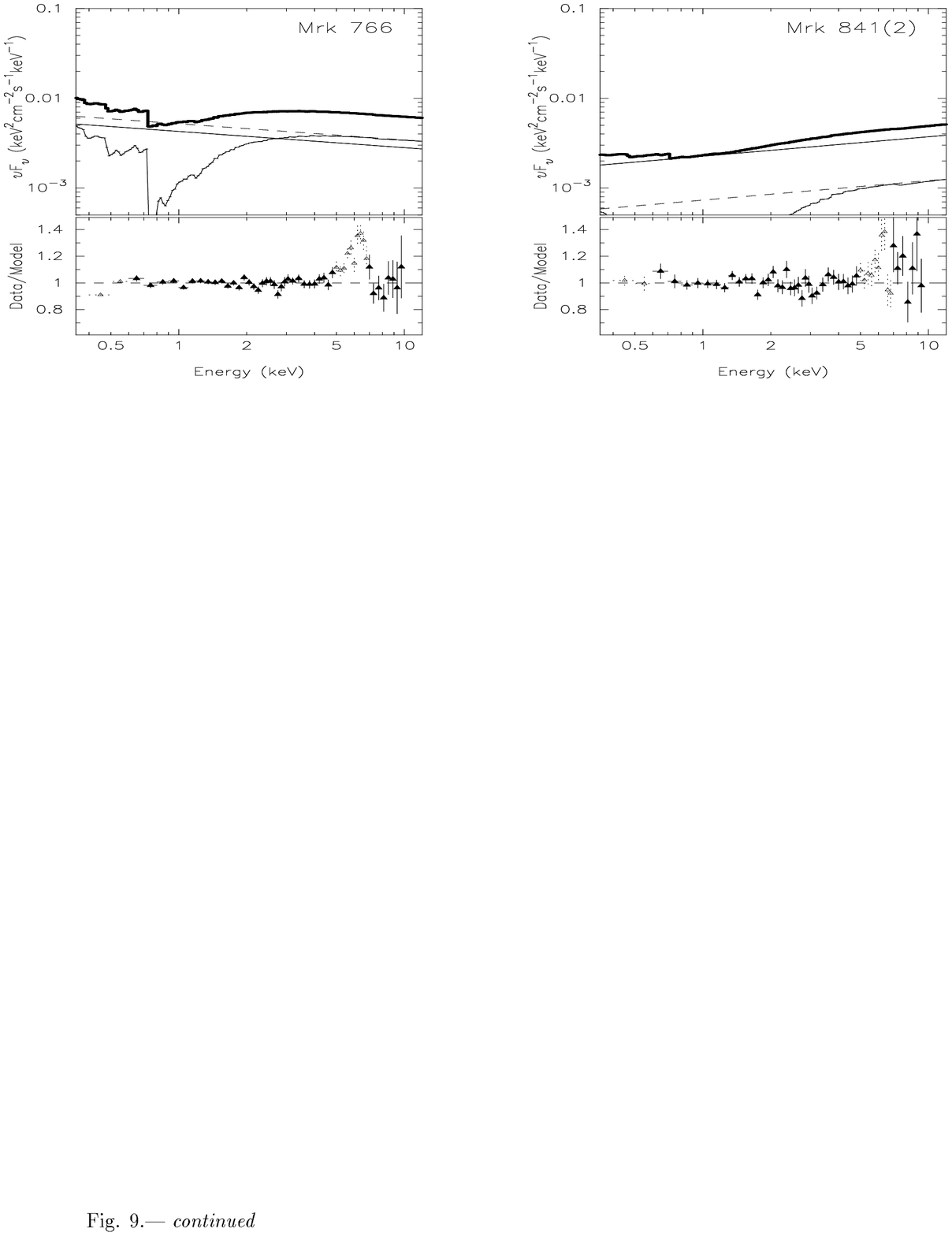}
\end{figure*}
\clearpage

\begin{figure*}[h]
\plotone{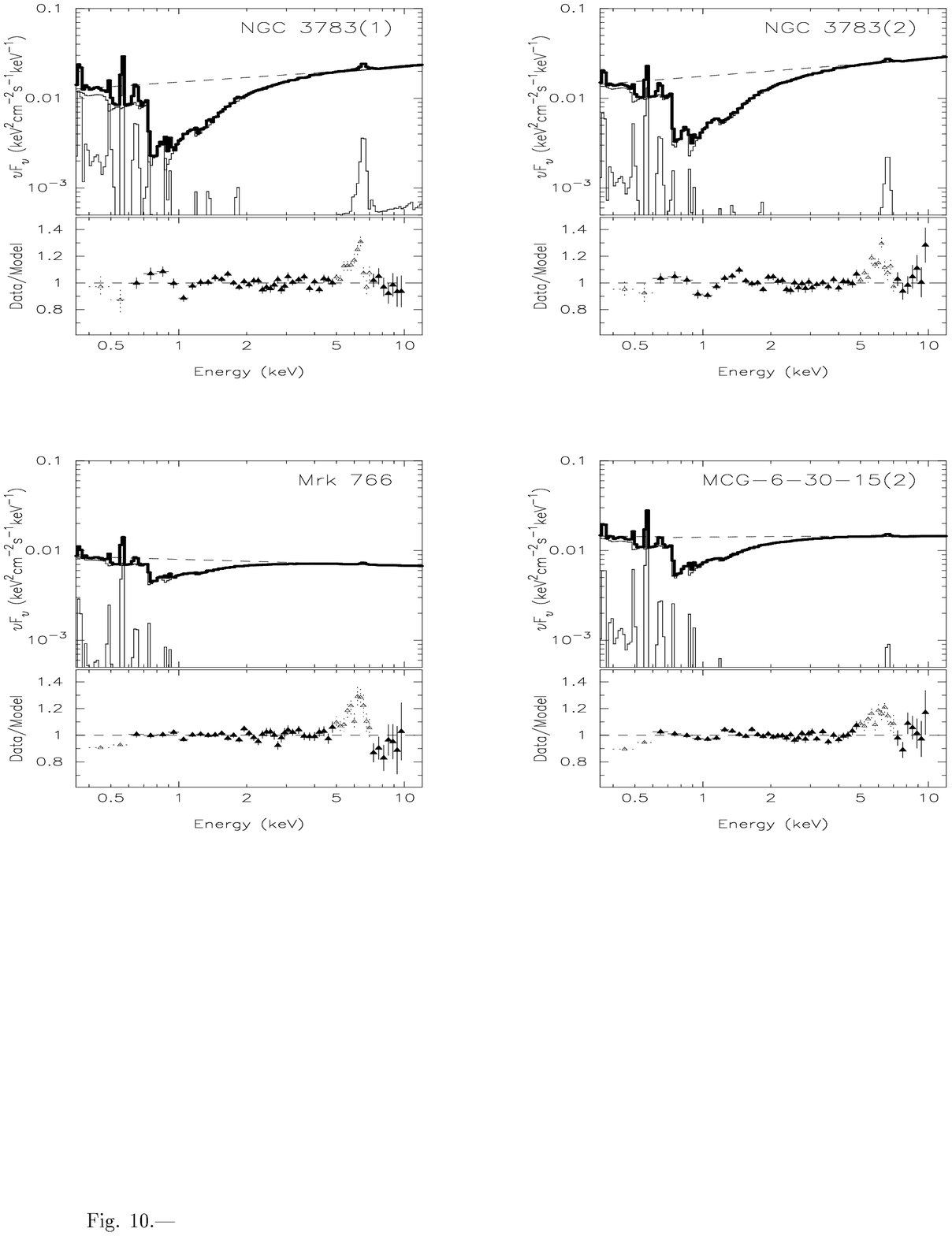}
\end{figure*}
\clearpage

\begin{figure*}[h]
\plotone{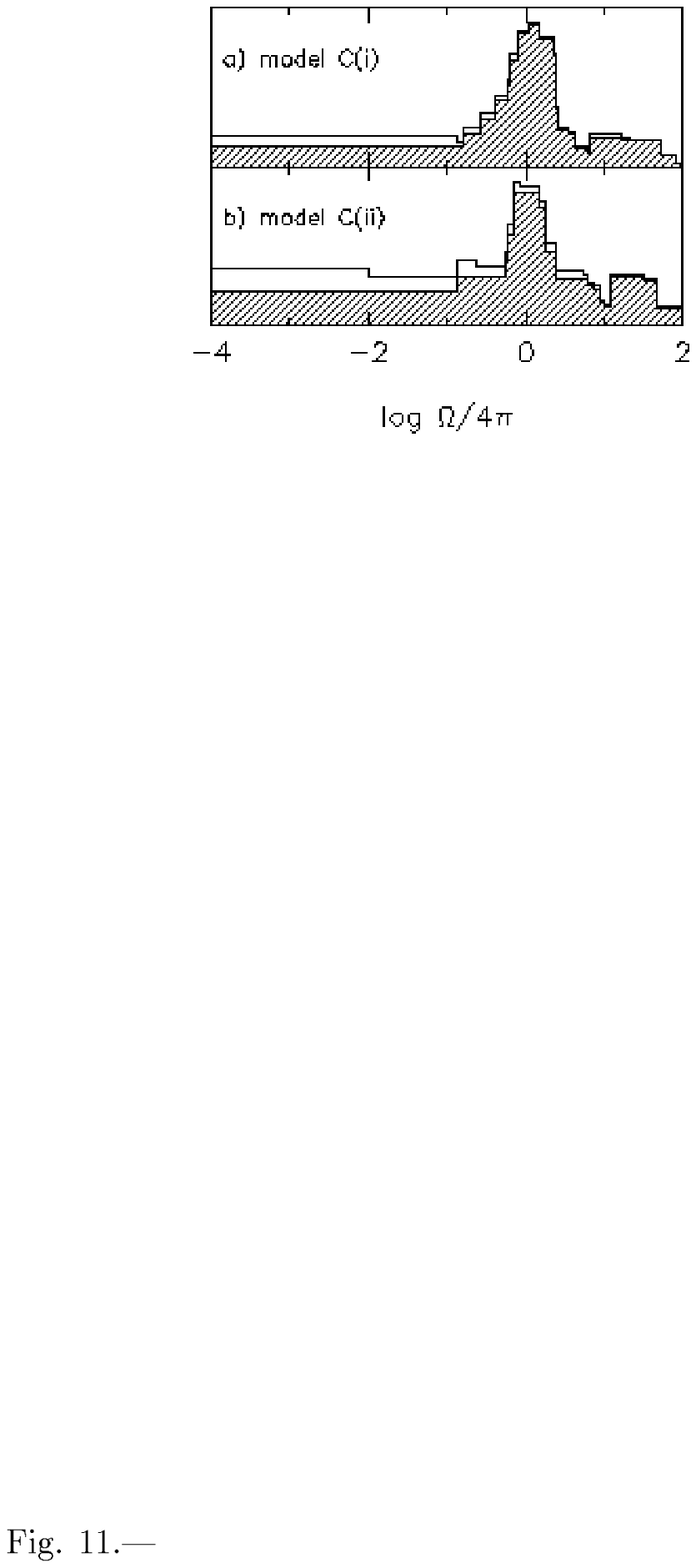}
\end{figure*}
\clearpage

\begin{figure*}[h]
\plotone{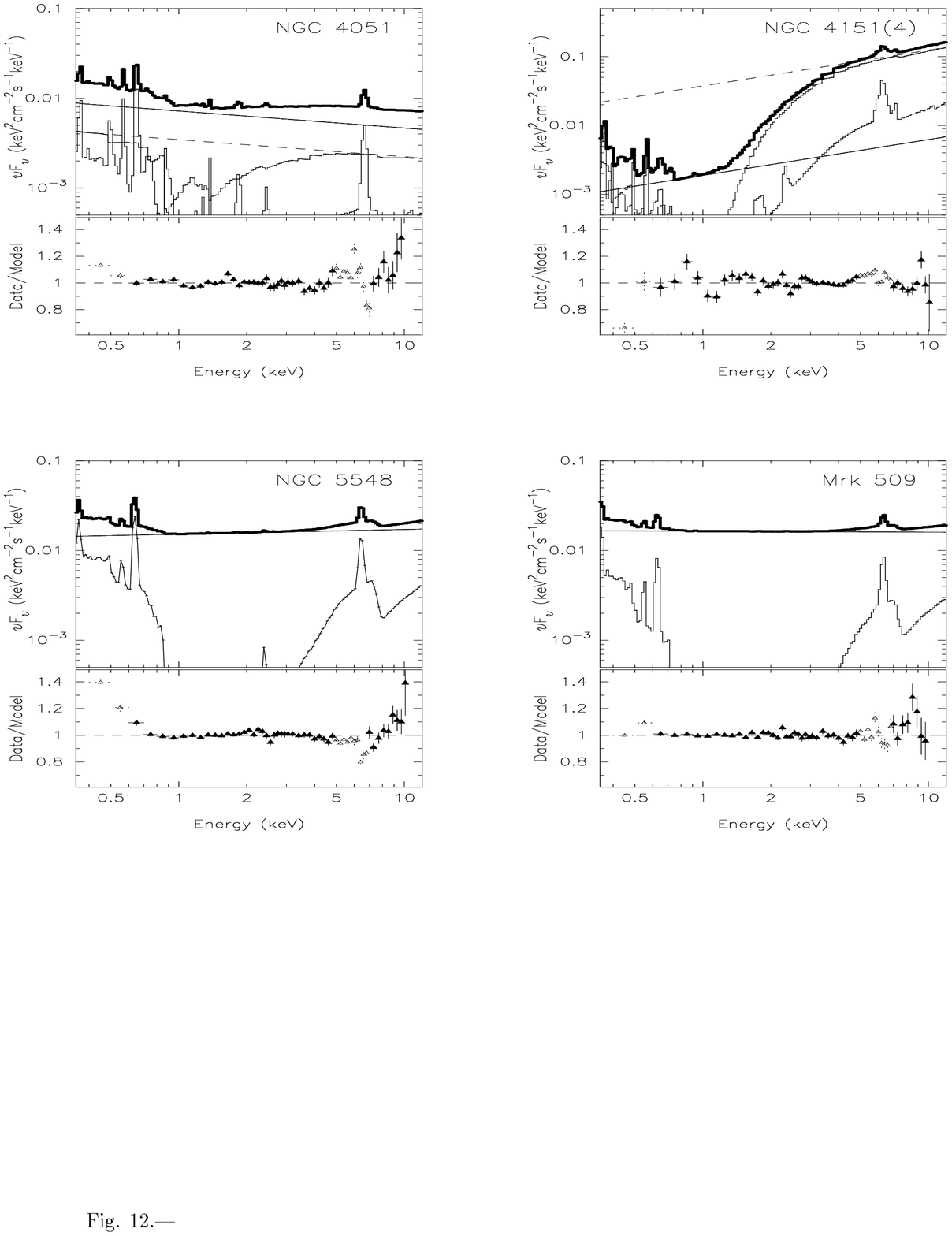}
\end{figure*}
\clearpage

\begin{figure*}[h]
\plotone{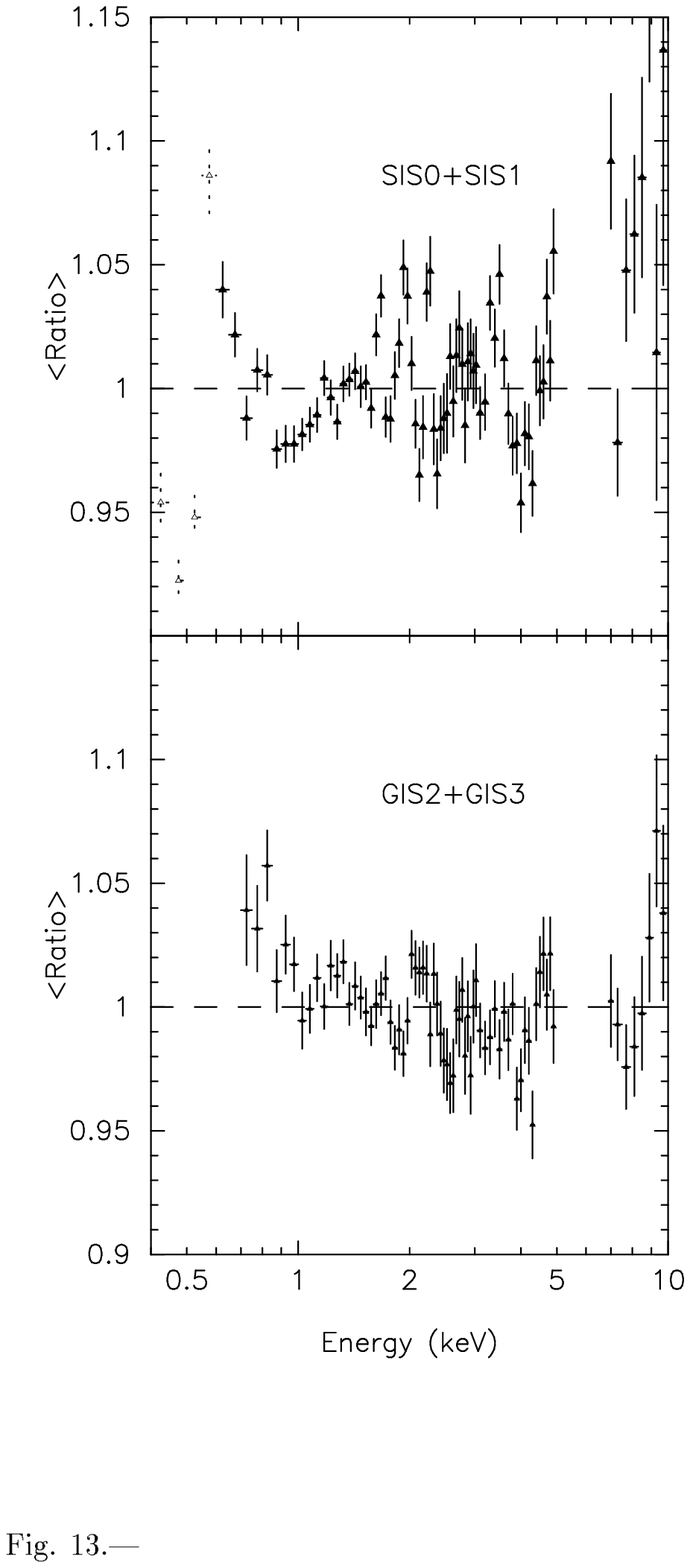}
\end{figure*}
\clearpage

\begin{figure*}[h]
\plotone{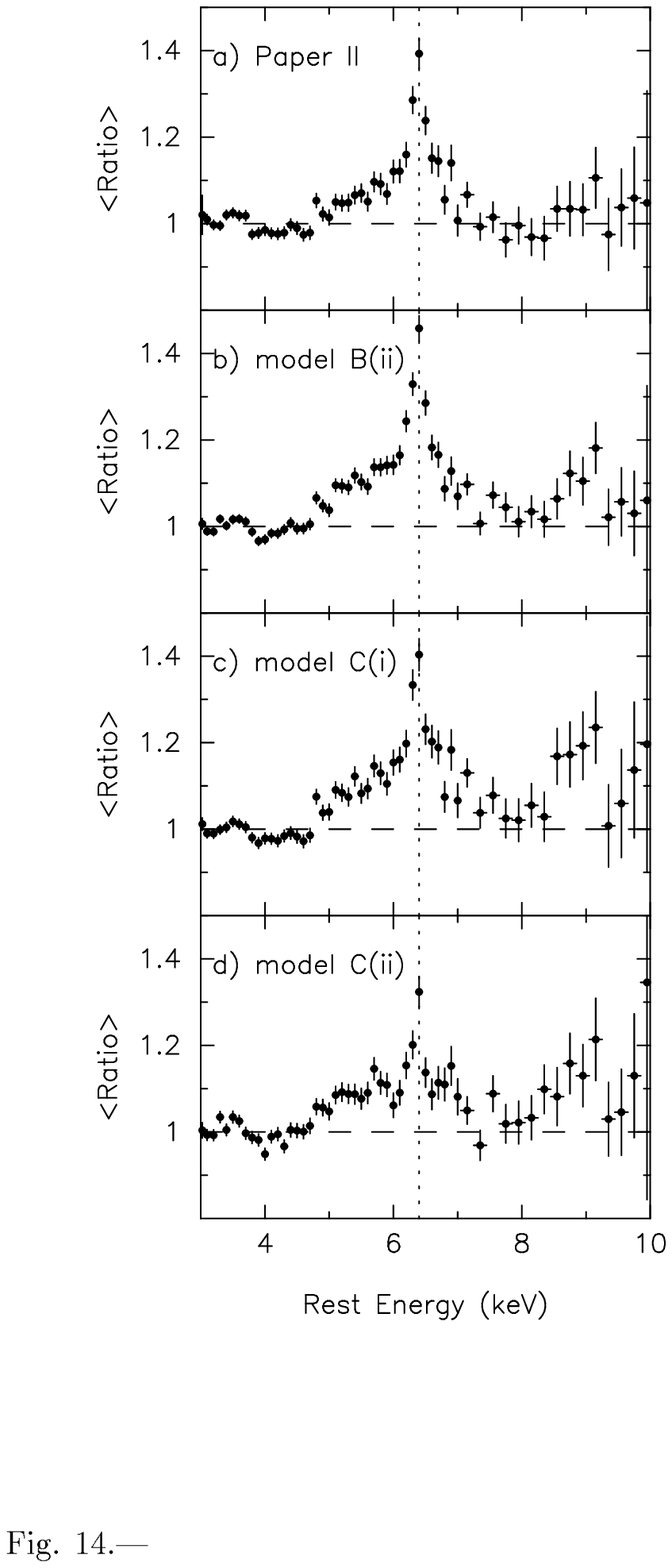}
\end{figure*}
\clearpage

\begin{figure*}[h]
\plotone{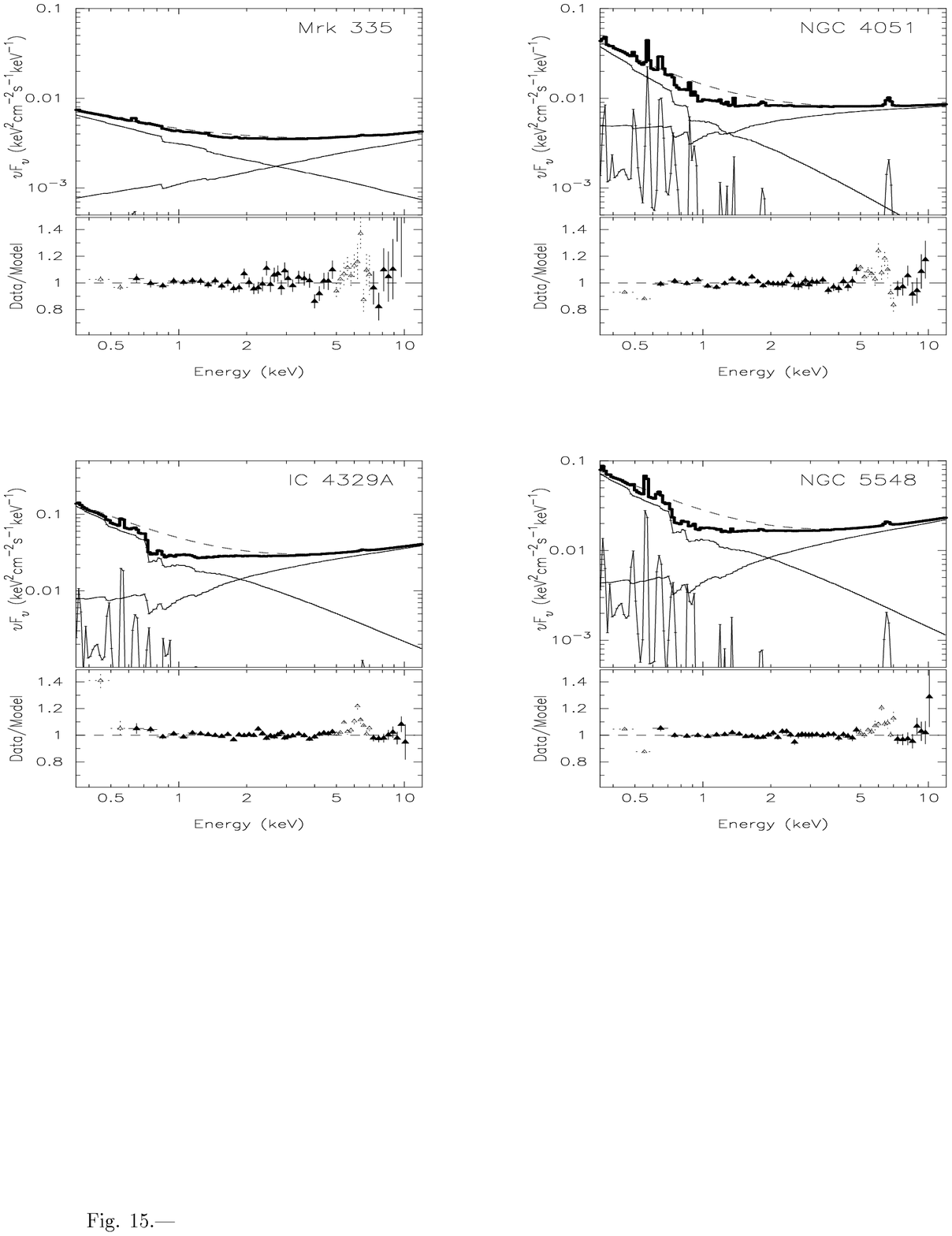}
\end{figure*}
\clearpage

\begin{figure*}[h]
\plotone{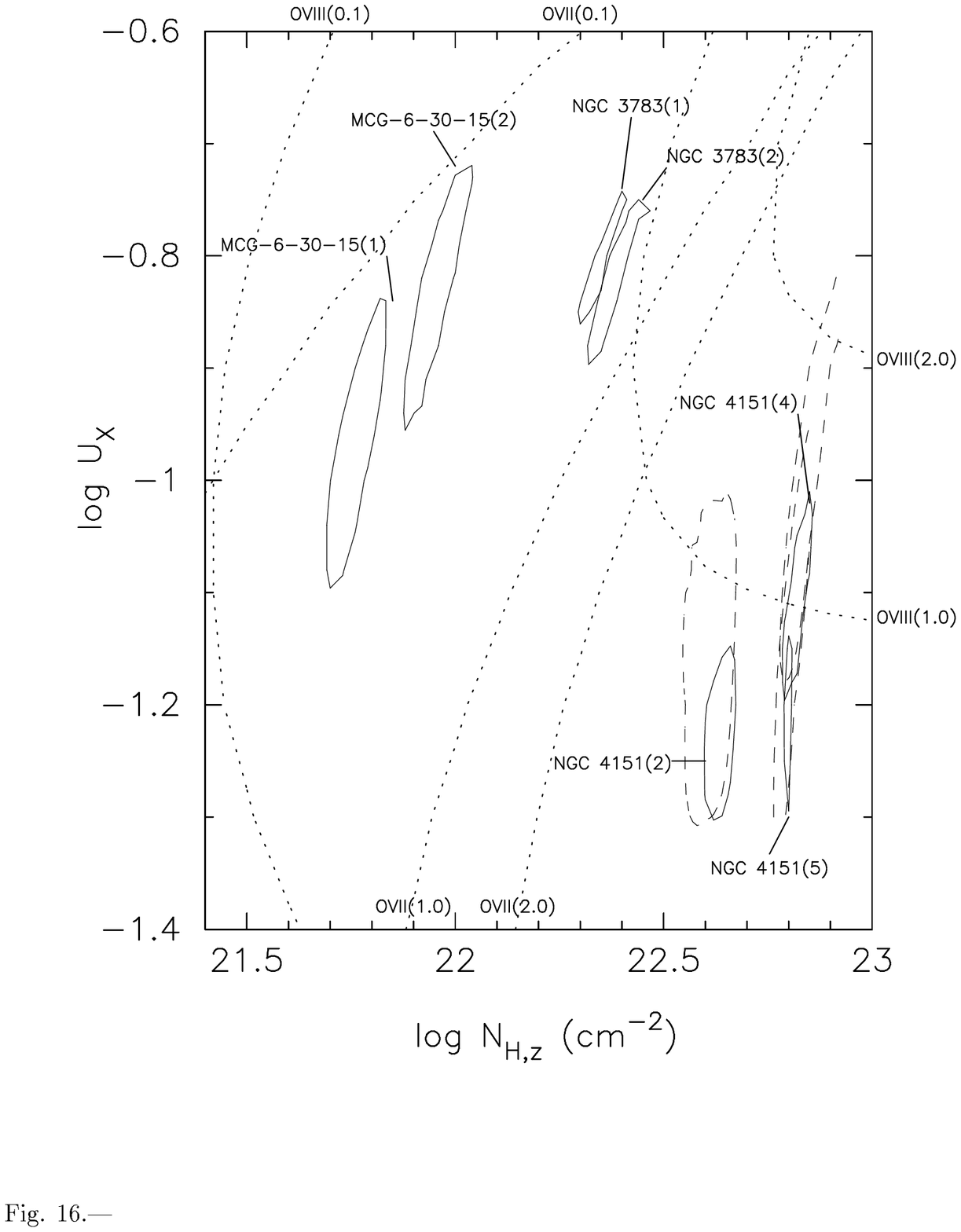}
\end{figure*}
\clearpage

\begin{figure*}[h]
\plotone{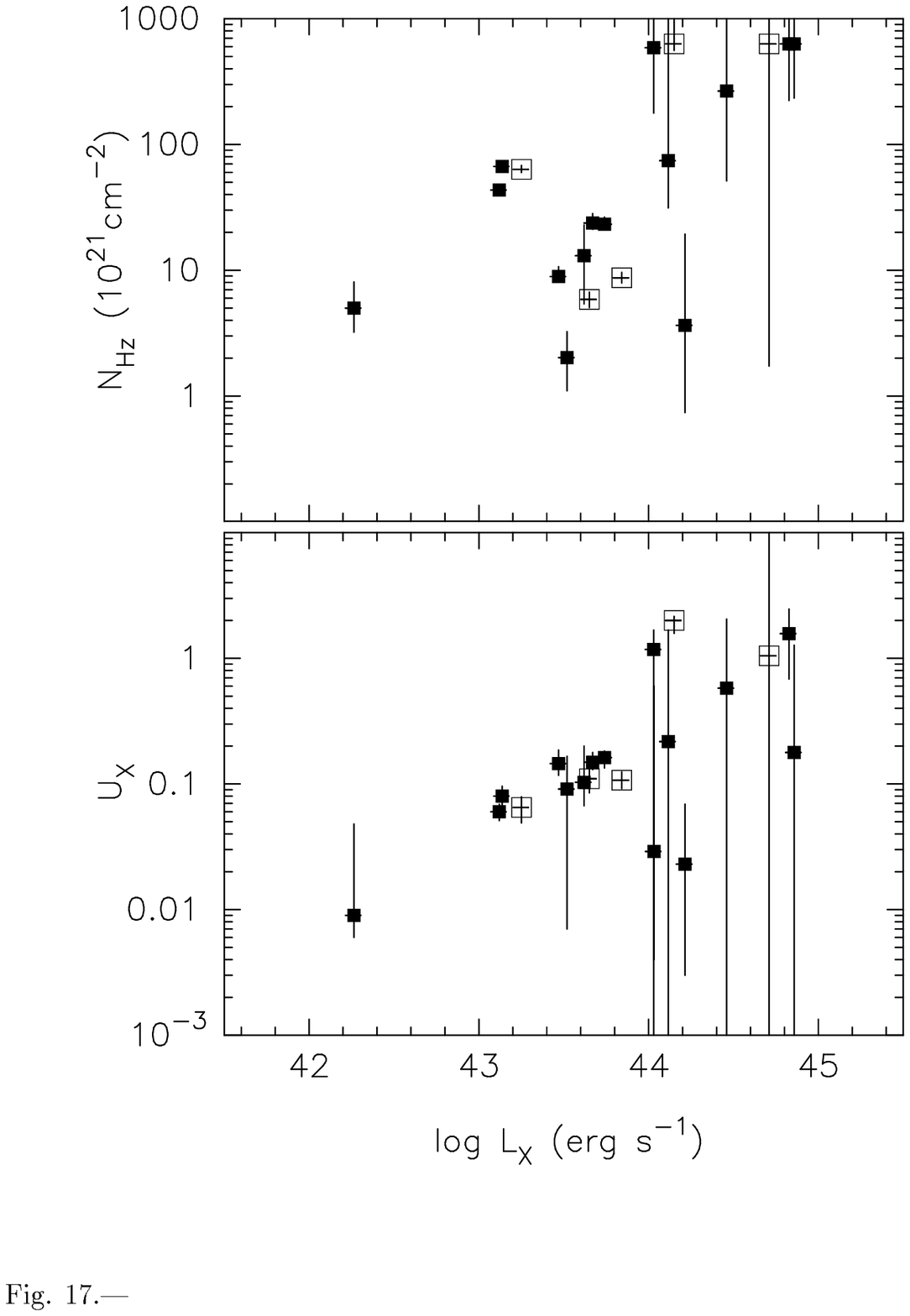}
\end{figure*}
\clearpage

\begin{figure*}[h]
\plotone{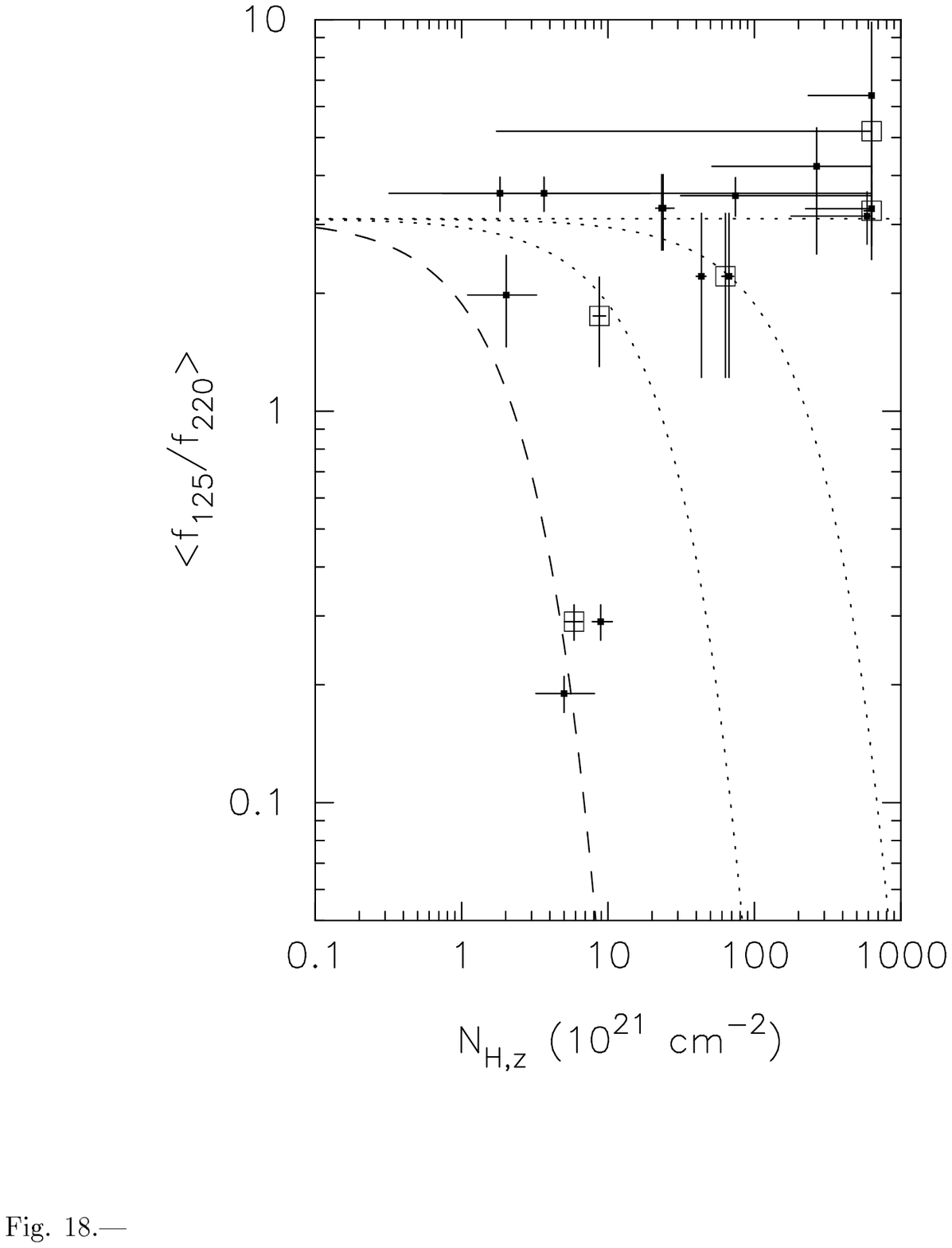}
\end{figure*}
\clearpage

\end{document}